\documentclass[10pt,journal,compsoc]{IEEEtran}

\ifCLASSOPTIONcompsoc
  \usepackage[nocompress]{cite}
\else
  \usepackage{cite}
\fi

\ifCLASSINFOpdf
\else
\fi

\usepackage[dvipsnames,prologue]{xcolor}
\usepackage{pstricks,pst-node}
\usepackage{amsmath}
\usepackage{graphicx}
\usepackage[numbers,sort&compress]{natbib}

\begin{document}
\title{Incremental Multiple Longest Common Sub-Sequences}

\author{Lu\'is~M.~S.~Russo, Alexandre~P.~Francisco, Tatiana~Rocher
  \IEEEcompsocitemizethanks{
    \IEEEcompsocthanksitem All authors are associated with INESC-ID.
    \IEEEcompsocthanksitem Lu\'is and Alexandre
    are with the Computer Science and Engineering
    Department of Instituto Superior T\'ecnico, Universidade de
    Lisboa.}}

\markboth{Journal of \LaTeX\ Class Files,~Vol.~14, No.~8, August~2015}%
{Shell \MakeLowercase{\textit{et al.}}: Bare Demo of IEEEtran.cls for Computer Society Journals}

\IEEEtitleabstractindextext{%
\begin{abstract}
  We consider the problem of updating the information about multiple
  longest common sub-sequences. This kind of sub-sequences is used to
  highlight information that is shared across several information
  sequences, therefore it is extensively used namely in bioinformatics and
  computational genomics. In this paper we propose a way to maintain this
  information when the underlying sequences are subject to modifications,
  namely when letters are added and removed from the extremes of the
  sequence. Experimentally our data structure obtains significant
  improvements over the state of the art.

\end{abstract}

}

\maketitle

\IEEEdisplaynontitleabstractindextext

\IEEEpeerreviewmaketitle

\IEEEraisesectionheading{\section{Introduction}\label{sec:introduction}}

\IEEEPARstart{I}{n} this paper we consider the problem of updating multiple longest common
subsequences (MLCS). Figure~\ref{fig:problem} illustrates this problem. The
top box shows an MLCS of four strings. The particular MLCS is the string
\texttt{AAAAABACA}. This string occurs as a subsequence in all four
strings, this is illustrated by gray line segments. Therefore this string
is a multiple common subsequence. Moreover any other multiple common
subsequence has at most nine letters. Hence this particular string is
actually a longest such subsequence.

We consider the problem of determining the size of an MLCS, when the
underlying strings are modified. In particular we consider the following
modifications:
\begin{itemize}
\item \texttt{Pop()}, removes the first letter from one of the
  strings.
\item \texttt{Append()}, adds a letter to the end of some of the strings.
\end{itemize}
\newcommand{\conl}[4]{
  \ncline{2,#1}{3,#2}\ncline{3,#2}{4,#3}\ncline{4,#3}{5,#4}
}

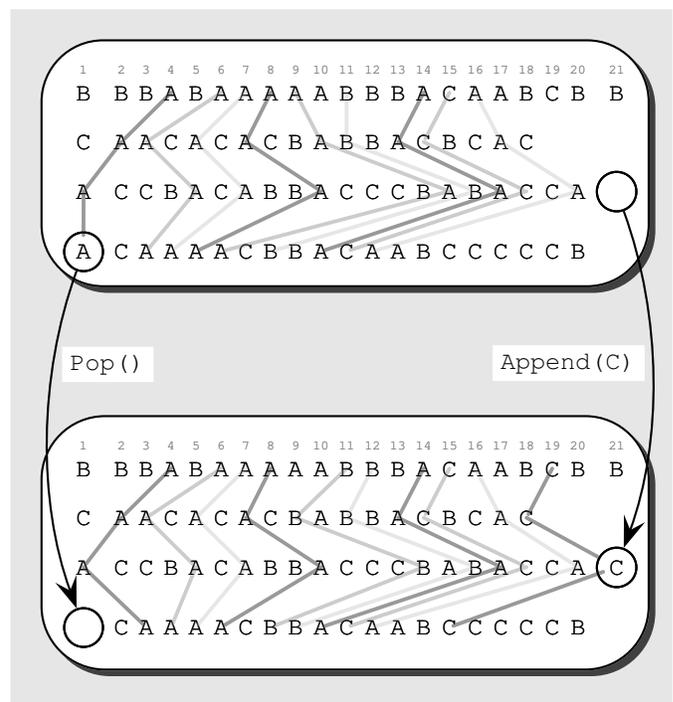
\begin{figure}
  \begin{center}
    \begin{pspicture}[showgrid=false](-0.5,-5.8)(8.3,3.5)
      \begin{ttfamily}
        \psframe[fillcolor=black!10,linecolor=black!10,fillstyle=solid](-0.5,-5.8)(8.3,3.5)
        \psframe[fillcolor=white,fillstyle=solid,shadow=true,framearc=0.5](-0.1,-0.2)(8.0,3.1)
        \psframe[fillcolor=white,fillstyle=solid,shadow=true,framearc=0.5](-0.1,-5.3)(8.0,-1.9)
        \def\psrowhooki{\tiny \color{gray} \vspace{-0.017cm}}
        \rput[bl](0,0){
          \begin{psmatrix}[rowsep=0.25,colsep=0.13]
            1&2&3&4&5&6&7&8&9&10&11&12&13&14&15&16&17&18&19&20&21\\
            B&B&B&A&B&A&A&A&A&A&B&B&B&A&C&A&A&B&C&B&B\\
            C&A&A&C&A&C&A&C&B&A&B&B&A&C&B&C&A&C\\
            A&C&C&B&A&C&A&B&B&A&C&C&C&B&A&B&A&C&C&A&\circlenode{Y}{\textcolor{white}C}\\
\circlenode{X}{A}
             &C&A&A&A&A&C&B&B&A&C&A&A&B&C&C&C&C&C&B
            \psset{linewidth=0.05}
            \psset{linecap=1}
            \psset{nodesepA=-0.1}
            \psset{nodesepB=-0.05}
            \psset{linecolor=black!10}
            \conl{16}{17}{20}{12}
            \psset{linecolor=black!20}
            \conl{15}{14}{18}{11}
            \psset{linecolor=black!40}
            \conl{14}{13}{17}{10}
            \psset{linecolor=black!10}
            \conl{11}{11}{16}{8}
            \psset{linecolor=black!20}
            \conl{9}{10}{15}{6}
            \psset{linecolor=black!40}
            \conl{8}{7}{10}{5}
            \psset{linecolor=black!10}
            \conl{7}{5}{7}{4}
            \psset{linecolor=black!20}
            \conl{6}{3}{5}{3}
            \psset{linecolor=black!40}
            \conl{4}{2}{1}{1}
          \end{psmatrix}
        }
        \rput[bl](0,0){
          \begin{psmatrix}[rowsep=0.25,colsep=0.13]
            1&2&3&4&5&6&7&8&9&10&11&12&13&14&15&16&17&18&19&20&21\\
            B&B&B&A&B&A&A&A&A&A&B&B&B&A&C&A&A&B&C&B&B\\
            C&A&A&C&A&C&A&C&B&A&B&B&A&C&B&C&A&C\\
            A&C&C&B&A&C&A&B&B&A&C&C&C&B&A&B&A&C&C&A&\circlenode{Y}{\textcolor{white}C}\\
\circlenode{X}{A}
             &C&A&A&A&A&C&B&B&A&C&A&A&B&C&C&C&C&C&B
          \end{psmatrix}
        }

        \rput[bl](0,-5.0){
        \begin{psmatrix}[rowsep=0.25,colsep=0.13]
          1&2&3&4&5&6&7&8&9&10&11&12&13&14&15&16&17&18&19&20&21\\
          B&B&B&A&B&A&A&A&A&A&B&B&B&A&C&A&A&B&C&B&B\\
          C&A&A&C&A&C&A&C&B&A&B&B&A&C&B&C&A&C\\
          A&C&C&B&A&C&A&B&B&A&C&C&C&B&A&B&A&C&C&A&\circlenode{YY}{C}\\
\circlenode{XX}{\textcolor{white}A}
          &C&A&A&A&A&C&B&B&A&C&A&A&B&C&C&C&C&C&B
          \psset{linewidth=0.05}
          \psset{linecap=1}
          \psset{nodesepA=-0.1}
          \psset{nodesepB=-0.05}
          \psset{linecolor=black!40}
          \conl{19}{18}{21}{15}
          \psset{linecolor=black!10}
          \conl{16}{17}{20}{12}
          \psset{linecolor=black!20}
          \conl{15}{14}{18}{11}
          \psset{linecolor=black!40}
          \conl{14}{13}{17}{10}
          \psset{linecolor=black!10}
          \conl{12}{11}{16}{9}
          \psset{linecolor=black!20}
          \conl{11}{9}{14}{8}
          \psset{linecolor=black!40}
          \conl{8}{7}{10}{6}
          \psset{linecolor=black!10}
          \conl{7}{5}{7}{5}
          \psset{linecolor=black!20}
          \conl{6}{3}{5}{4}
          \psset{linecolor=black!40}
          \conl{4}{2}{1}{3}
        \end{psmatrix}
      }
      \rput[bl](0,-5.0){
        \begin{psmatrix}[rowsep=0.25,colsep=0.13]
          1&2&3&4&5&6&7&8&9&10&11&12&13&14&15&16&17&18&19&20&21\\
          B&B&B&A&B&A&A&A&A&A&B&B&B&A&C&A&A&B&C&B&B\\
          C&A&A&C&A&C&A&C&B&A&B&B&A&C&B&C&A&C\\
          A&C&C&B&A&C&A&B&B&A&C&C&C&B&A&B&A&C&C&A&\circlenode{YY}{C}\\
\circlenode{XX}{\textcolor{white}A}
           &C&A&A&A&A&C&B&B&A&C&A&A&B&C&C&C&C&C&B
        \end{psmatrix}
        }
        \ncarc[arrowsize=0.3,arcangle=-20]{->}{X}{XX} \naput*[npos=0.27]{Pop()}
        \ncarc[arrowsize=0.3,arcangle=20]{->}{Y}{YY} \nbput*[npos=0.45]{Append(C)}
      \end{ttfamily}
    \end{pspicture}
  \end{center}
\caption{Illustration of updating a multiple longest common
  sub-sequence. The MLCS on the top box is updated to the MLCS on the
  bottom box.}
\label{fig:problem}
\end{figure}

This paper studies how to perform this computation efficiently. Our main
contributions are the following:
\begin{itemize}
\item We describe a data structure to represent the MLCS information,
  Section~\ref{sec:idea}. We describe the appropriate update algorithms
  that are applied when the underlying strings are modified. The
  \texttt{Append()} operation is described in Section~\ref{sec:idea} and
  the \texttt{Pop()} operation is described in
  Section~\ref{sec:textttpop-operation}. We discuss appropriate data
  structures for implementing these operations
  Section~\ref{sec:dynam-orth-range}.
\item We present our data structure's implementation and experimental
  results that highlight the performance of our approach,
  Section~\ref{sec:experimental-results}. The experimental results show
  that our structure is efficient for a large spectrum of MLCS parameters
  and, in some cases, it outperforms state of the art algorithms.
\end{itemize}
\section{The Problem} 
\label{sec:problem}
For our particular purpose we adopt the view that algorithms for solving
the MLCS problem are a sort of match classification process where a match
is a tuple that corresponds to the same letter in all four strings. The
gray line segments in Figure~\ref{fig:problem} represent this precise concept.
For example the first match in the top box
corresponds to the tuple $(4,2,1,1)$, the underlying letter is
\texttt{'A'}. This letter occurs in position $4$ in the first string,
position $2$ in the second string and so on. We also refer to these tuples
as multi-dimensional points, thus exploring the geometrical nature of the
problem. A critical geometric property is point dominance. We say that the
point $(6,3,5,3)$ dominates the point $(4,2,1,1)$ because in every
coordinate the values of the first point are strictly larger than the values
of the second point, i.e., $6 > 4; 3 > 2; 5 > 1$ and $3 > 1$. This is a
desirable property. When a match point dominates another match, both matches
are compatible for the same multiple common subsequence
(MCS). In particular both the points we mentioned are part of the MLCS in
the top box of Figure~\ref{fig:problem}. On the contrary the match
$(2,9,4,8)$, corresponding to letter \texttt{'B'}, does not dominate the
match $(4,2,1,1)$, because in the first coordinate we have $2 <
4$. These two matches are not compatible and therefore may
not both occur in a MCS.

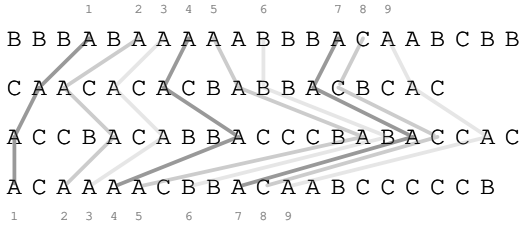
\begin{figure}
  \begin{center}
    \begin{pspicture}[showgrid=false](7.1,3.0)
      \begin{ttfamily}
        \def\psrowhooki{\tiny \color{gray} \vspace{-0.017cm}}
        \def\psrowhookv{\vspace{-0.013cm}}
        \def\psrowhookvi{\tiny \color{gray}}
        \rput[bl](0,0){
          \begin{psmatrix}[rowsep=0.25,colsep=0.13]
             & & &1& &2&3&4&5& &6& & &7&8&9\\
            B&B&B&A&B&A&A&A&A&A&B&B&B&A&C&A&A&B&C&B&B\\
            C&A&A&C&A&C&A&C&B&A&B&B&A&C&B&C&A&C\\
            A&C&C&B&A&C&A&B&B&A&C&C&C&B&A&B&A&C&C&A&C\\
            A&C&A&A&A&A&C&B&B&A&C&A&A&B&C&C&C&C&C&B\\
            1& &2&3&4&5& &6& &7&8&9
            \psset{linewidth=0.05}
            \psset{linecap=1}
            \psset{nodesepA=-0.1}
            \psset{nodesepB=-0.05}
            \psset{linecolor=black!10}
            \conl{16}{17}{20}{12}
            \psset{linecolor=black!20}
            \conl{15}{14}{18}{11}
            \psset{linecolor=black!40}
            \conl{14}{13}{17}{10}
            \psset{linecolor=black!10}
            \conl{11}{11}{16}{8}
            \psset{linecolor=black!20}
            \conl{9}{10}{15}{6}
            \psset{linecolor=black!40}
            \conl{8}{7}{10}{5}
            \psset{linecolor=black!10}
            \conl{7}{5}{7}{4}
            \psset{linecolor=black!20}
            \conl{6}{3}{5}{3}
            \psset{linecolor=black!40}
            \conl{4}{2}{1}{1}
          \end{psmatrix}
        }
        \rput[bl](0,0){
          \begin{psmatrix}[rowsep=0.25,colsep=0.13]
             & & &1& &2&3&4&5& &6& & &7&8&9\\
            B&B&B&A&B&A&A&A&A&A&B&B&B&A&C&A&A&B&C&B&B\\
            C&A&A&C&A&C&A&C&B&A&B&B&A&C&B&C&A&C\\
            A&C&C&B&A&C&A&B&B&A&C&C&C&B&A&B&A&C&C&A&C\\
            A&C&A&A&A&A&C&B&B&A&C&A&A&B&C&C&C&C&C&B\\
            1& &2&3&4&5& &6& &7&8&9
          \end{psmatrix}
        }
      \end{ttfamily}
    \end{pspicture}
  \end{center}
\caption{Illustration of level classification of matches.}
\label{fig:levels}
\end{figure}

Using this formulation an MCS is simply a sequence of matches
$p_1, \ldots, p_\lambda$ where for every $i<\lambda$ we have that $p_{i+1}$
dominates $p_i$. If such a sequence is not a sub-sequence of any other
MCS, it is maximal and corresponds to an MLCS. Instead of comparing several
sub-sequences, which would be infeasible,
we  assign levels to each match. A match belongs to level
$i$ if it occurs as $p_i$ in some MCS, i.e., it is the i-th match of some
MCS. For example $(4,2,1,1)$ belongs to level $1$, which is not surprising because
every match belongs to at least level $1$. The match $(6,3,5,3)$ on the other hand
belongs to levels $1$ and $2$ as it may appear in the second position of an
MCS. Therefore it is enough to identify the maximum level of a match, as it
must necessarily also contain all smaller levels. Figure~\ref{fig:levels}
shows several matches and their corresponding maximum levels, the top and
bottom small gray numbers. In general to determine the maximum levels we
use the rule that a match's maximum level is $\ell+1$ if it dominates a
match with maximum level $\ell$.

The simplest and most well known approach to determine maximum levels is a
dynamic programming algorithm that computes the maximum level of every
single point, not only matches. The main advantage of this approach is that
for any point only a few neighbor points need to be verified. The value
$D[a,b,c,d]$, that
represents the maximum level of point $(a,b,c,d)$, is computed recursively
as follows:
\begin{equation*}
  \label{eq:1}
              \begin{array}{l}
                0 \mbox{, if some coordinate is $0$}\\
                D[a-1,b-1,c-1,d-1] + 1 \mbox{, if $(a,b,c,d)$ is a match} \\
                \max\left\{
                \begin{array}{l}
                  D[a-1,b,c,d], \\
                  D[a,b-1,c,d], \\
                  D[a,b,c-1,d], \\
                  D[a,b,c,d-1]
                \end{array}
                \right\} \mbox{, Otherwise}
              \end{array}
\end{equation*}
However the total amount of points that are processed may be extremely
large: for $k$ strings of size $m$ this amounts to $O(m^k)$
points. Moreover this approach is not compatible with modification
operations. Obtaining the MLCS size after a \texttt{pop()} operation
requires rebuilding the whole table from scratch. For the \texttt{Append()}
operation the table can actually be extended, instead of re-computed, but
the corresponding performance is still $O(m^{k-1})$.

\section{Our Approach} 
\label{sec:idea}
To avoid having to process such a huge amount of points we focus only on
points that correspond to matches. In general this significantly reduces
the number of points to process. However some bad configurations have
$O(m^k)$ matches. For example when all the strings consist of repetitions
of the same letter, i.e, $\mathtt{A}^m$. To avoid this problem we use the
notion of covered matches.  The
general definition is that a match $p$ covers a match $q$ if both have the
same maximum level and for every coordinate its value in $p$ is less than
or equal to the value of the same coordinate in $q$, moreover for at least
one of the coordinates the value is equal. The equal coordinate condition
follows from the fact that both $p$ an $q$ belong to the same maximum level, as
otherwise $q$ would dominate $p$ and have a larger maximum level.
In our example, the match
$(6, 3, 5, 3)$ covers the match $(6, 5, 5, 4)$ because both have the same
maximum level of $2$ and $6 = 6 ; 3 \leq 5; 5 \leq 5$ and $3 \leq 4$.

We can now restrict our attention to non covered matches. In the example
where all the strings are $\mathtt{A}^m$ the number of non covered matches
is only $O(m)$ and thus much smaller than the total number of matches.
Table~\ref{tab:points} shows the classification of all the non covered
matches that exist in the strings of the top box of
Figure~\ref{fig:problem}. The matches that are drawn in the
Figure are shown inside gray boxes.

\begin{table}
\renewcommand{\arraystretch}{1.5}
  \begin{tabular}{r|p{8cm}}
1 & (1,9,4,8); \colorbox{black!40}{(4,2,1,1)}; (15,1,2,2)\\ \hline
2 & (2,11,8,9); (4,10,5,10); (5,9,4,8); \colorbox{black!20}{(6,3,5,3)};
    (15,4,2,2); (16,2,5,3) \\ \hline
3 & (3,12,9,14); (4,13,10,10); (5,11,8,14); (6,10,5,10);
    \colorbox{black!10}{(7,5,7,4)}; (15,4,6,7); (16,5,5,3); (17,3,7,4);
    (18,9,4,8); (19,6,3,7) \\ \hline
4 & (5,15,14,14); (6,17,15,12); (7,13,7,12)
    \colorbox{black!40}{(8,7,10,5)}; (11,9,8,8);  (15,6,11,7);
    (15,14,6,11); (16,5,7,10);  (17,7,7,4); (19,4,11,7); (19,6,6,7);
    (20,9,4,8); \\ \hline
5 & (8,17,10,13); \colorbox{black!20}{(9,10,15,6)}; (11,9,14,8);
    (11,15,8,14); (12,11,9,9); (14,10,10,10); (15,8,11,7); (16,7,15,10);
    (16,17,7,12); (17,7,10,12); (18,9,8,8); (19,6,11,11); \\ \hline
6 & (10,13,17,10); \colorbox{black!10}{(11,11,16,8)}; (13,12,14,14);
    (14,10,15,10); (14,13,10,10); (15,14,18,7); (18,9,14,8); (19,8,11,15);
    (19,8,18,11); (20,11,9,9); \\ \hline
7 & \colorbox{black!40}{(14,13,17,10)}; (15,14,11,11); (18,11,16,14);
    (20,9,14,20); (20,11,16,9); (21,12,14,14); \\ \hline
8 & \colorbox{black!20}{(15,14,18,11)}; (16,17,15,12); (18,15,14,14); (19,16,12,15)\\ \hline
9 & \colorbox{black!10}{(16,17,20,12)}; (19,16,18,15)
  \end{tabular}
  \caption{Table containing the classification of uncovered matches
    according to their maximum level values. The matches that are used in the MLCS
    of the top box of Figure~\ref{fig:problem} are shown in gray boxes.}
  \label{tab:points}
\end{table}

Our data structure stores a list of points for each level
value. The $i$-th list contains the non covered matches whose maximum level
is $i$. We can now focus on how this structure changes when the underlying
strings are modified. The \texttt{Append()} operation is simpler and faster
than the \texttt{Pop()} operation. Hence let us start with the
\texttt{Append()} operation.

In the example of Figure~\ref{fig:problem} we considered an
\texttt{Append()} operation that added the letter \texttt{'C'} to the end
of the third string. This generates a set of new matches that we should
insert into our lists. One approach would be to generate all the new
matches, i.e., the set
$\{15,19\}\times\{1,4,6,8,14,16,18\}\times\{21\}\times\{2,7,11,15,16,17,18,19\}$. Note
that for each string we considered the set of positions of \texttt{'C'},
except for the third string where we considered only the new position,
i.e., $21$. We would therefore need to consider $112$ matches, most of
which are covered. Instead we use the notion of match generation. We can
use a non covered match to generate a new match, that is potentially non
covered, but that may turn out to be covered. Consider for instance the
match $(16,17,20,12)$ in Table~\ref{tab:points}. To generate a match using
letter \texttt{'C'} we find the next occurrence of this letter, in each
string. In the first string the next occurrence of \texttt{'C'} after
position $16$ is in position $19$. In the second string the next occurrence
after position $17$ is in position $18$. Continuing this process generates
the match $(19,18,21,15)$. The new match is added to the list of matches of
the next level. In this case it is added to the list of level $10$, because
the point $(16,17,20,12)$ was on the list of level $9$.

We used the point $(16,17,20,12)$ to generate this match, but there was no
particular reason to single this point out from all the points in
Table~\ref{tab:points}. The remaining points might also yield new uncovered
matches. Hence let us consider the point $(21,12,14,14)$ from the list of
level $7$. Trying to use letter \texttt{'C'} to generate a new match raises
a tricky problem. As position $21$ is the last position of the first string,
there is no occurrence of the letter \texttt{'C'} after it, therefore it is not necessary
to consider point $(21,12,14,14)$. This defines our first criteria: we only
need to consider the points in Table~\ref{tab:points} that are dominated by
the point $(19,18,21,19)$, where this point consists of the positions of
the last occurrences of \texttt{'C'} in the respective strings. This avoids
considering points that can not generate matches with the given letter.

The point $(9,10,15,6)$, from the list of level $5$, passes this first
criteria. Looking for the letter \texttt{'C'} generates the point
$(15,14,18,7)$. However this point already exists at level $6$, see
Table~\ref{tab:points}. We conclude that it was not necessary to consider
the point $(9,10,15,6)$. This problem occurred because it
uses position $15$ for the third string. This means that the generated
point used position $18$ for the third string and not $21$. Therefore the
generated match is not one of the new matches that arise from the
\texttt{Append()} operation. Hence it is not useful to consider the point
$(9,10,15,6)$. Therefore our second criteria is to consider only the points
that have the value of $19$ or more for the third string, where $19$ is the
position of the last occurrence of \texttt{'C'} in the third string before
the \texttt{Append()} operation. In other words we only consider the
matches that dominate the point $(0,0,19-1,0)$.

Using this second criteria on the matches in Table~\ref{tab:points} yields
only the point $(16,17,20,12)$, which also satisfies the first criteria
because it is dominated by $(19,18,21,19)$. Hence we conclude that to
update our data structure after the \texttt{Append()} operation it is only
necessary to add a list for level $10$ containing the point
$(19,18,21,15)$. Hence only one of the $112$ new matches is non covered. In
this case the generated match is the only point at level $10$ and no
further processing is necessary, but in general we still have to verify if
the generated point is not covered by any one of the points at level
$10$. In which case it would not be accepted in the list. Moreover if any
point of the list is covered by the point being inserted then the covered
point should be removed from the list. This is a general check that is
always performed when inserting matches into a list. Whenever we mention
inserting a point into a list we assume that this procedure is being
performed, in particular it will be necessary for the \texttt{Pop()}
operation.

This concludes the description on how to update our data structure after an
\texttt{Append()} operation. In the next Section we describe the analogous
process for the \texttt{Pop()} operation and discuss several technical and
practical considerations.
\section{The Details} 
\label{sec:details}
\subsection{The \texttt{Pop()} operation}
\label{sec:textttpop-operation}
Updating the information in our data structure after a \texttt{Pop()}
operation is fairly more elaborated because it involves transferring
matches from one list to another and searching for covered points that
suddenly become uncovered.

Let us consider the \texttt{Pop()} operation in Figure~\ref{fig:problem}
that removes the first letter of the fourth string. This implies that all
the matches in the set
$\{4,6,7,8,9,10,14,16,17\}\times\{2,3,5,7,10,13,17\}\times\{1,5,7,10,15,17,20\}\times\{1\}$
that occur in our data structure (Table~\ref{tab:points}) need to be
removed. Notice that we selected these matches by using the sets of
positions of letter \texttt{'A'} in the first three strings and the
position $1$ in the fourth string. Because the last coordinate is fixed to
the first letter of the fourth string, these matches can not dominate other
matches hence all of them must be at level $1$. Therefore all matches are
covered matches except possibly the first match $(4,2,1,1)$.

Instead of generating all of these $441$ matches we only need to consider
the first match $(4,2,1,1)$, that corresponds to the first position of
\texttt{'A'} in each string. This first match does occur in
Table~\ref{tab:points} in the list of level $1$. We therefore remove this
point from the list. This has an impact in the structure. First some
covered points may become non covered, this happens with the point
$(4,2,1,3)$. For this particular case uncovering is simple, we consider the
point that indicates the first positions of letter \texttt{'A'} in each of
the strings. In this case the position of the first occurrence of
\texttt{'A'} in the fourth string is $3$ thus yielding this particular
point. The uncovered point gets inserted into the list of level $1$ before
the next verification. Note that before inserting $(4,2,1,3)$ into the list
of level $1$ it is necessary to verify that it is not covered by any match
on that list, noting that at this time the point $(4,2,1,1)$ is not part of
the list. Therefore the point $(4,2,1,3)$ gets accepted. Moreover if
$(4,2,1,3)$ covered any points on the list those points would have to be
removed, but this is not the case.

The second important consequence of removing the match $(4,2,1,1)$ from the
list of level $1$ is that some matches at level $2$ may be relying on this
match and may have to be transferred to level $1$. In particular this is
the case of $(6,3,5,3)$ and $(15,4,2,2)$. Since neither of them dominates
$(4,2,1,3)$, nor the other points in level $1$, they do have to be
moved. In this process $(6,3,5,3)$ is eliminated because it is covered by
$(4,2,1,3)$ and $(15,4,2,2)$ is eliminated because it is covered by
$(15,1,2,2)$. Hence they are removed and disappear. We now need to further
update the structure to account for the fact that these points got moved.
For both of these matches we need to check if they uncover points and if
there are matches at level $3$ that relied on them and that have to be
pulled to level $2$. This process continues until there are no more matches
to move to a lower level. The final result of this process is shown in
Table~\ref{tab:new-points}, where the highlighted changes are shown with
gray boxes. In this particular case the process ends at level $6$.

\begin{table}
\renewcommand{\arraystretch}{1.5}
  \begin{tabular}{r|p{8cm}}
    1 &  (1,9,4,8); \colorbox{black!20}{(4,2,1,3)}; (15,1,2,2)\\ \hline
    2 &   (2,11,8,9); (4,10,5,10); (5,9,4,8);
        \colorbox{black!20}{(6,3,5,4)}; \colorbox{black!20}{(15,4,2,7)};
        (16,2,5,3);\\ \hline
    3 & (3,12,9,14); (4,13,10,10); (5,11,8,14); (6,10,5,10);
        \colorbox{black!20}{(7,5,7,5)}; (15,4,6,7);
        \colorbox{black!20}{(16,5,5,10)}; (17,3,7,4); (18,9,4,8);
        \colorbox{black!20}{(19,6,3,11)}; \\ \hline
    4 & (5,15,14,14); (6,17,15,12); (7,13,7,12);
        \colorbox{black!20}{(8,7,10,6)}; (11,9,8,8); (15,6,11,7);
        (15,14,6,11); (16,5,7,10); (19,4,11,7);
        \colorbox{black!20}{(19,6,6,11)}; \colorbox{black!20}{(20,9,4,14)};
    \\ \hline
    5 &  (8,17,10,13); \colorbox{black!20}{(9,10,15,10)}; (11,9,14,8);
        (11,15,8,14); (12,11,9,9); (14,10,10,10); (15,8,11,7);
        (16,7,15,10); (16,17,7,12); (17,7,10,12);
        \colorbox{black!20}{(18,9,8,14)}; (19,6,11,11);\\ \hline
    6 &  (10,13,17,12); \colorbox{black!20}{(11,11,16,14)};
        \colorbox{black!20}{(12,11,16,9)}; (13,12,14,14); (14,10,15,10);
        (14,13,10,10); (18,9,14,8); (19,8,11,15); (19,8,18,11);
        \colorbox{black!20}{(20,11,9,20)}; \\ \hline
    7 &  (14,13,17,10); (15,14,11,11); (18,11,16,14); (20,9,14,20);
        (20,11,16,9) \\ \hline
    8 & (15,14,18,11); (16,17,15,12); (18,15,14,14); (19,16,12,15)\\ \hline
    9 & (16,17,20,12); (19,16,18,15)\\ \hline
    10 & (19,18,21,15)
  \end{tabular}
  \caption{Table containing the classification of uncovered matches
    according to their maximum level values. In this table the gray boxes
    highlight the differences with Table~\ref{tab:points}.}
  \label{tab:new-points}
\end{table}

Moving points to a lower level is fairly straight forward, as described
before it requires checking that the point is not covered in the lower
level and removing any points that it covers. However the uncovering
process is a bit more complex than what was previously exemplified. To
describe the complete process consider the match $(11,11,16,8)$ at level
$6$. This point is moved to level $5$ but in that list it is covered by the
point $(11,9,14,8)$ and disappears. However before that it will leave the
point $(12,11,16,9)$ as an uncovered match at level $6$. The exact process
to uncover this match is the following. First notice that the letter that
corresponds to the match $(11,11,16,8)$ is \texttt{'B'}. We start of by
considering the point indicating the last positions of \texttt{'B'}, in this case
$(21,15,16,20)$. We combine this point with
$(\mathbf{11},\mathbf{11},\mathbf{16},\mathbf{8})$ using only one value
from the match and the remaining values from the point of last
positions. In this case the resulting points are
$(\mathbf{11},15,16,20);(21,\mathbf{11},16,20);(21,15,\mathbf{16},20)$ and
$(21,15,16,\mathbf{8})$. The reason for generating these four points is
that a covered match must be equal in at least one coordinate. Therefore we
test all of them.

For each one of these points we determine the points at level $5$ that are
dominated by at least one of them. In this particular case it is enough to
check for the points dominated by $(21,15,16,20)$, as any point dominated
by the other three is also dominated by $(21,15,16,20)$. Consulting the
matches at level $5$ in Table~\ref{tab:new-points} we can observe a large
number of dominated points. In particular there is the match
$(11,9,14,8)$. From each one of these matches we generate a new match by
using the letter \texttt{'B'}. Hence from $(11,9,14,8)$ we obtain
$(12,11,16,9)$ as expected. Also note that the match $(9,10,15,10)$ is also
dominated by $(21,15,16,20)$ and it generates the match $(11,11,16,14)$,
which also becomes a new uncovered match.

It may seem that this elaborated process is unrelated to the previous
description we gave for uncovering. In particular there should be no level
$0$ list. However we include one such list, containing only the point
$(-1,-1,-1,-1)$. This point does not correspond to an actual match, but it
behaves as a sentinel in the data structure. This reduces the elaborated
uncover procedure to the simple description we gave before, as once the
sentinel point is selected it then is used to generate the first match of a
given letter. Moreover for the same reason this sentinel is useful for the
\texttt{Append()} operation.

The only remaining detail about our data structure is how to represent the
lists of points. We could use actual lists, but instead we used orthogonal
range trees. The reason being that an orthogonal range tree allows us to
compute dominance in $O(\log^k n)$ time, where $k$ is the number of
strings, or dimension of the point space, and $n$ is the number of points
in the list. Hence the steps in the previous description that mentioned
determining the matches of a list dominated by a given point $p$ can be
determined in this time. The underlying algorithmic primitive is known as
an orthogonal range query and it can also be used to support the double
condition described in the procedure of the \texttt{Append()}
operation. This primitive allows us to filter a list of points by
specifying intervals that each coordinate should respect. In our example of
Section~\ref{sec:idea} we concluded that to perform the \texttt{Append()}
operation we need to filter the points in Table~\ref{tab:points} that where
dominated by the point $(19,18,21,19)$ and that dominated the point
$(0,0,18,0)$. From these two points we can then consider the orthogonal
range $[0+1,19-1]\times[0+1,18-1]\times[18+1,21-1]\times[0+1,19-1]$,
defined as the Cartesian product of the intervals that restrict the
coordinates. Hence a query of this range in Table~\ref{tab:points} yields
the desired point $(16,17,20,12)$.

In the next Section we describe several implementation choices we made
and the resulting performance analysis. Namely we need to support the
dynamic nature of our lists, where points are inserted and removed during
the course of the algorithms. The orthogonal range queries are actually
interleaved among these modifications. Obtaining such a dynamic
implementation requires a particular implementation that we will now
describe.

\subsection{Dynamic Orthogonal Range Trees}
\label{sec:dynam-orth-range}
In this section we give a simple description of the implementation of
dynamic orthogonal range trees. Our approach to the problem is simple and
pragmatic. The description is also similar in spirit. Our implementation
still lacks several refinements that may yield further polylog speed-ups,
i.e., $O(\log^c n)$, for some constant $c$. We plan to implement
those techniques soon, but the current state of the software is sufficient
to establish the validity of the algorithm we describe in the paper.

The simplest instance of the problem is the one dimensional case, were the
data structure stores a set of points (single numbers) and the query
consists of an interval. An example of a query would be to search for all
the numbers contained in the interval $[3,7]$. The result would depend on
the points that are stored. If the list of points consisted of $\{1, 4, 5, 7,
10\}$ then the result of the query would be the set $\{4, 5, 7\}$.

To compute queries we store the points in a binary
search tree (BST). We can build a completely balanced BST in $O(n)$ time,
for a set with $n$ points. We assume that the points are pre-sorted and
that we only need to build the tree structure itself. As points are added
or removed this balanced property is likely to be lost. In particular, long
branches may arise. This is undesirable as traversing them takes longer
than the desired performance bound. To avoid this problem several balancing
schemes have been proposed in the literature. For our particular
application we favor an approach that does not use rotations, meaning
that once a node is considered unbalanced its entire sub-tree needs to be
rebuilt. This approach is simple and adequate for orthogonal range trees,
because rebuilds can not be avoided even when using rotations.

We use a weight balanced BST, similar to BB-$\alpha$ trees. We will use a
balancing scheme that is simple to implement and provides good
performance. Let us consider a node $u$ with weight $w(u)$. The weight of a
node is the number of points stored in its sub-tree. To make sure that
the tree is balanced we force the weight to decrease by a factor
$\alpha < 1$. Therefore if $v$ is a child of $u$ we must have that
Inequality~(\ref{eq:0}) holds.
\begin{equation}
  \label{eq:0}
  w(v) \leq \alpha w(u)
\end{equation}
When this condition on $v$ and similar condition on $v'$ are verified the
node $u$ is considered balanced, where $v'$ is the other child of $u$. If a
child node does not exist its weight is considered to be $0$. Therefore a
leaf is always balanced. A node $u$ that is not balanced is unbalanced.

In order for Equation~(\ref{eq:0}) to be meaningful we must have
$\alpha < 1$. Moreover we should also have $1/2 < \alpha$, as anything
below $1/2$ is impossible to ensure for both children. Even $1/2$ is
unreasonable as every modification might imply a rebuilt. In conclusion we
should choose $\alpha$ to respect $1/2 < \alpha < 1$.

Now let us consider a branch of length $h$ that starts at the root, with
weight $n$, and finishes at a leaf, with weight $1$. If we iterate
Equation~(\ref{eq:0}) along the branch, of length $h$ we obtain the
restriction that $1 \leq \alpha^h n$. This restriction implies $h \leq
\log_{1/\alpha} n$.

Let us turn to the analysis of the modification operations. The analysis is
amortized. We associate to each node $u$ of the tree a potential $\Phi(u)$
defined in Equation~(\ref{eq:6}).

\begin{equation}
  \label{eq:6}
  \Phi(u) = \left( \frac{2\max\{w(v),w(v')\} - w(u)}{2
      \alpha - 1} \right)
\end{equation}
The overall potential of the data structure is the sum of the $\Phi(u)$
values, for all the nodes. The important properties of $\Phi(u)$ are that
it is $0$ when the sub-tree below $u$ is completely balanced and that it
is at least $w(u)$ when the condition in Inequality~(\ref{eq:0})
fails. Hence when a node becomes unbalanced there are enough credits to
rebuilt the corresponding sub-tree.

When a point is inserted into the sub-tree rooted at $u$ the weight value
$w(u)$ increases by $1$. One of the weights $w(v)$ or $w(v')$ also increase
by $1$. The conclusion in terms of the analysis is that it is sufficient to
stock up $1/(2 \alpha - 1)$ credits for each node in the corresponding
branch. This branch is the path that the insertion procedure
traverses. Given the bound on the size of the branches $h$ we obtain an
$O((1/(2 \alpha -1))\log_{1/\alpha} n)$ amortized time for the insertion
operation. A similar argument, and bound, also holds for deletion.

In our experiments we use $\alpha = 3/4$. This is a reasonably good
value. Each node needs to stock up $2$ credits per node on modification
operations and the factor $1/\log(1/\alpha)$ is around $2.4$. Which is
reasonable, given that for AVL trees the corresponding factor is $1.44$ and
for red-blacks its $2.0$.

Finally we discuss general orthogonal range trees (ORT), for points in $k$
dimensions. General ORTs are defined recursively. As we have seen, a 1D ORT
is simply a BST. A $k$ dimensional ORT is also a BST, ordered by the last
coordinate of the points. Moreover each node in this BST stores a $d-1$ ORT
which contains the same points as its sub-tree, but the points are
projected to a $k-1$ space by removing the last coordinate. Computing a
query on the trees consists in traversing a branch on the $k$ dimensional
tree and for each node in this branch traversing the corresponding branches
on the $k-1$ tree and so on. The total amount of nodes visited is at most
$h$ for dimension $k$, at most $h^2$ for dimension $k-1$ and so on up to
$h^k$ nodes for dimension $1$. Adding the resulting geometric series yields
a total of $h(h^k)/(h-1) = O(h^k)$ nodes. Hence the resulting time bound
for the $k$ dimensional query operations is $O(\log_{1/\alpha}^{k}n)$.

A similar series is used for the modification operations. To build a $k$
dimensional ORT with $n$ points we use a procedure that requires
$O(n \log^{k-1} n)$ time. For any $k > 1$ we find the median in $O(n)$ time
and split the points into two sets of $n/2$ points. We also need to build
the tree with the same $n$ points in dimension $k-1$. Let $T(n,k)$
represent the time necessary to build a $k$ dimensional tree with $n$
points. We have that $T(n,1) = O(n)$. For the general case with $k > 1$,
the procedure we explained can be accounted by the recurrence in
Equation~(\ref{eq:5}).
\begin{equation}
  \label{eq:5}
  T(n,k) = 2 T(n/2,k) + T(n, k-1) + O(n)
\end{equation}
This recurrence yields a bound of $T(n,k) = O(n \log^{k-1} n)$ construction
time.

Now we update the potential function of Equation~(\ref{eq:6}) by including
an $\log^{k-1} n$ factor. The resulting amount of credits to stock in a
node of a $k$ dimensional tree is $(1/(2 \alpha - 1)) \log^{k-1} n$. Hence
the total amount of credits necessary is given by the following geometric
series:
\begin{equation}
  \frac{1}{2 \alpha - 1} \left( h \log^{k-1} n + h^2 \log^{k-2} n + \ldots +
    h^k \right)
\end{equation}
We can now conclude that the amortized time bound of modification
operations is
$O((1/(2 \alpha - 1))h^{k}) = O((1/(2 \alpha - 1))(\log_{1/\alpha}^{k}n))$.

There is a final refinement to our implementation, which has a very
significant impact in the practical performance of the structure. We choose
a parameter $c$, that we estimate is close to the height $h$. In the one
dimensional case whenever the weight $w(u)$ of a node is smaller than $c$
we can skip the sub-tree of $u$ and store the points in an array. In a
general dimension $k$ we check if $k \times w(u) < c^k$, when this
condition is true we store the points in an array. To be absolutely precise
we check for ${c+k \choose k}$ instead of $c^k$ as this seems a more
precise estimate of operation complexity. The intuition is that this number
counts the number of possible root to leaf paths, where we only need to
choose when to move to a lower dimension. The resulting data structure is a
sort of graft tree that along is branch is an orthogonal range tree up to a
certain threshold and after that is an array of points.

\subsection{Experimental Results}
\label{sec:experimental-results}
We implemented the data structure that we described in the previous
sections. We measured both execution time and memory usage. We used a
virtual machine running \texttt{Debian 4.9.144-3.1} the \texttt{x86\_64}
version with 8Gb of Ram and an \texttt{Intel(R) Xeon(R) CPU E5-2630 v3 @
  2.40GHz} with 4 cores. The implementation was developed in C and compiled
with gcc with optimization flag -O3.

We also implemented, in C, the Dynamic programming algorithm mentioned in
Section~\ref{sec:problem} and the Quick-DP algorithm\footnote{We bealive
  our Quick-DP implementation is as efficient as possible, it includes an
  $O(n \log^{k-2}n)$ algorithm for multidimensional minima.} that will be
briefly described in Section~\ref{sec:related}. The source code is
available at \verb|https://github.com/LuisRusso-INESC-ID/IMLCS|. Since the
Naive and Quick-DP algorithms are not designed for the \texttt{Pop()}
operation execute them from scratch whenever this operation is applied. For
the \texttt{Append()} operation we simply do nothing. Meaning that the
algorithms are only executed when a \texttt{Pop()} operation is
applied. However the DP and Quick-DP algorithms are still forced to execute
at least once before the prototypes terminate. This simulates the fact that
the DP and Quick-DP algorithms could be adapted to support
\texttt{Append()} efficiently and that in fact \texttt{Pop()} is the
intricate and inefficient operation. This means that the comparison of the
experimental results is in favor of the DP and Quick-DP
algorithms. Therefore whenever our IMLCS algorithm is more efficient than
these algorithms the comparison is significant.

We generated random sequences of \texttt{Append()} and \texttt{Pop()}
operations. Each generated sequence is defined by a few parameters:
\begin{description}
\item [S] the size of the alphabet.
\item [k] the number of strings.
\item [m] the size of the strings.
\end{description}

\begin{figure}[htb]
  \includegraphics{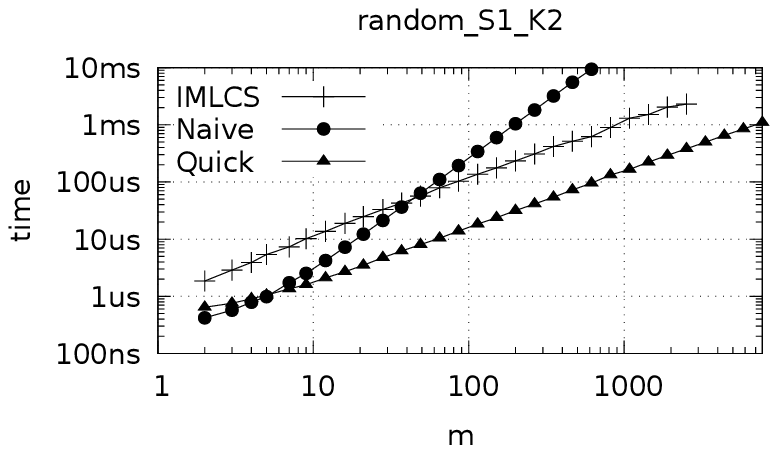}
  \includegraphics{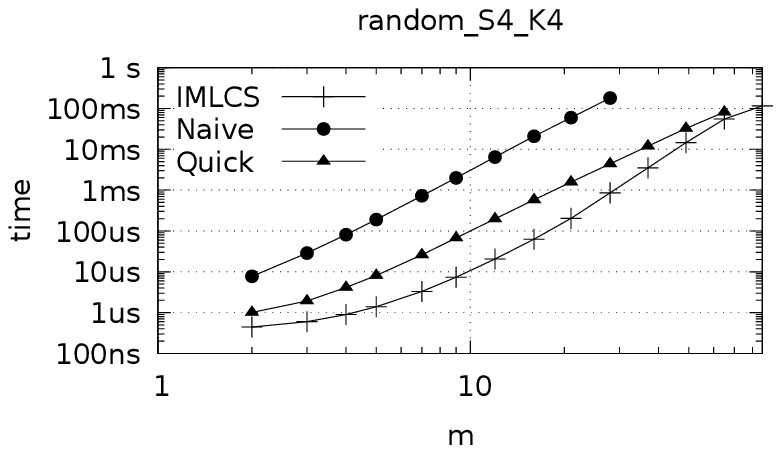}
  \includegraphics{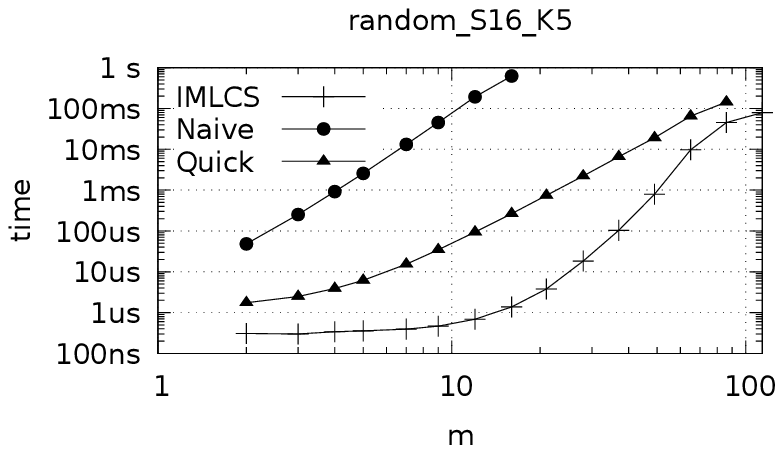}
\caption{Average value of $m$ versus average operation time for different
  combinations of $k$ strings and alphabets $S$.}
\label{fig:gen1}
\end{figure}

The general generation procedure is the following. First one of the $k$
sequences is chosen uniformly at random. Then we consider the size of that
string. If the size of the string is smaller than $m$ an \texttt{Append()}
operation is selected. If the size of the string is larger than $2*m$ then
a \texttt{Pop()} operation is selected. Otherwise either \texttt{Pop()} or
\texttt{Append()} is selected with $50\%$ probability each. If an
\texttt{Append()} operation is selected the letter is chosen uniformly from
the alphabet. An experiment executes several operations and reports the
average time per operation by dividing total time by number of
operations. We selected three plots to present in Figure~\ref{fig:gen1}. In
these plots we used a double log scale. These scales turn polynomials into
straight lines, after a certain point. The inclination of the lines
indicates the respective polynomial degree. It also means that gaps of what
appears to be one unit actually represent a factor of $10$ slowdown. The
remaining plots are shown in the Appendix. Moreover the memory requirements
of these procedures are small, we also show memory requirement plots in the
Appendix.

From these plots we can observe that our approach obtains the best results
for larger alphabets and when the number of strings $k$ increases. However
in these cases the size of the underlying MLCS decreases. To partially
factor out this condition we devised a second generator, that produces
longer MLCS. This time if the operation selected is \texttt{Append()} we
decide with a probability of $1/k$ to insert the same letter in all $k$
string. We present similar the plots from this generator
(Figure~\ref{fig:gen2}) and the full set of plots in the Appendix.
\begin{figure}[htb]
  \includegraphics{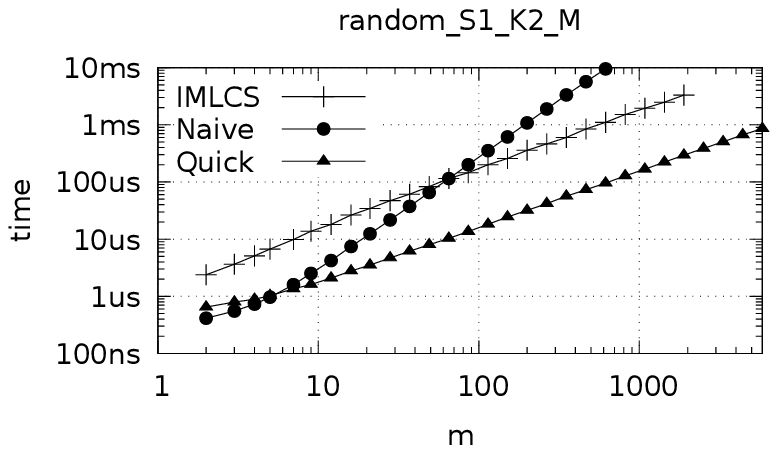}
  \includegraphics{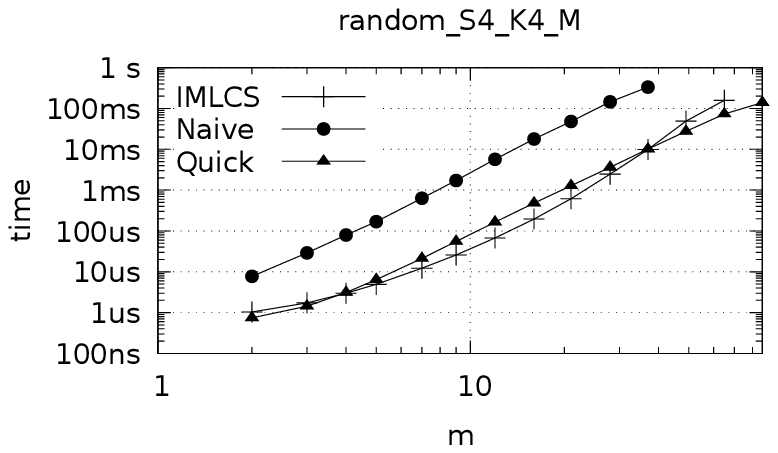}
  \includegraphics{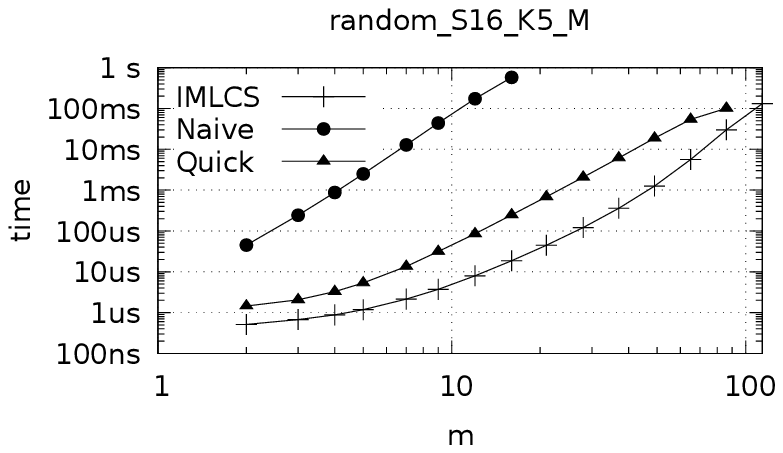}
\caption{Average value of $m$ versus average operation time for different
  combinations of $k$ strings and alphabets $S$.}
\label{fig:gen2}
\end{figure}

As expected, the longer MLCS does have an impact in the resulting
performance. Still, for the larger alphabets, our approach still achieves
significantly better performance.

\begin{figure}[htb]
  \includegraphics{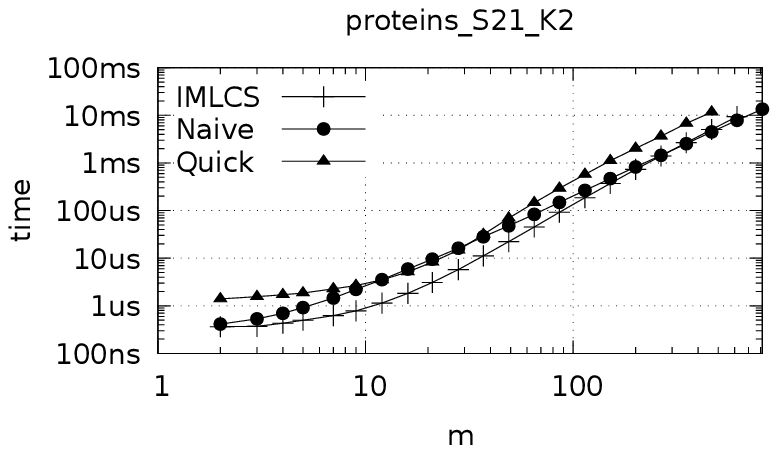}
  \includegraphics{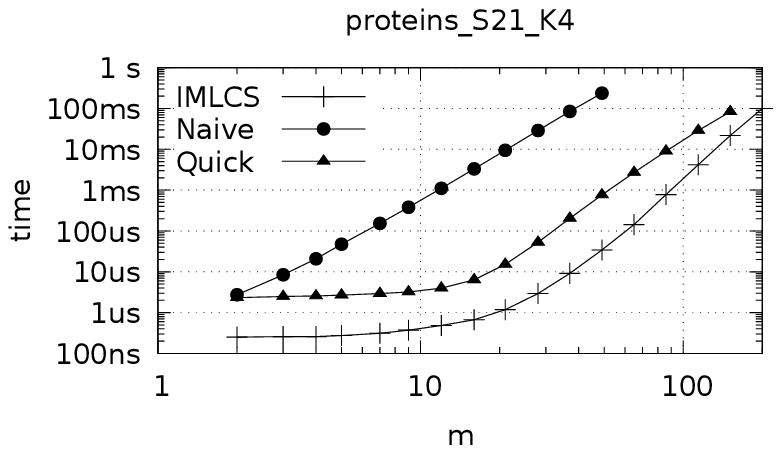}
  \includegraphics{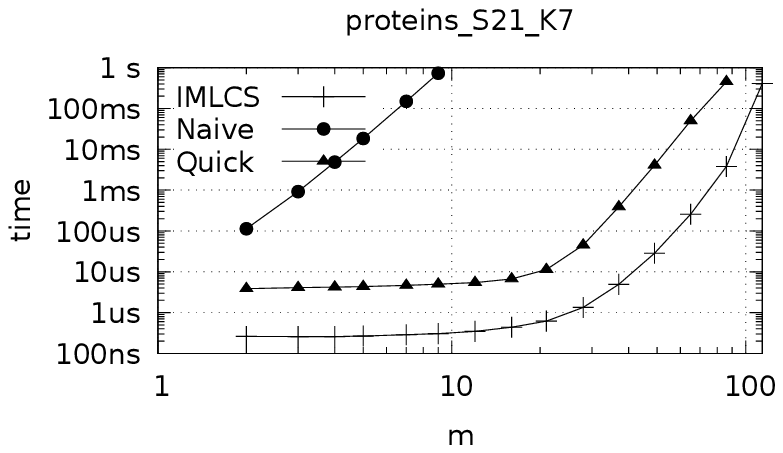}
\caption{Sliding window of size $m$ versus average operation time for different
  combinations of $k$ protein sequences.}
\label{fig:prot}
\end{figure}

As a final test we experimented with actual protein sequences. Therefore
the underlying alphabet has a fixed size of $21$. We obtained the sequences
from PFAM, corresponding to the \texttt{AP\_endonuc\_2} family
(\texttt{A0AMF1}, \texttt{A0B937}, \texttt{A0AF79}, \texttt{A0JS78},
\texttt{A0LNH5}, \texttt{A0K062}, \texttt{A0JT54}, \texttt{A0NI89}). To
generate the tests we chose a fixed size of $m$. The generator uses $k$ 
sliding windows of size $m$ that passes through the sequences. The sliding
windows of all the $k$ sequences move in sync. Whenever a sliding window
reaches the end of a sequence it wraps around to its first letter. The
resulting performance is shown in Figure~\ref{fig:prot}.

Clearly in these tests our prototype largely benefits from the alphabet
size and generally out performs the alternatives. Thus our data structure
is particularly well suited for this kind of application. More tests and
memory results are given in the Appendix.

\section{Related Work} 
\label{sec:related}
Let us now revise previous work. As far as we are aware this is the first
work to consider the dynamic multiple longest common subsequence
problem. Previous worked considered either dynamic longest common
substrings, i.e., with only two strings, or the multiple longest common
subsequence in the static setting. We revise both these lines of research.

The work on dynamic longest common substrings, denominated incremental
string comparison, was initiated by~\citet*{landau1998incremental}, which
obtained an $O(m)$ time algorithm to implement a \texttt{Prepend()}
operation, where is the letter is added to the beginning of the strings. A
simpler version, with the same performance was presented
by~\citet*{kim2000dynamic}, which is simultaneously incremental and
decremental. This solution was presented for the edit
distance.~\citet*{ishida2005fully} presented an algorithm which reduced the
time complexity from $O(m)$ to $O(\lambda)$ and was fully incremental,
where $\lambda$ is the size of the LCS. The
algorithms was presented for the LCS. The space requirements were also
reduced from $O(m^2)$ to $O(m \lambda)$.

\citet*{landau2004two} studied the problem of consecutive suffix alignment
problem, which obtained the size of the LCS between all the suffixes of a
string $A$ and a string $B$, the final version of the paper appeared
in~\citeyearpar{LANDAU20071095}. The authors presented two algorithms for
this problem, which required $O(m \lambda)$ and
$O(m \lambda + m \log \sigma)$ time, where $\sigma$ is the size of the
alphabet of the underlying strings. Their approach uses a structure similar
to the $T_k$ lists from the Hunt-Szymanski
algorithm~\citet*{hunt1977fast}. Moreover their structure is not
decremental. Because of these nuances the relation to LTSS is not immediate
which justifies the algorithm of~\citet*{kosowski2004efficient}, in the
same year.

A corner stone of all these results is the algorithm
from~\citet*{hunt1977fast}, which contains several crucial ideas. One
fundamental idea was the focus on matches, instead of all the points in the
dynamic programming table. This idea carried over naturally to the multiple
LCS cases. Another crucial idea was the reduction from the LCS to the
Longest Increasing Subsequence, although this was not immediately clear in
the original presentation. It was partially identified
by~\citet*{apostolico1986improving},~\citet*{apostolico1987longest} and made
explicit by~\citet*{jacobson1992heaviest} and independently
by~\citet*{pevzner1992matrix}. Interestingly the original presentation
of~\citet*{hunt1977fast} reported an $O((m+ n) \log n)$ time bound, where
$n$ is the number of matches. This is a significant improvement over the
plain dynamic programming algorithm, which always requires $O(m^2)$
time. Although in the worst case $n$ may be $O(m^2)$, in general it may be
significantly smaller.  The original complexity was not always faster than
the plain algorithm, because $n$ may be $\Omega(m / \log m)$. This issue
was addressed by~\citet{apostolico1986improving} which obtained $O(m^2)$
time worst case guarantees. Improvements of the Hunt-Szymanski algorithm
based on bitwise operations where proposed
by~\citet*{crochemore2003speeding}.

The general Multiple Longest Common Subsequence problem is not restricted
to only two strings, instead it considers $k$ strings at a time. This
problem was shown to be NP-Hard by~\citet{Maier_1978}. Considering $k > 2$,
but fixed, several efficient algorithms were
proposed~\citet{Chen_2006,Hakata_1998,Korkin_2008,Wang_2011}. Most of these
methods use the Hunt-Szymanski approach of processing only match
points. Moreover the increasing subsequence problem is generalized nicely
to the notion of point domination, which we also adopted in
Section~\ref{sec:problem}. This line of research obtained increasingly more
efficient results, culminating in the Quick-DP algorithm
of~\citet*{Korkin_2008}. Therefore we compared against this algorithm in
Section~\ref{sec:experimental-results}. Moreover this algorithm contains
several key insights that we used in designing our procedures for the
modification operations. First they divided the matches according to their
maximum levels, like in Table~\ref{tab:new-points}. In Quick-DP these
matches are generated incrementally by level, meaning that all the level
$2$ matches are generated before the level $3$ and so on. After all the
matches of a level are generated they are filtered by efficient algorithm
to compute geometrical minima. The resulting matches are the ones we
defined as non-covered matches in Section~\ref{sec:idea}. One key
achievement of the algorithm by~\citet*{Wang_2011} is that determining
non-covered matches was possible without having to process all possible
matches, as they pointed out in Section 3.2. We used a similar approach in
our procedure for the \texttt{Append()} operation, essentially the notion
of point generation we mentioned.

Given the nature of the \texttt{Append()} operation we can not use such a
tight level separation as in Quick-DP. Instead to obtain a procedure that
follows the same principles we need to use the orthogonal range trees to
replace the multidimensional minima algorithm. This overhead actually leads
to one advantage as we use one letter to generate points, whereas Quick-DP
used the complete alphabet.

\section{Conclusions and further work}
\label{sec:conclusions} 

This paper presented the problem of computing the longest common
sub-sequence of multiple strings when these strings are subject to
modifications, in particular the \texttt{Append()} and \texttt{Pop()}
operations, Sections~\ref{sec:idea} and~\ref{sec:textttpop-operation}. We
proposed a data structure to represent the underlying MLCS that is well
suited to the modification operations we considered. This structured relied
on dynamic orthogonal range trees, described in
Section~\ref{sec:dynam-orth-range} to support the modification algorithms.

We tested these algorithms experimentally,
Section~\ref{sec:experimental-results}, and confirmed that these algorithms
can outperform state of the art alternatives. Particularly when the
alphabet size is big and/or the number of strings to consider is big. This
makes it particularly well suited for the bioinformatics, in particular we
tested subsequences of proteins.

We plan to further develop this line of research, in particular by
extending the amount of operations, for example \texttt{Prepend()} and
\texttt{Trim()}. Where \texttt{Prepend()} inserts a letter at the beginning
of a sequence and \texttt{Trim()} removes the last letter from a
sequence. Like the \texttt{Append()} operation the \texttt{Trim()}
operation seems to be the simplest, it should only require deleting some
matches from the structure. The \texttt{Prepend()} operation on the other
hand seems similar to the \texttt{Pop()} operation and will involve
transferring matches among lists. We also intend to study much more general
operations, such as \texttt{Insert()} and \texttt{Delete()} that inserts or
removes a letter from an arbitrary position of the string.

\appendix
In this section we show all the plots we obtained from our experimental
results. The first set of plots are obtained from the first generator. The
second set of plots are obtained from the second generator, the title of
these second plots ends in \verb|_M|. The final results, with $S=21$, are
from the protein data set.
\includegraphics{random_S1_K2.eps}
\includegraphics{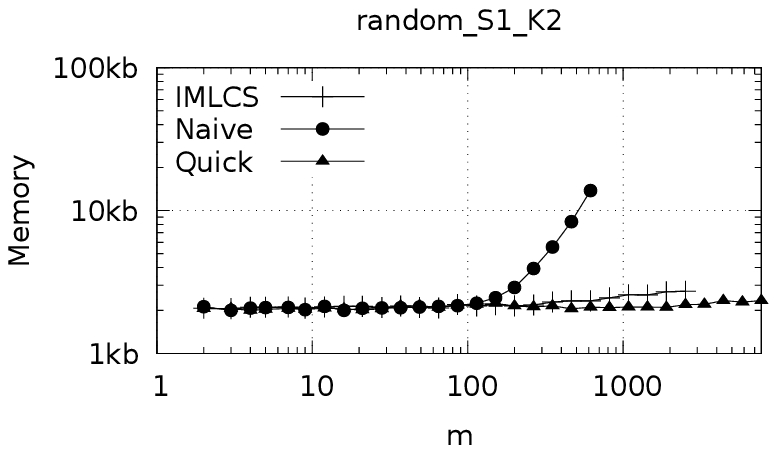}
\includegraphics{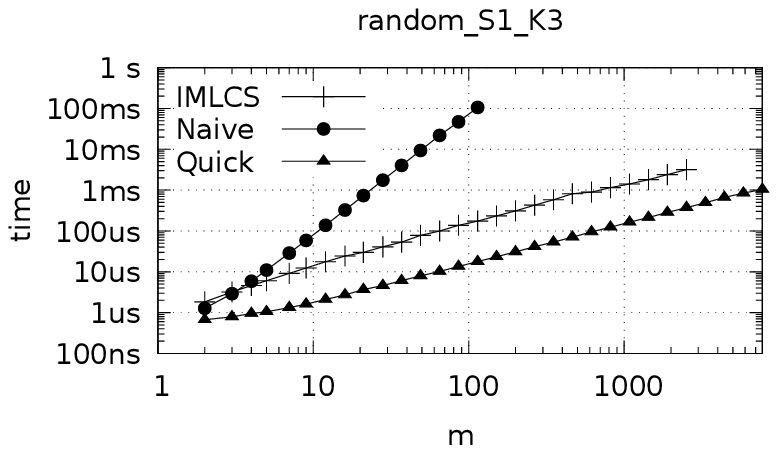}
\includegraphics{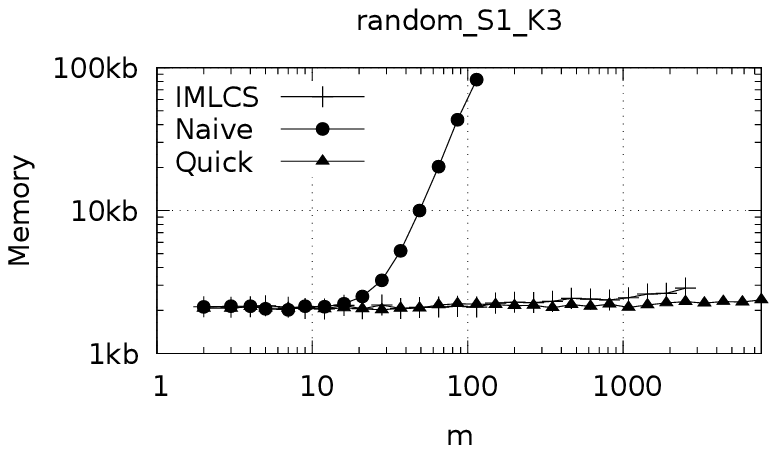}
\includegraphics{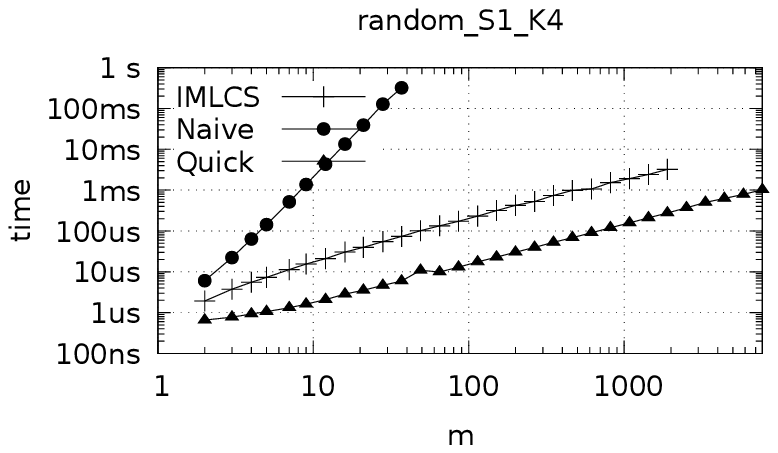}
\includegraphics{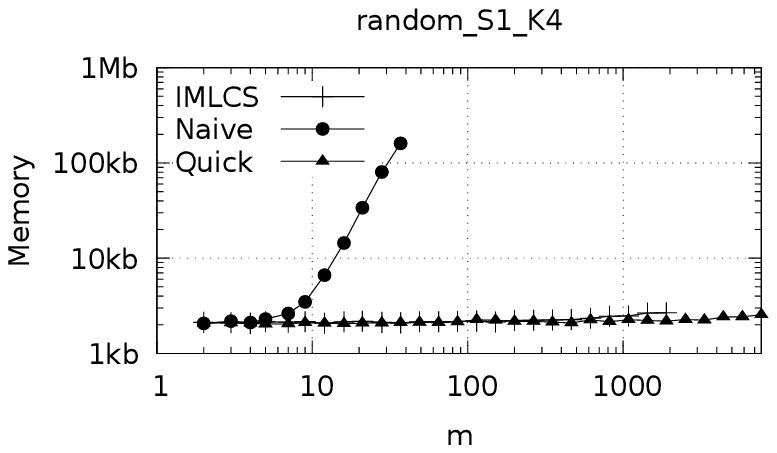}
\includegraphics{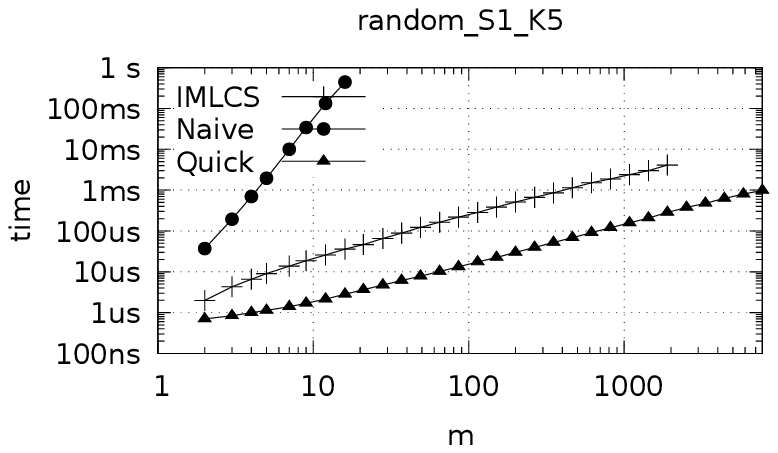}
\includegraphics{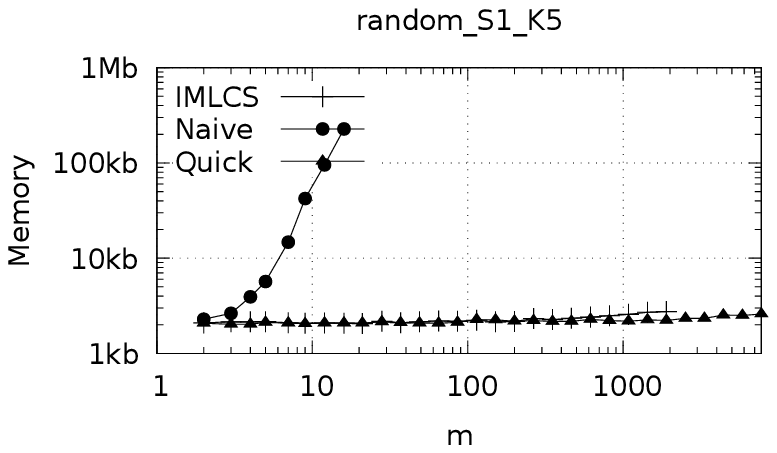}
\includegraphics{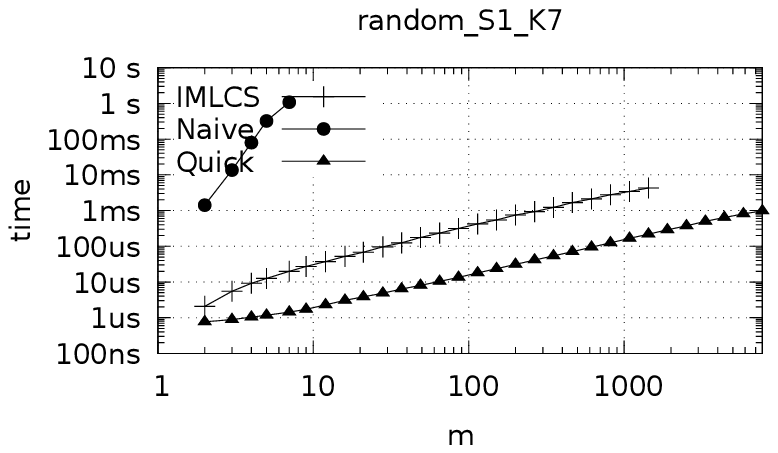}
\includegraphics{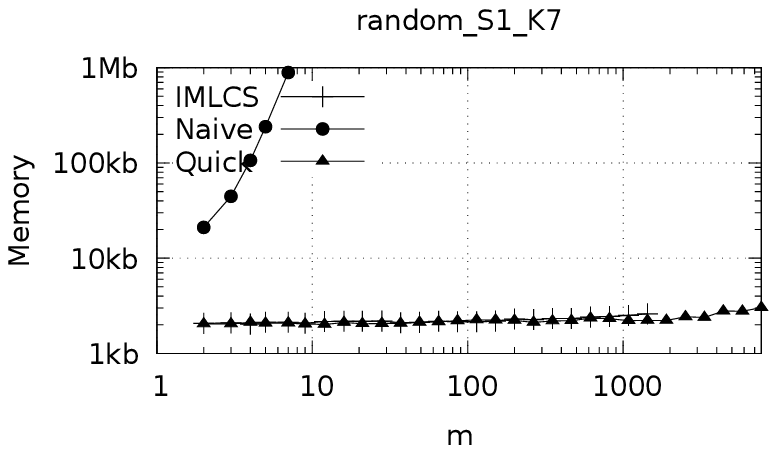}
\includegraphics{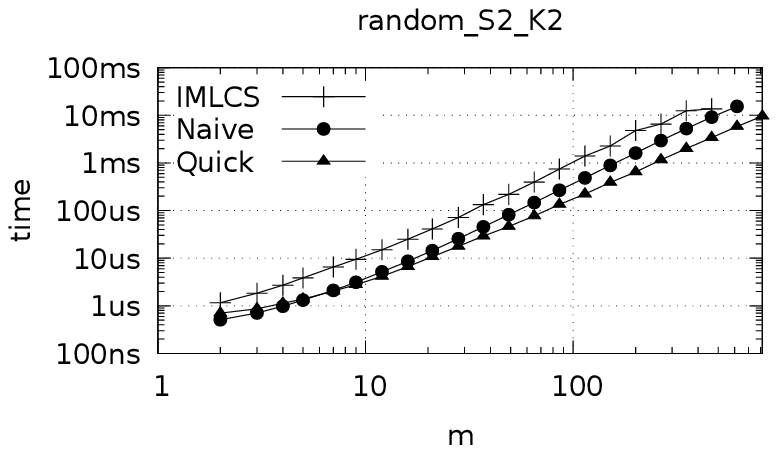}
\includegraphics{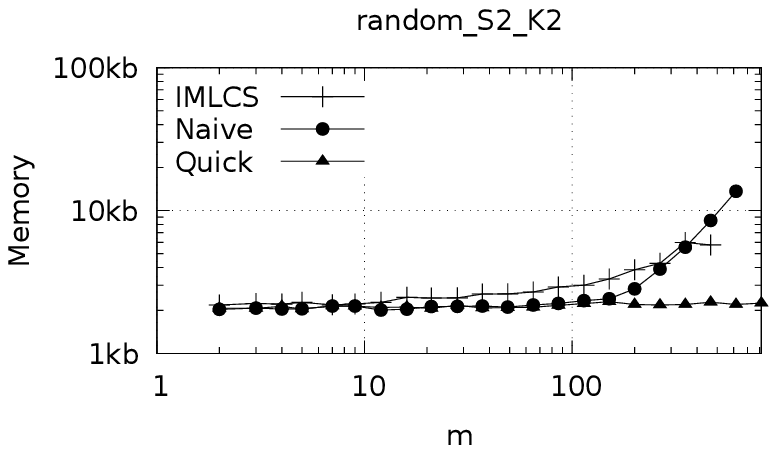}
\includegraphics{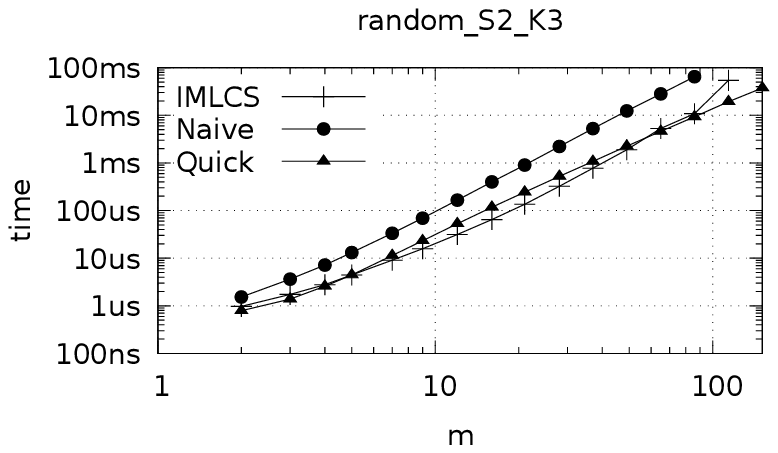}
\includegraphics{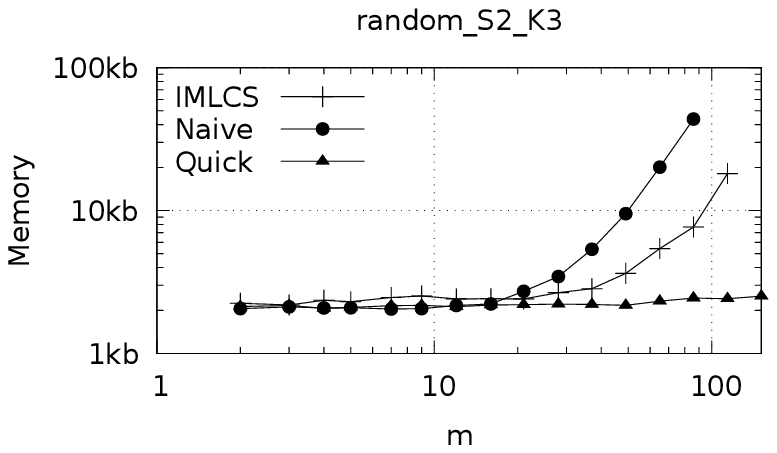}
\includegraphics{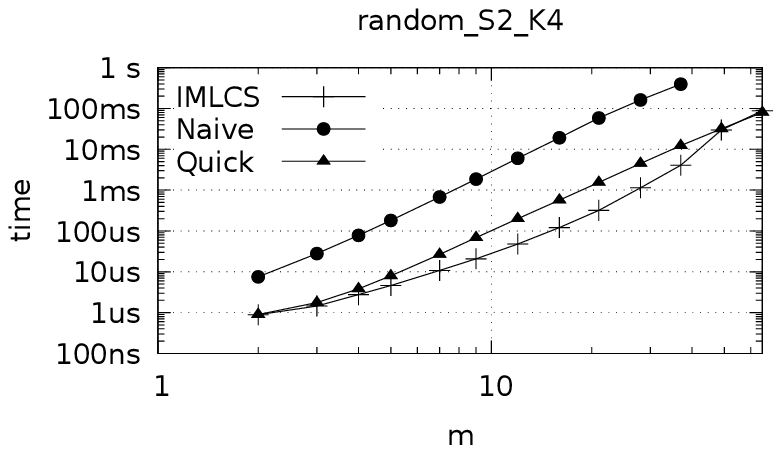}
\includegraphics{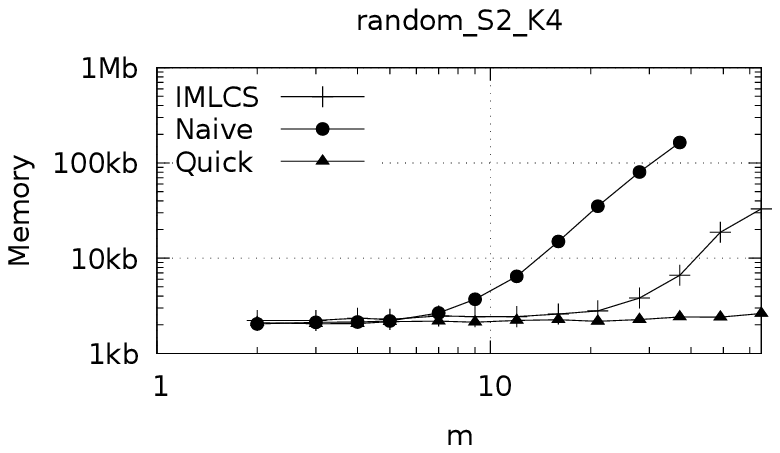}
\includegraphics{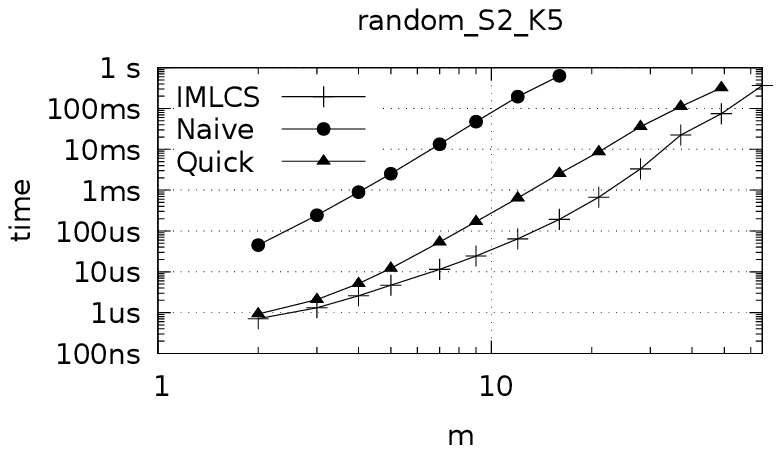}
\includegraphics{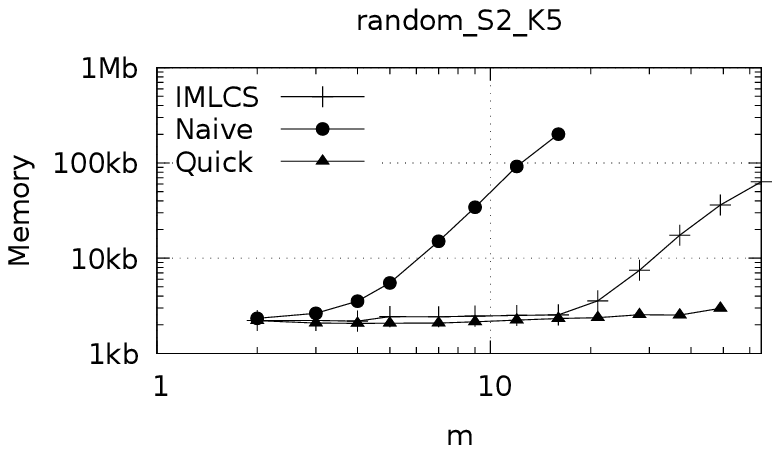}
\includegraphics{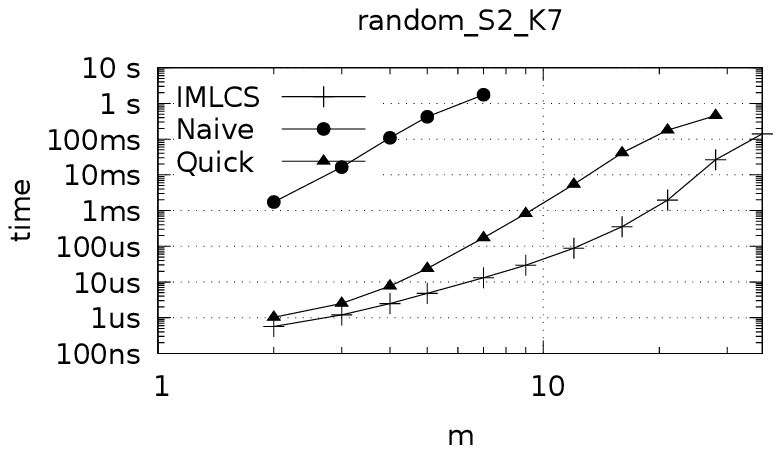}
\includegraphics{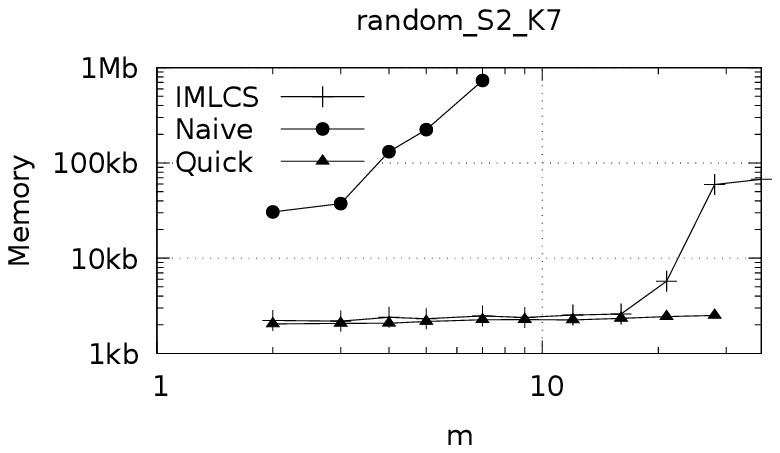}
\includegraphics{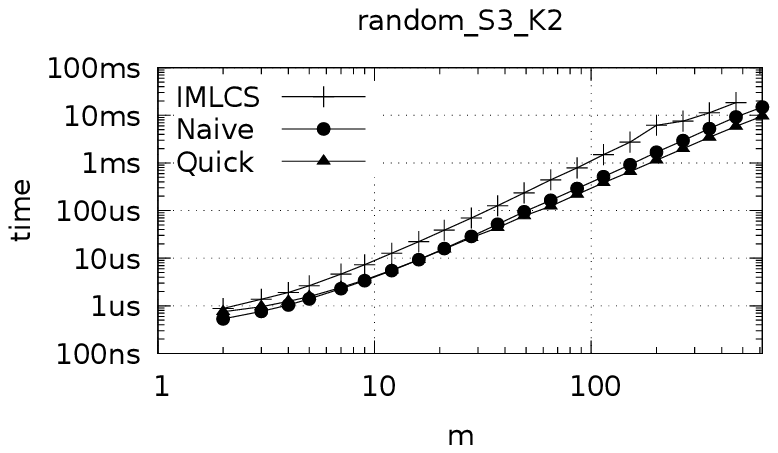}
\includegraphics{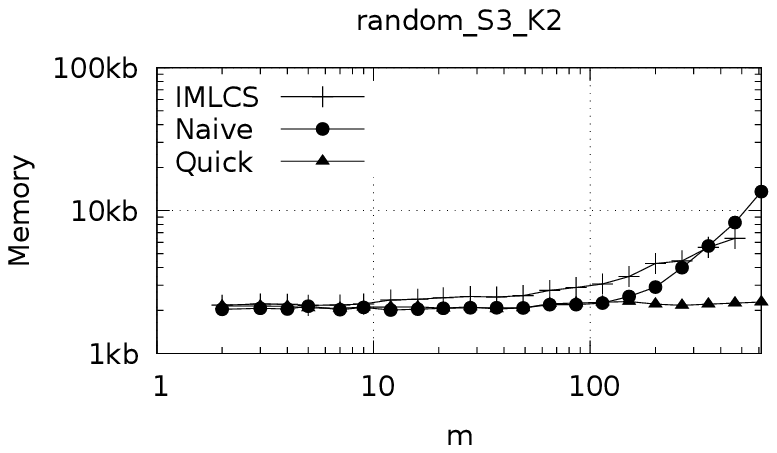}
\includegraphics{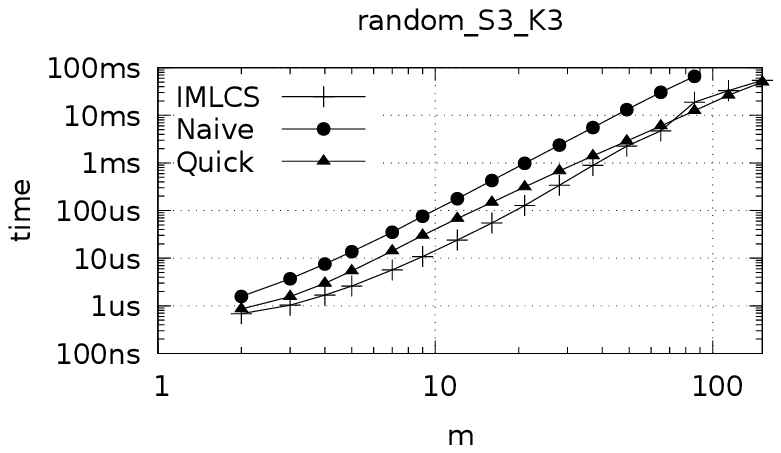}
\includegraphics{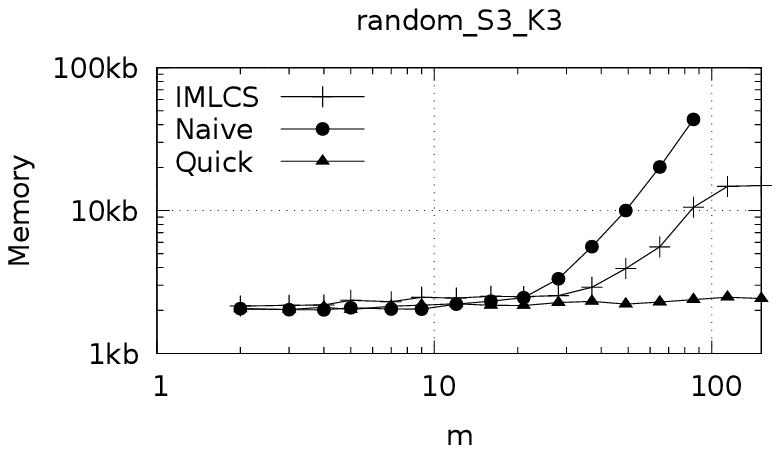}
\includegraphics{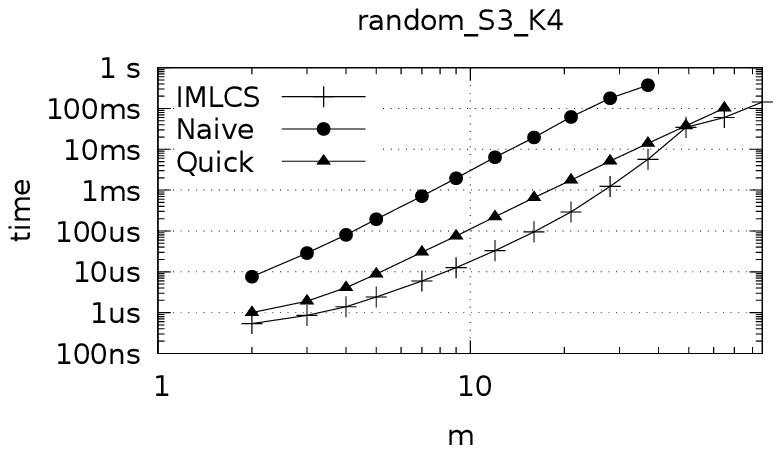}
\includegraphics{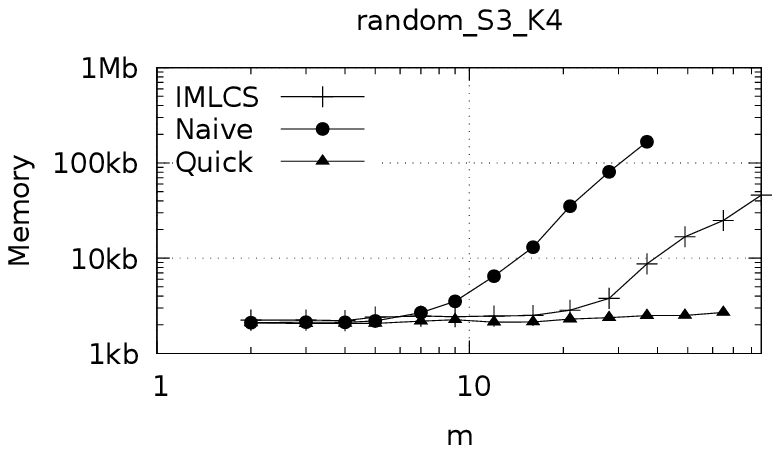}
\includegraphics{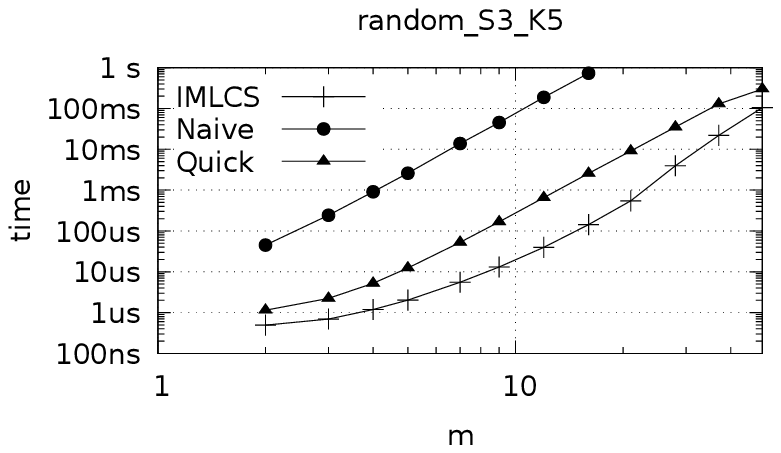}
\includegraphics{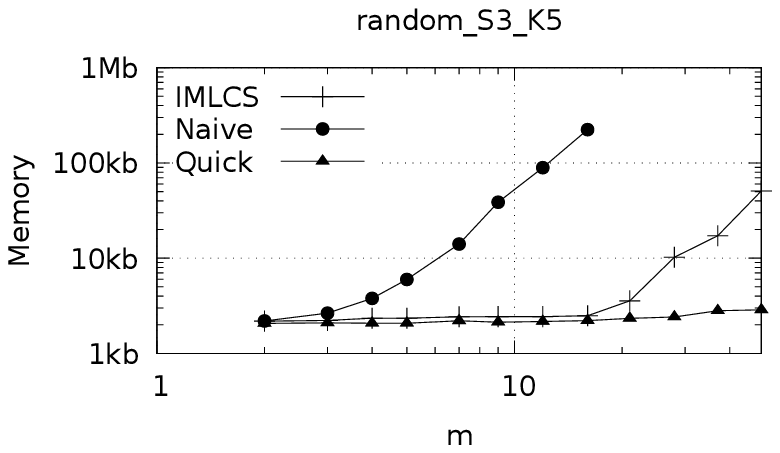}
\includegraphics{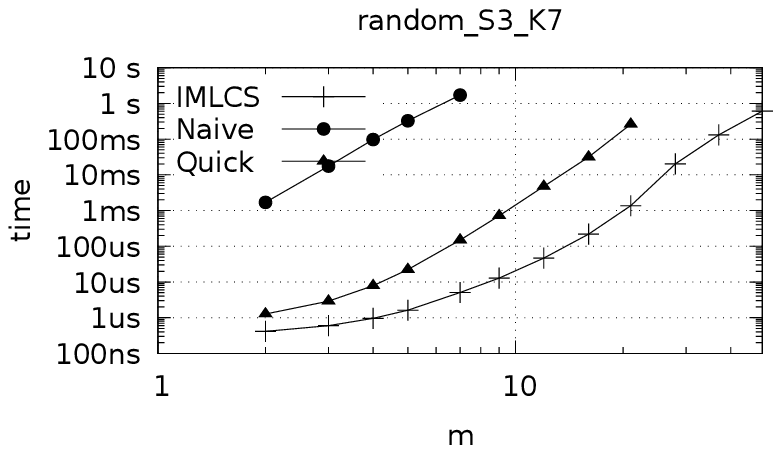}
\includegraphics{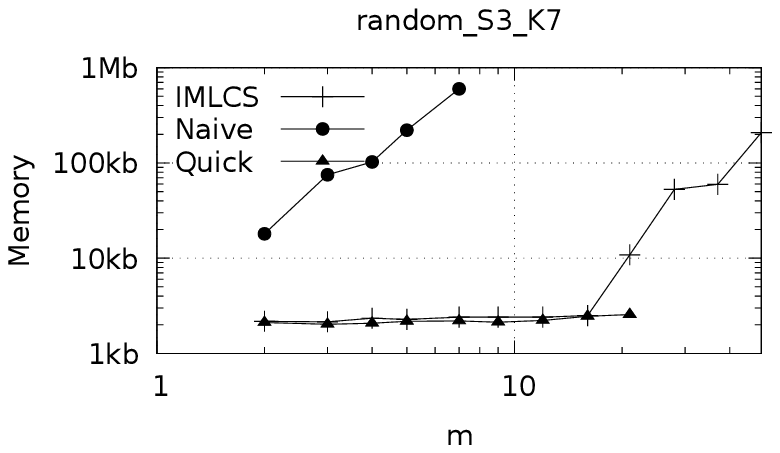}
\includegraphics{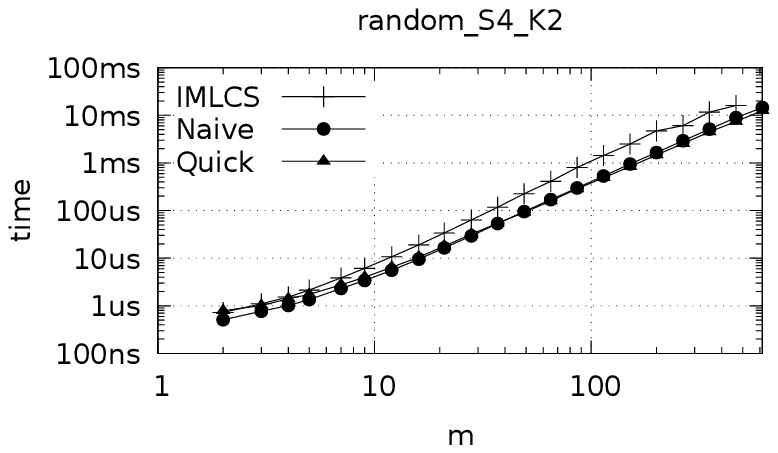}
\includegraphics{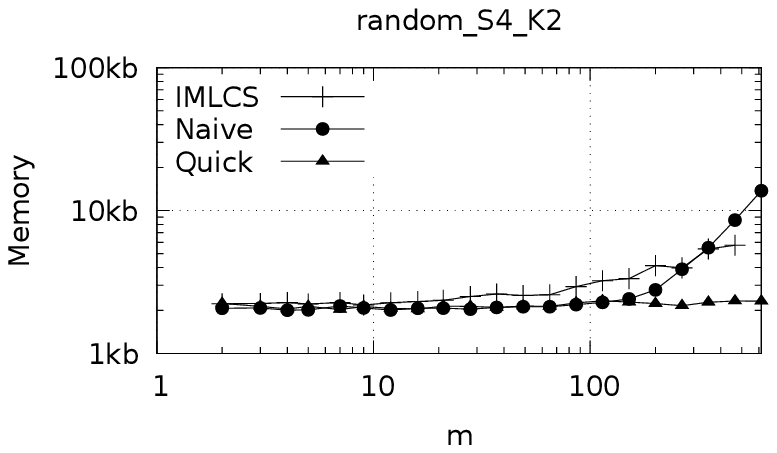}
\includegraphics{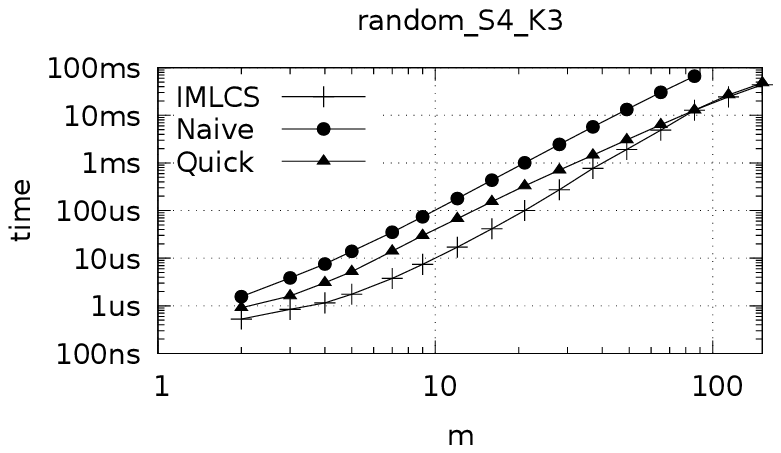}
\includegraphics{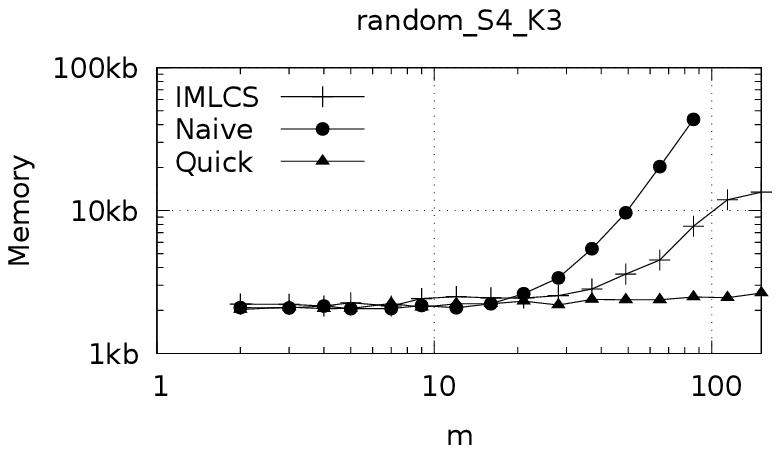}
\includegraphics{random_S4_K4.eps}
\includegraphics{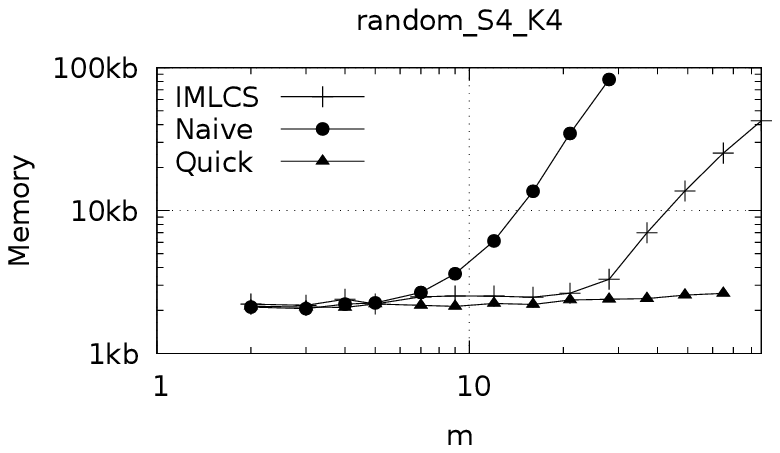}
\includegraphics{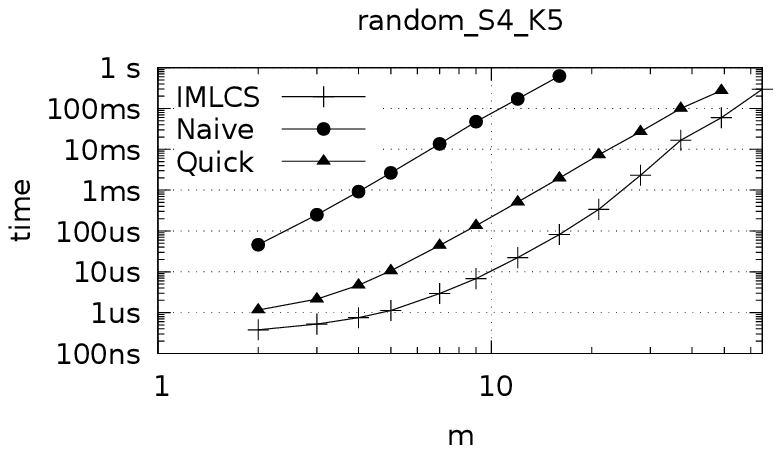}
\includegraphics{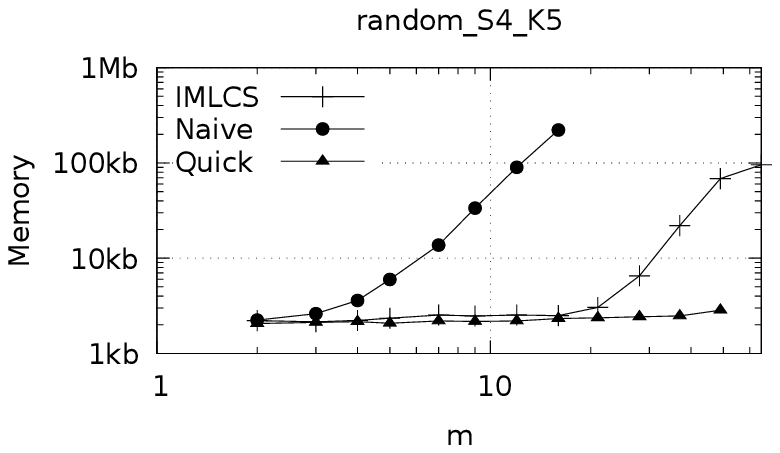}
\includegraphics{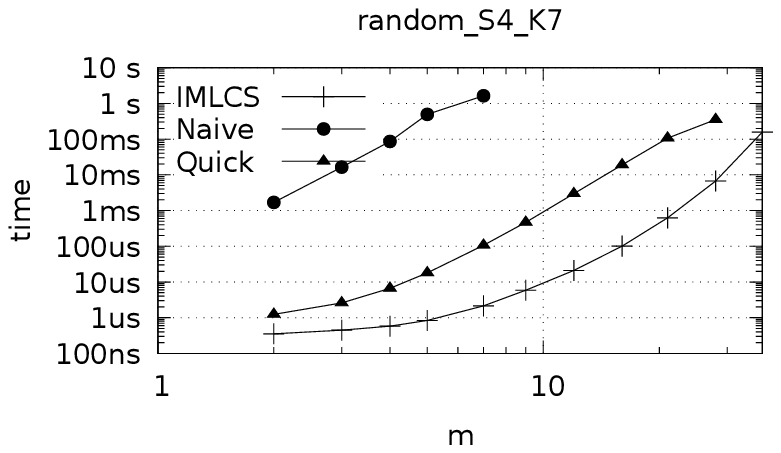}
\includegraphics{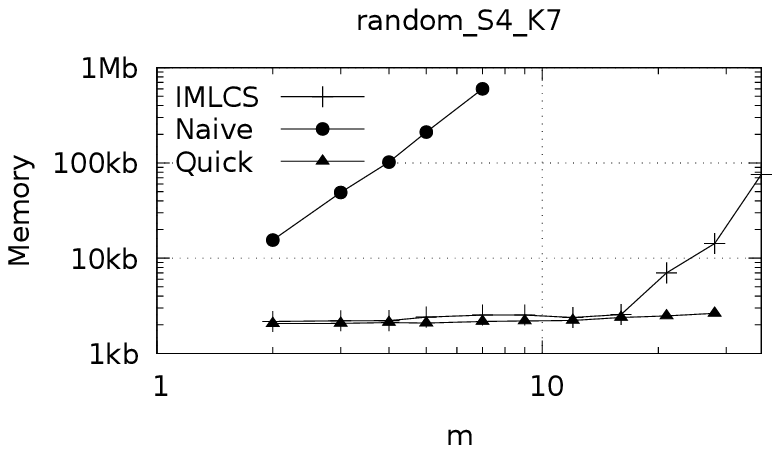}
\includegraphics{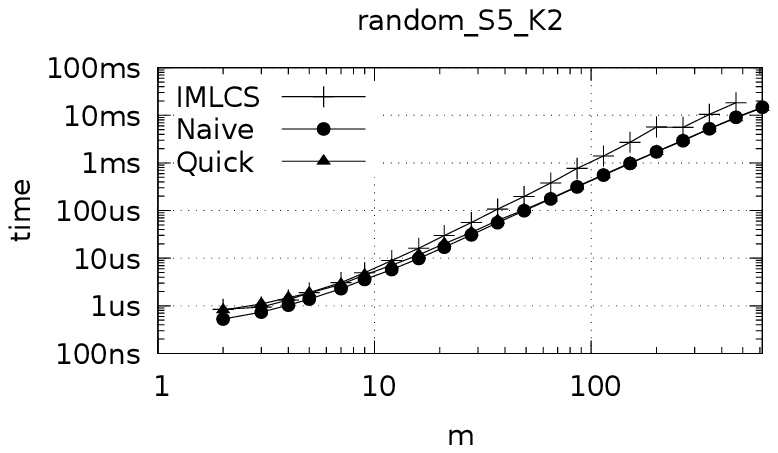}
\includegraphics{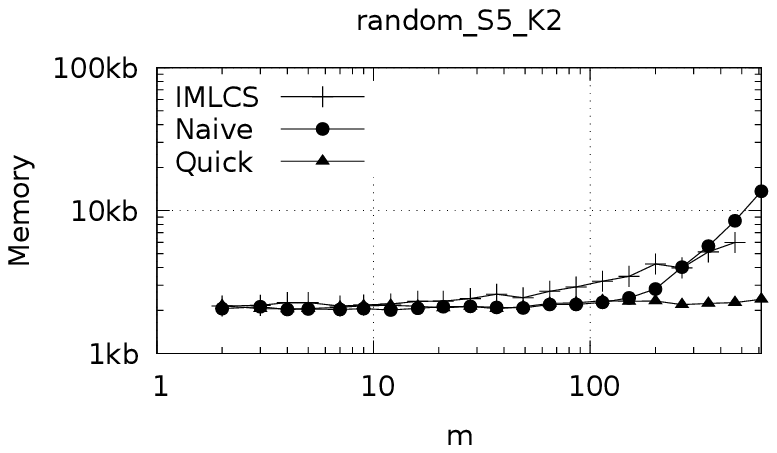}
\includegraphics{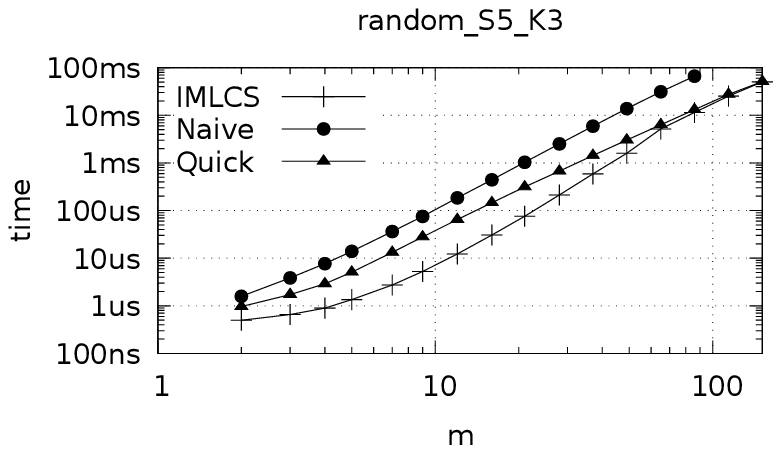}
\includegraphics{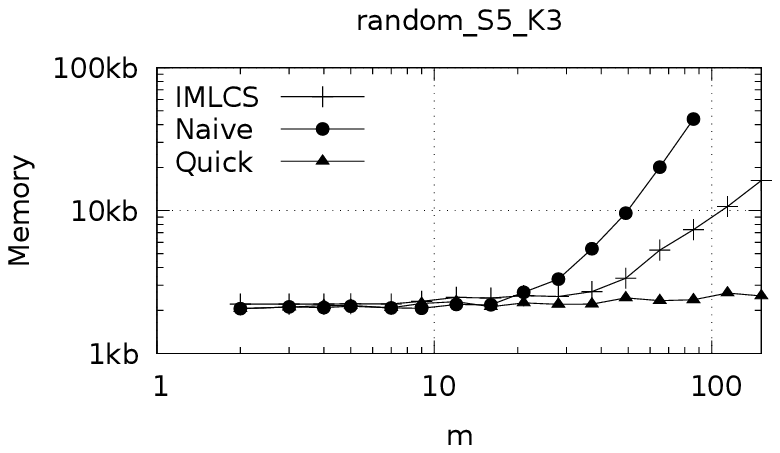}
\includegraphics{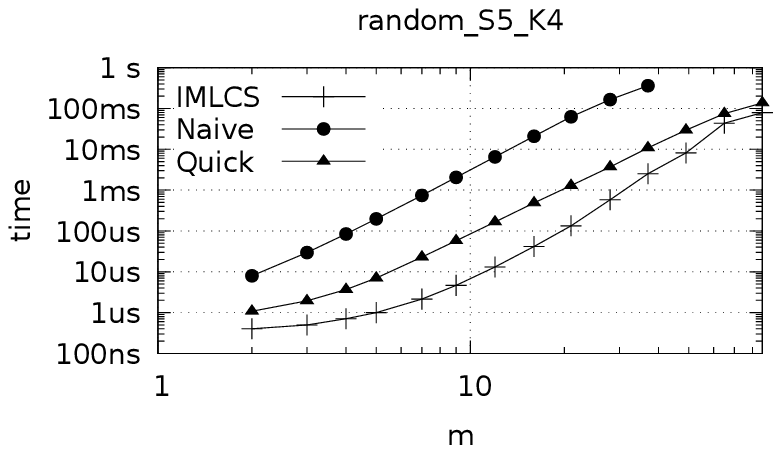}
\includegraphics{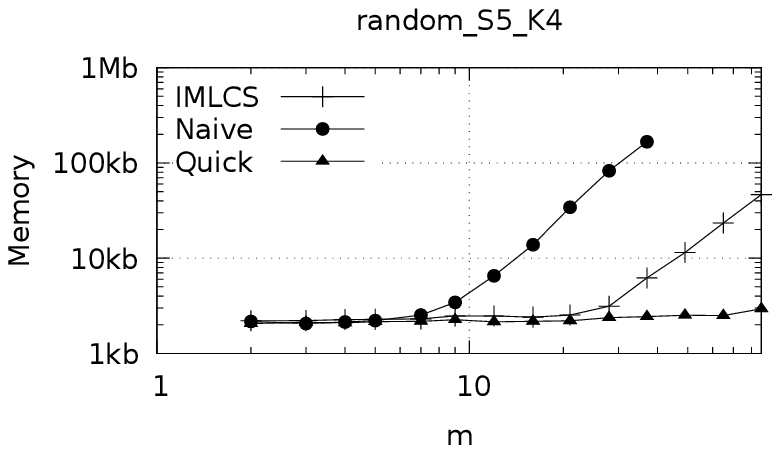}
\includegraphics{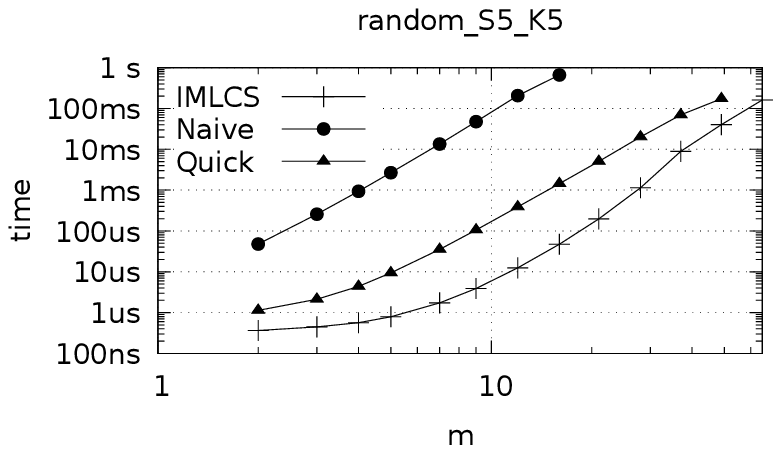}
\includegraphics{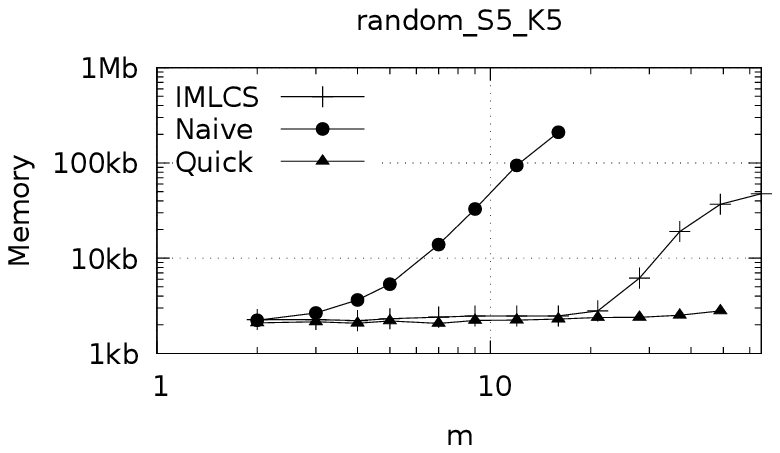}
\includegraphics{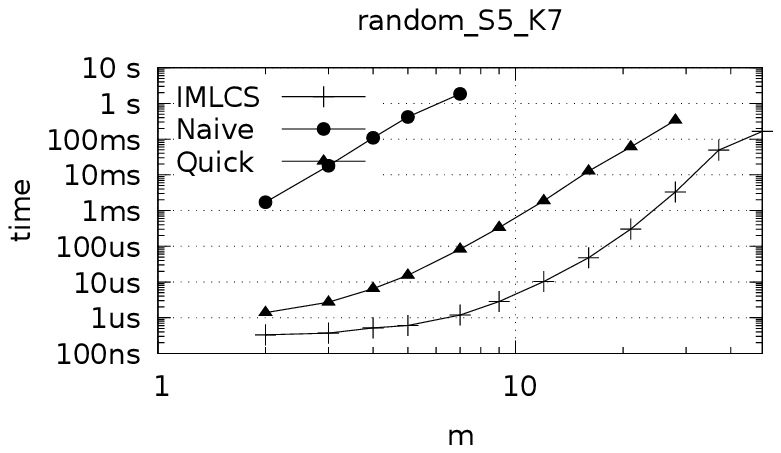}
\includegraphics{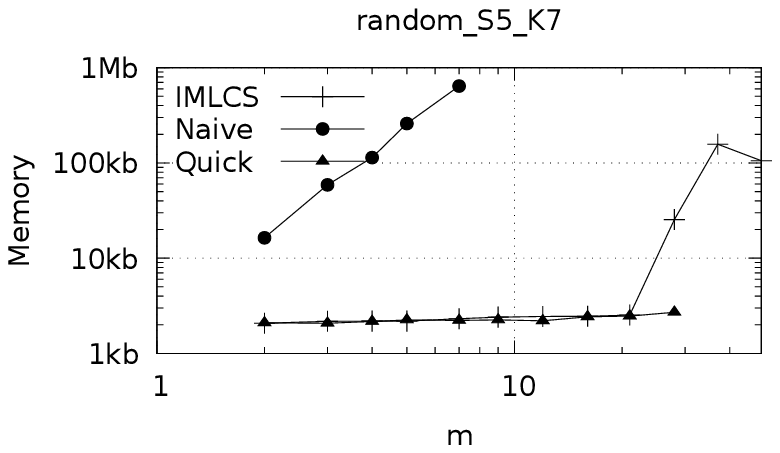}
\includegraphics{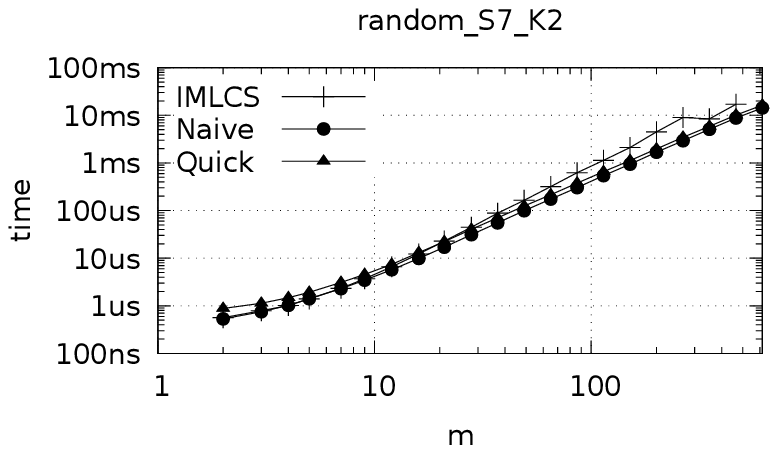}
\includegraphics{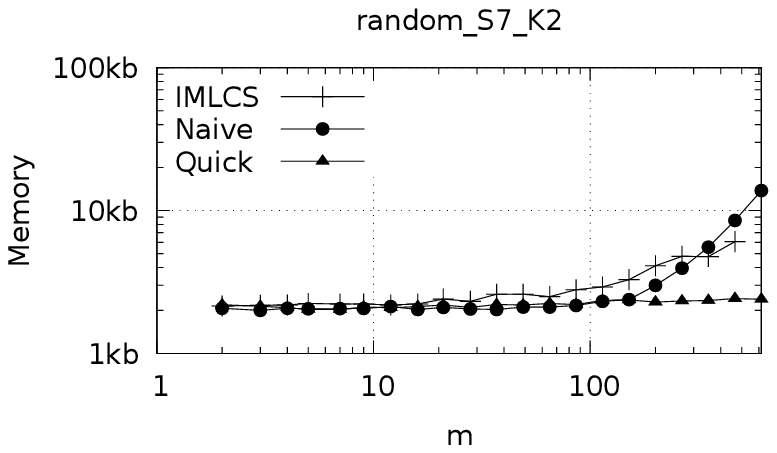}
\includegraphics{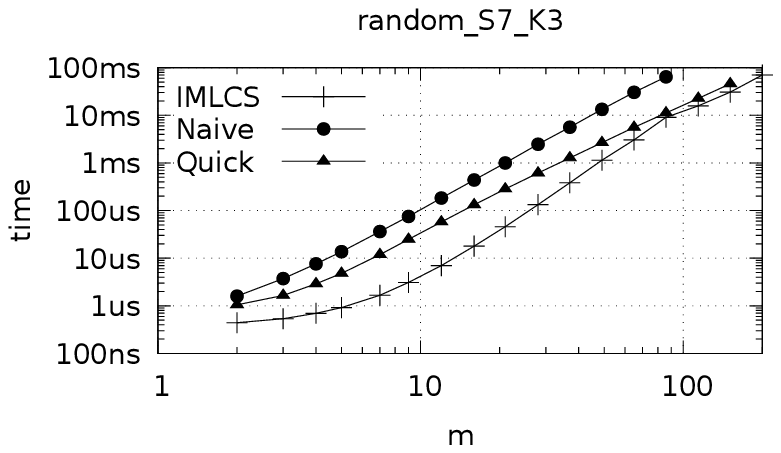}
\includegraphics{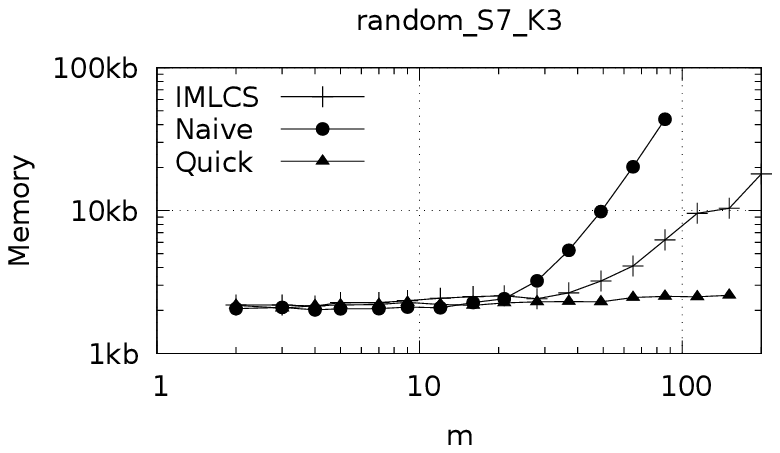}
\includegraphics{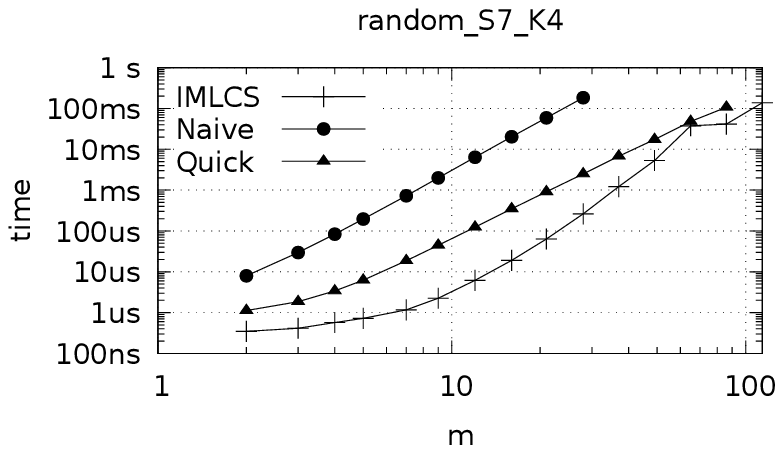}
\includegraphics{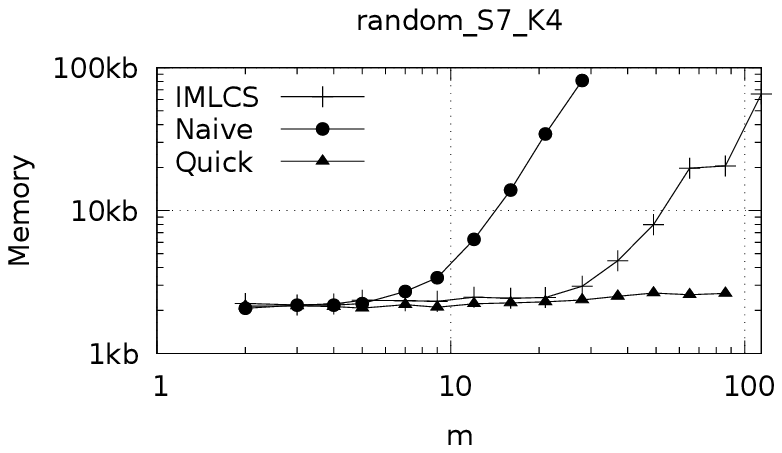}
\includegraphics{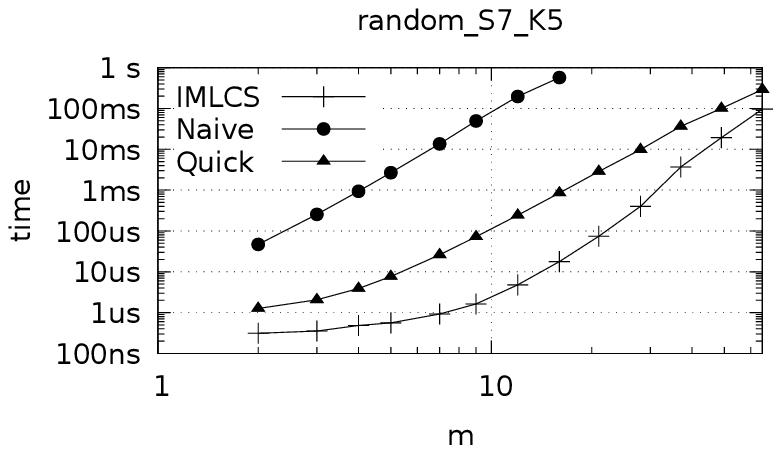}
\includegraphics{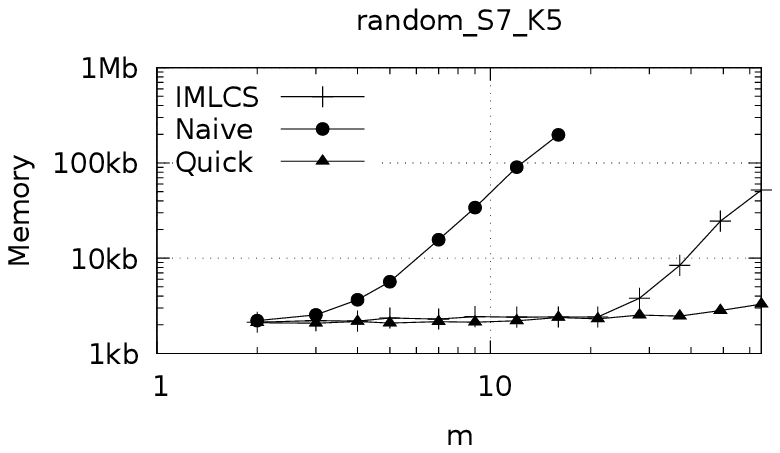}
\includegraphics{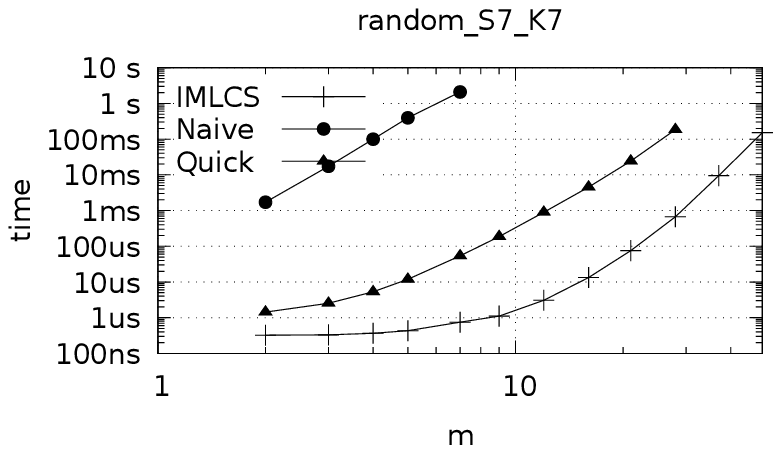}
\includegraphics{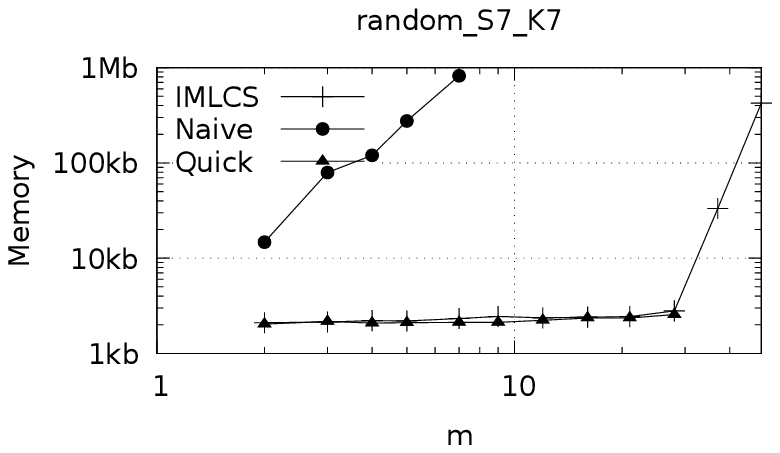}
\includegraphics{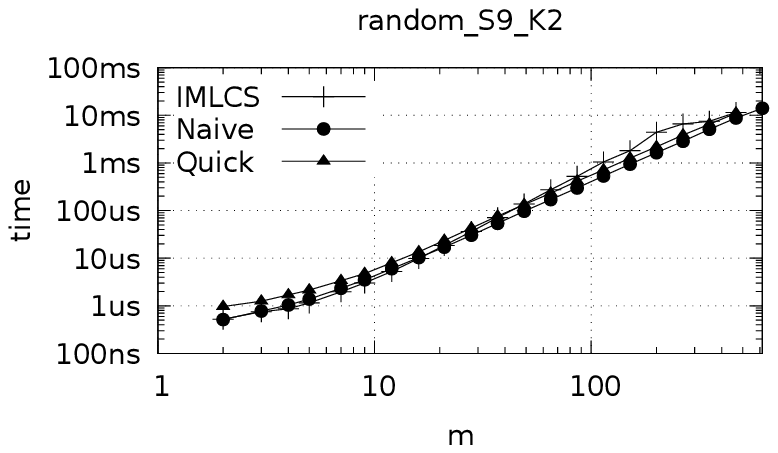}
\includegraphics{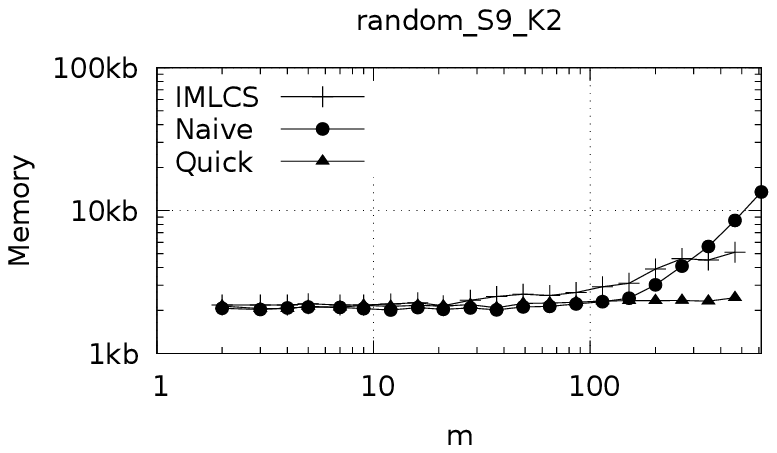}
\includegraphics{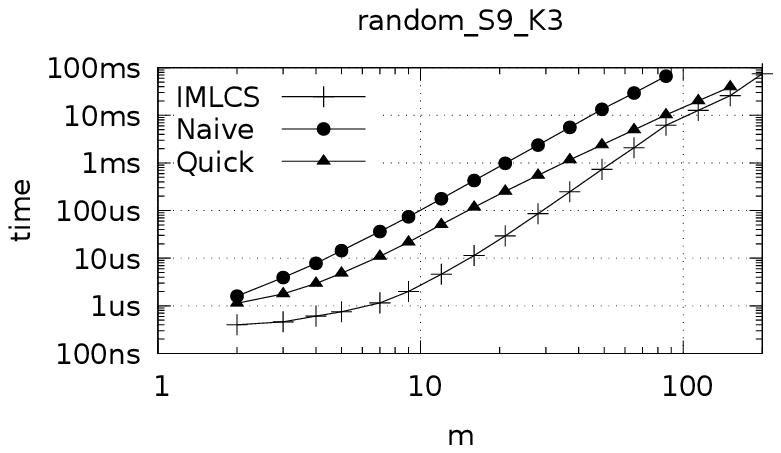}
\includegraphics{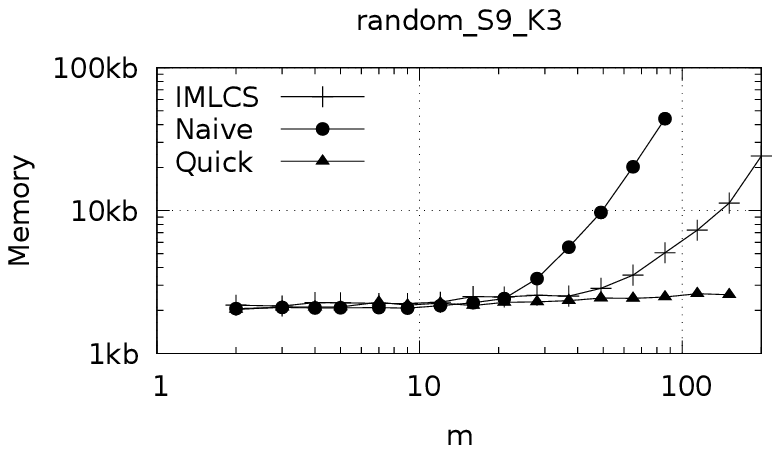}
\includegraphics{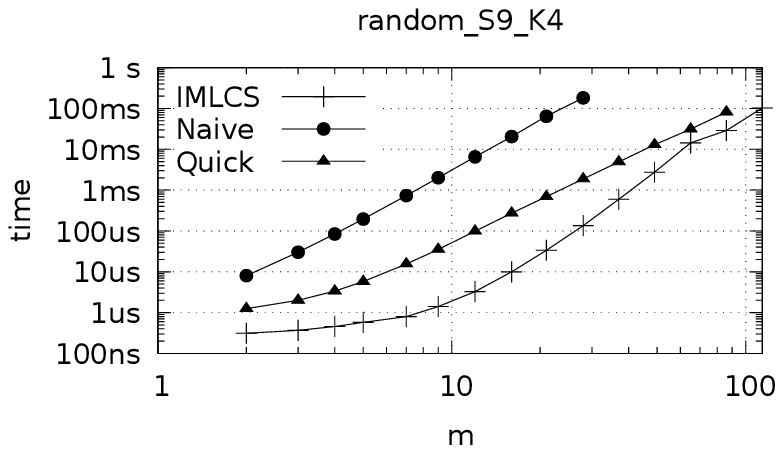}
\includegraphics{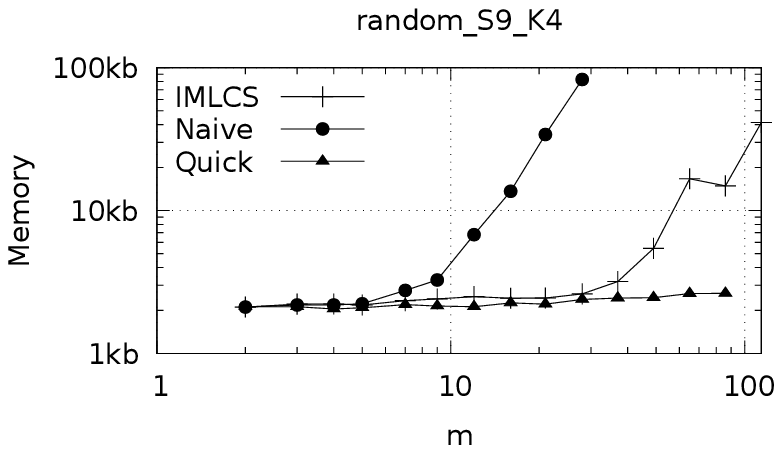}
\includegraphics{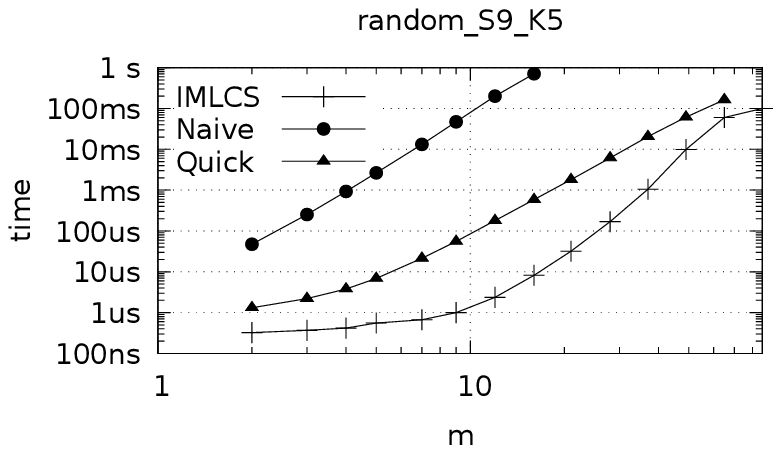}
\includegraphics{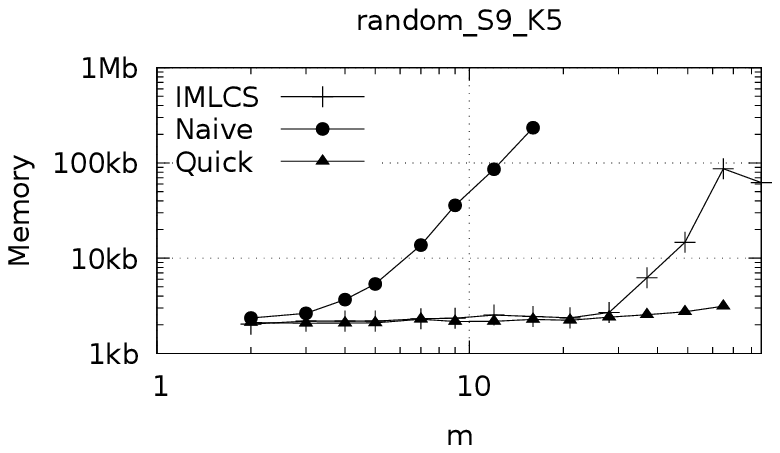}
\includegraphics{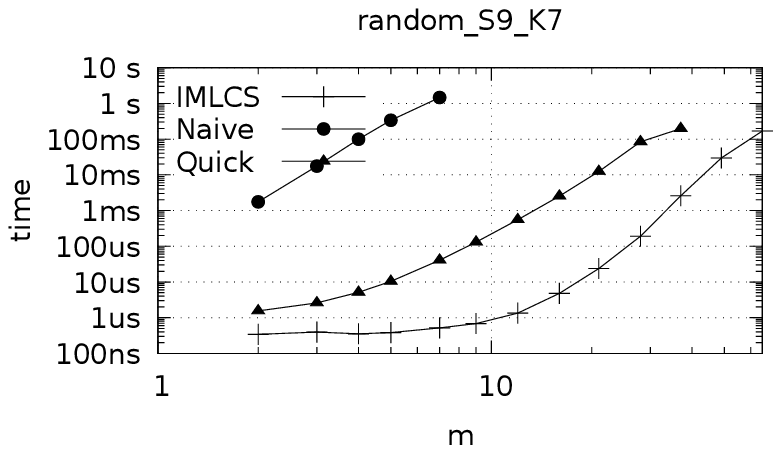}
\includegraphics{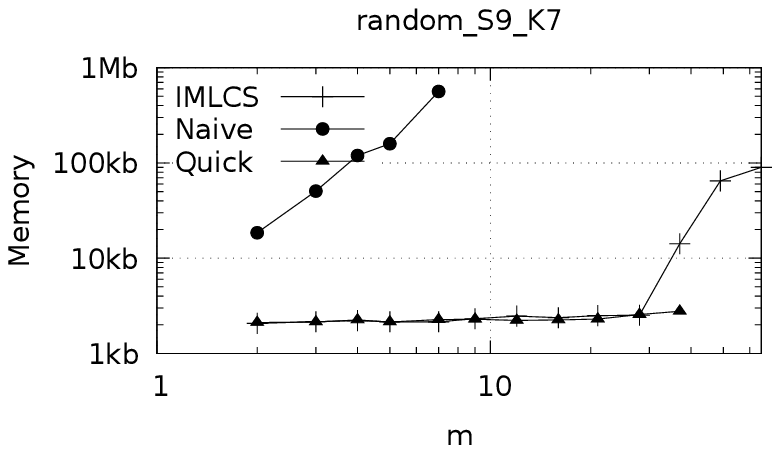}
\includegraphics{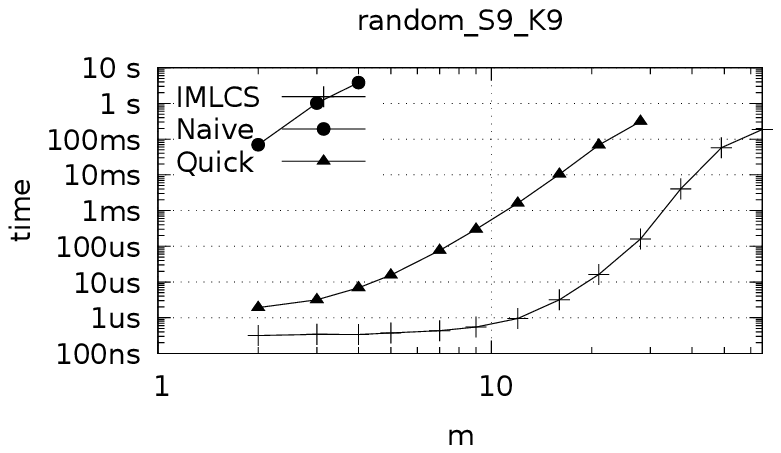}
\includegraphics{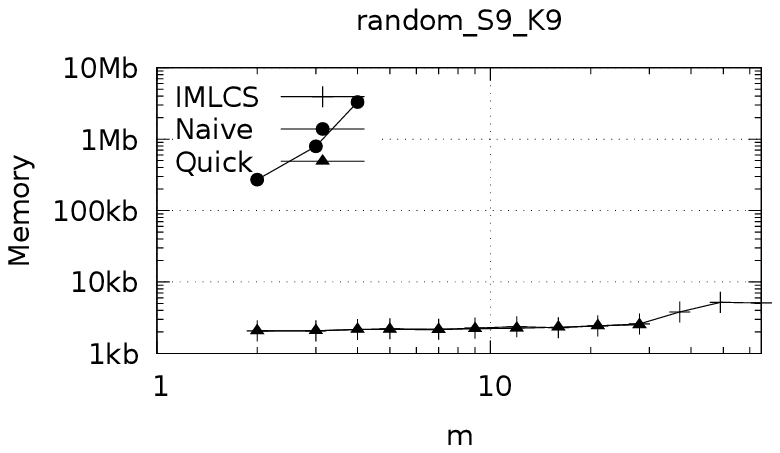}
\includegraphics{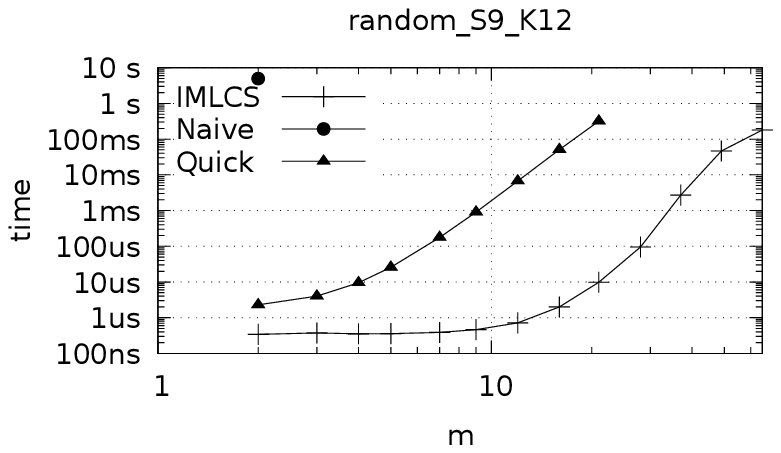}
\includegraphics{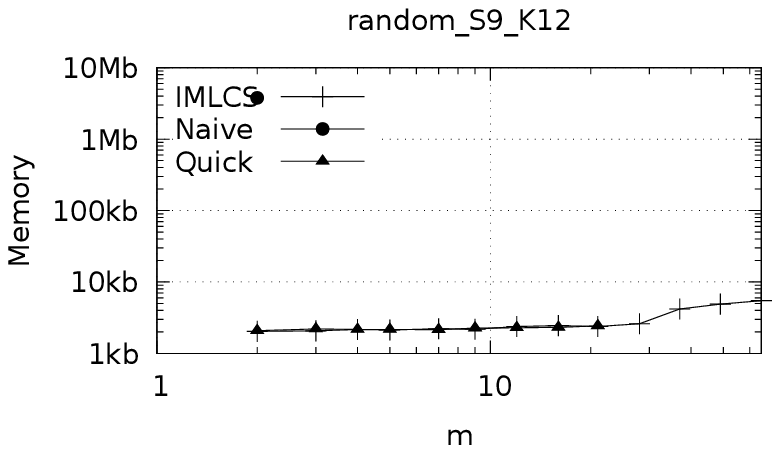}
\includegraphics{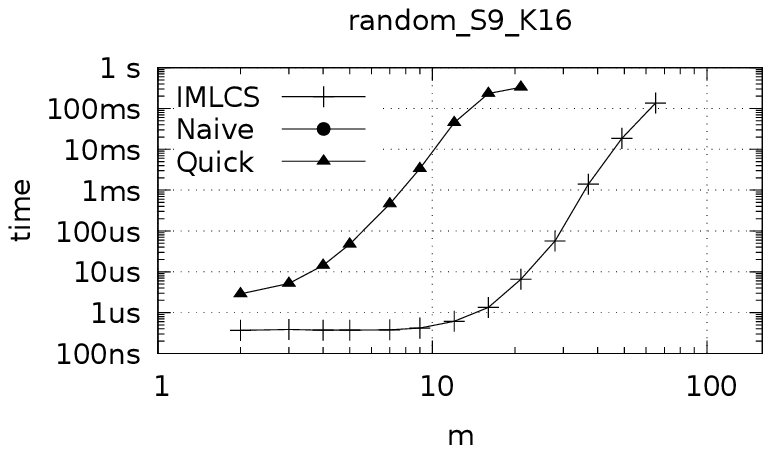}
\includegraphics{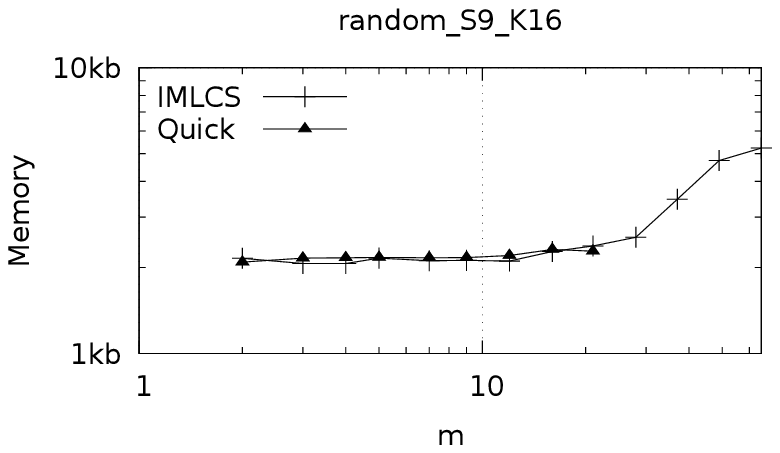}
\includegraphics{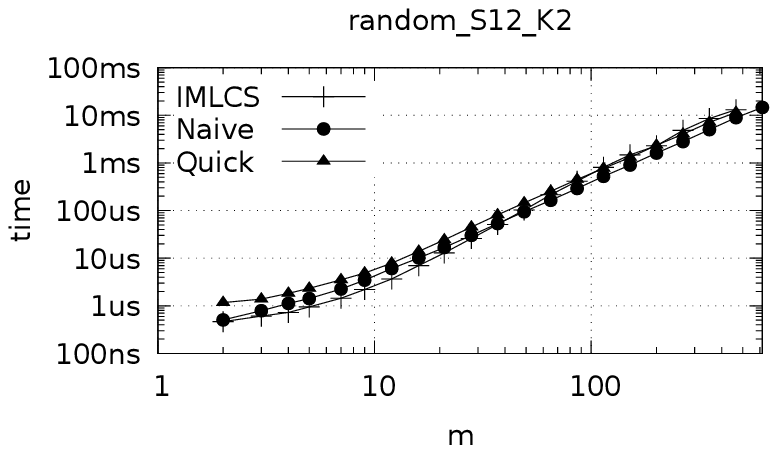}
\includegraphics{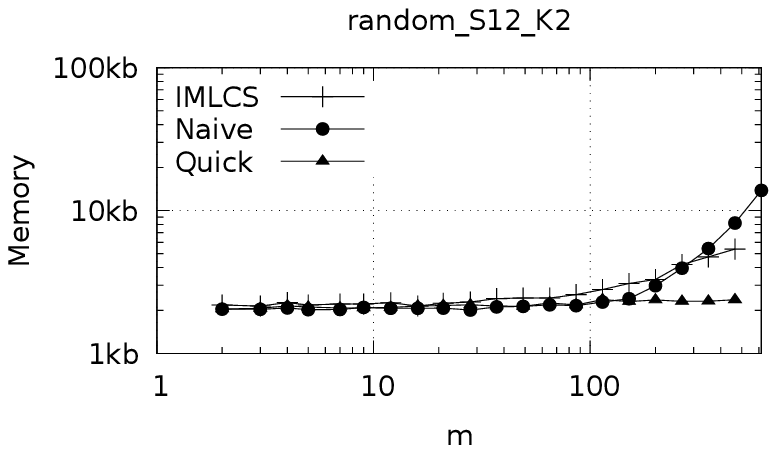}
\includegraphics{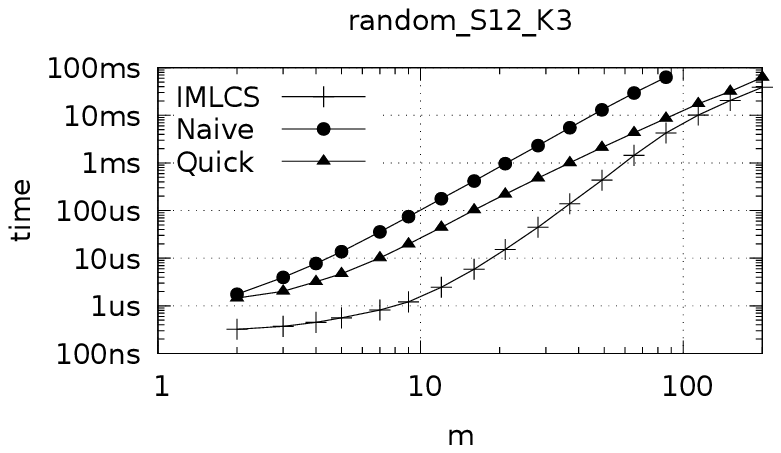}
\includegraphics{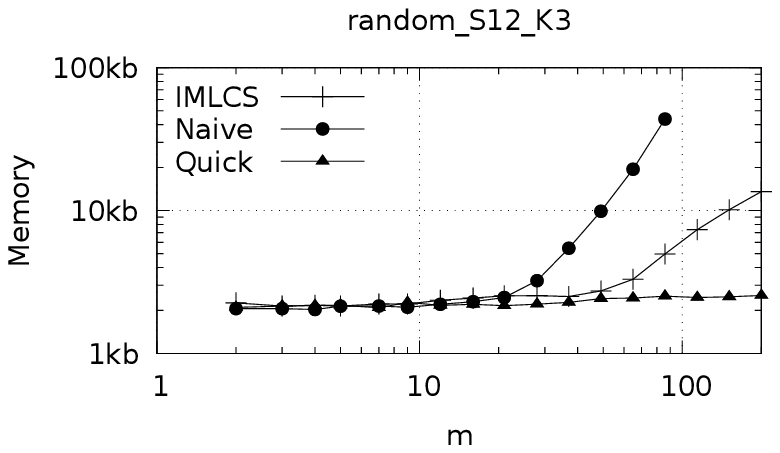}
\includegraphics{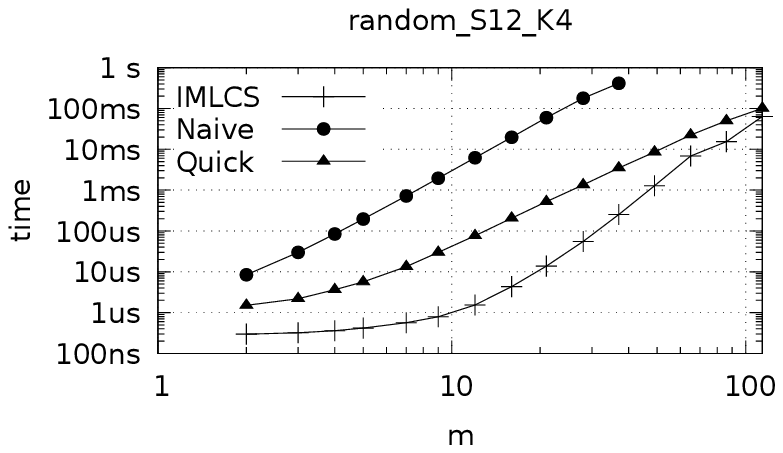}
\includegraphics{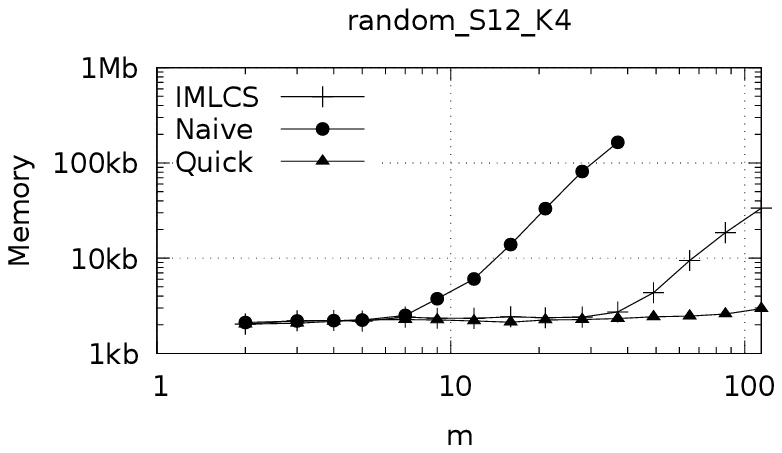}
\includegraphics{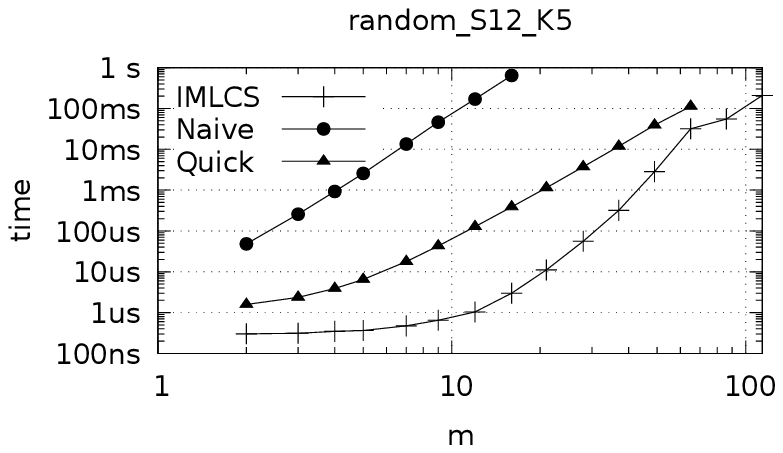}
\includegraphics{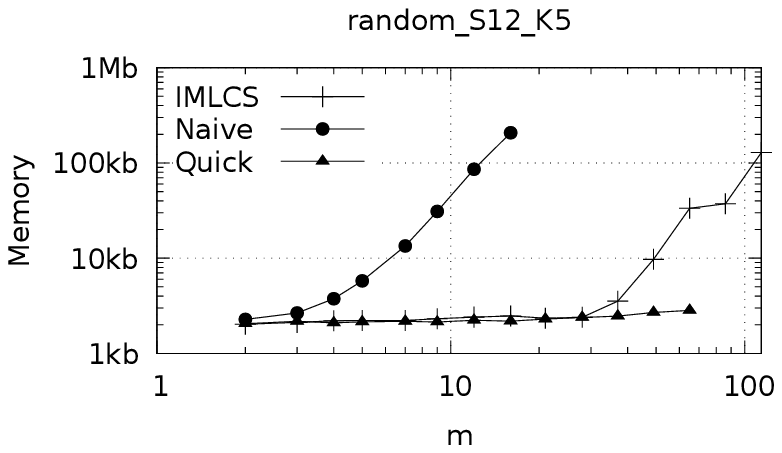}
\includegraphics{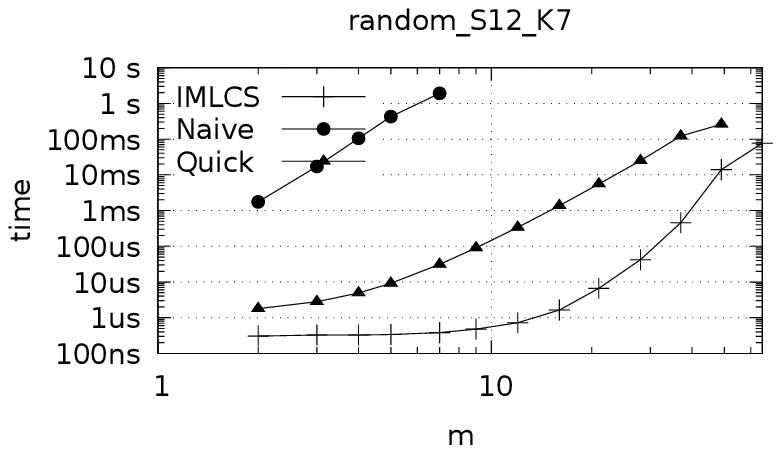}
\includegraphics{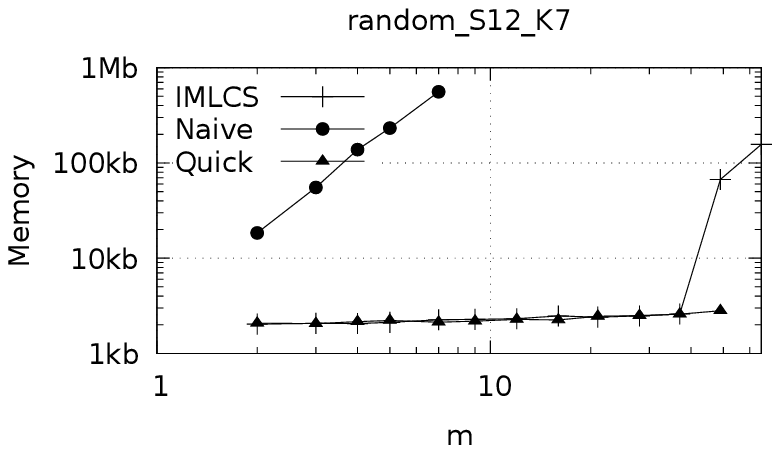}
\includegraphics{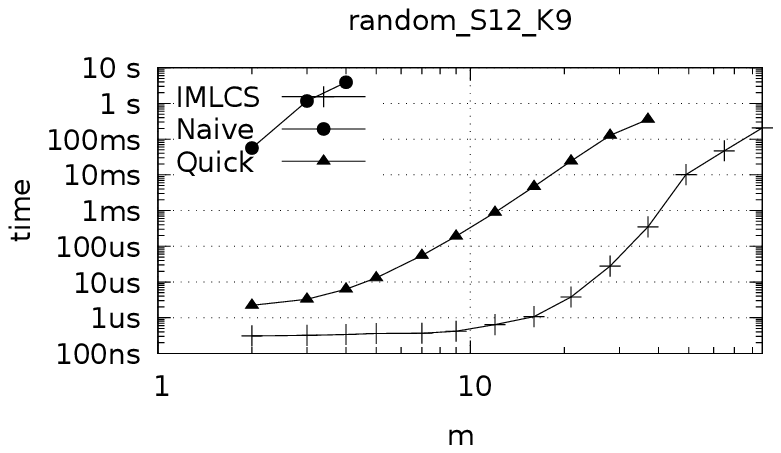}
\includegraphics{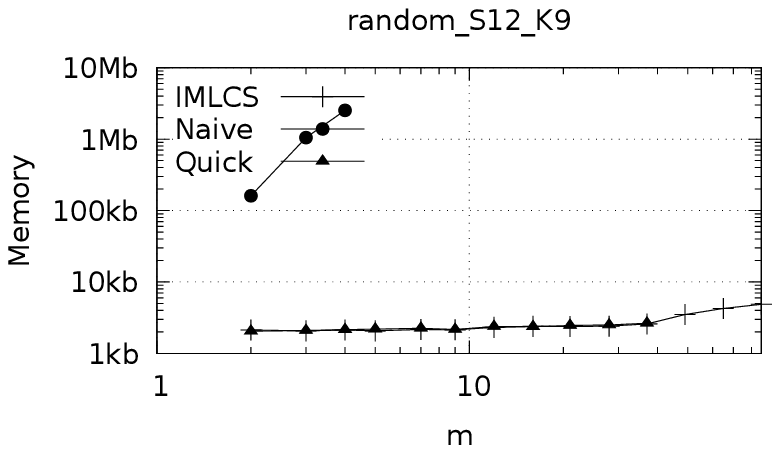}
\includegraphics{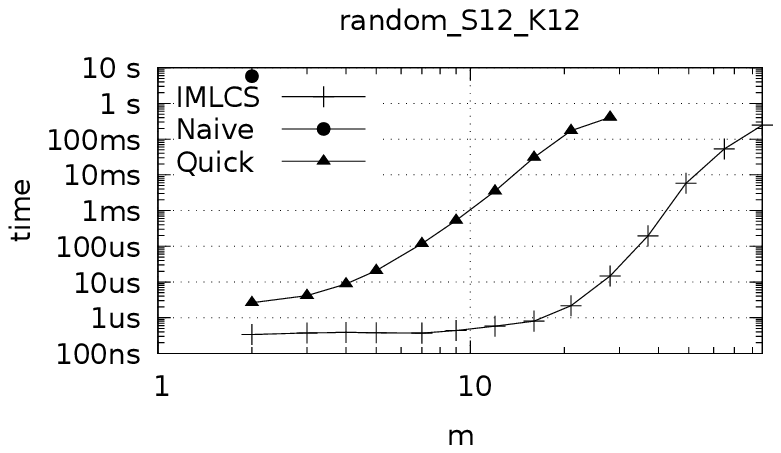}
\includegraphics{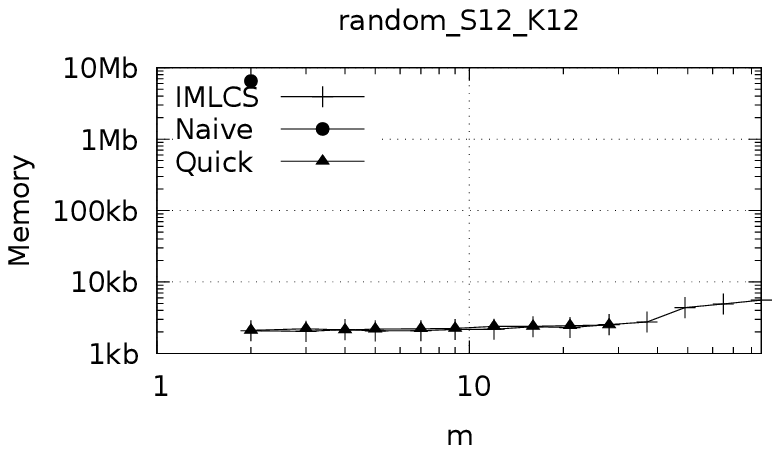}
\includegraphics{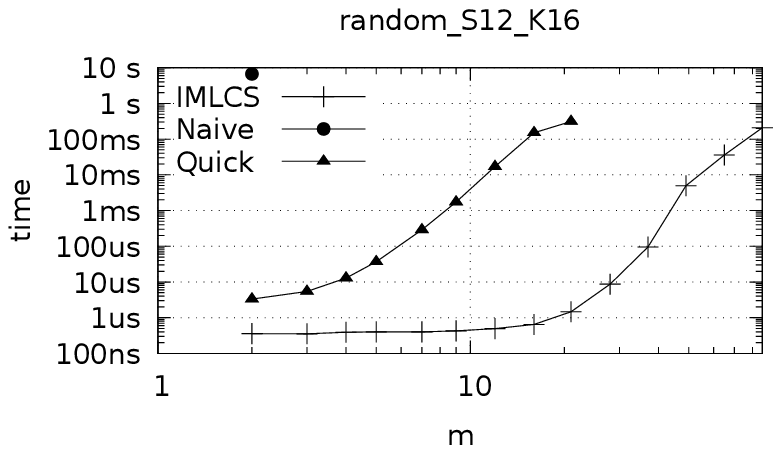}
\includegraphics{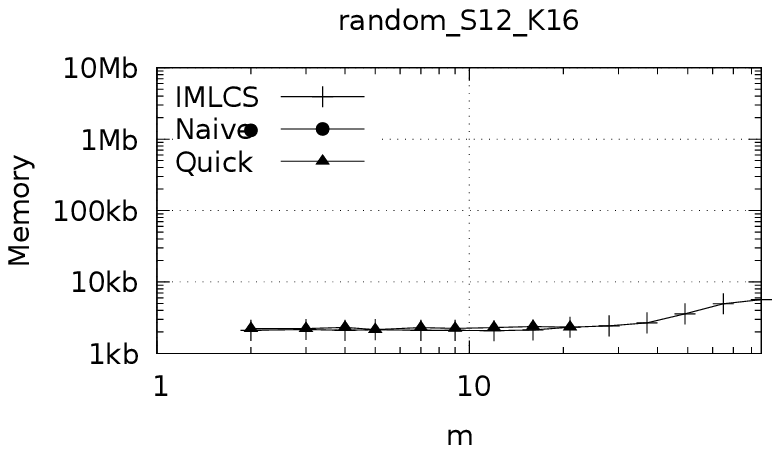}
\includegraphics{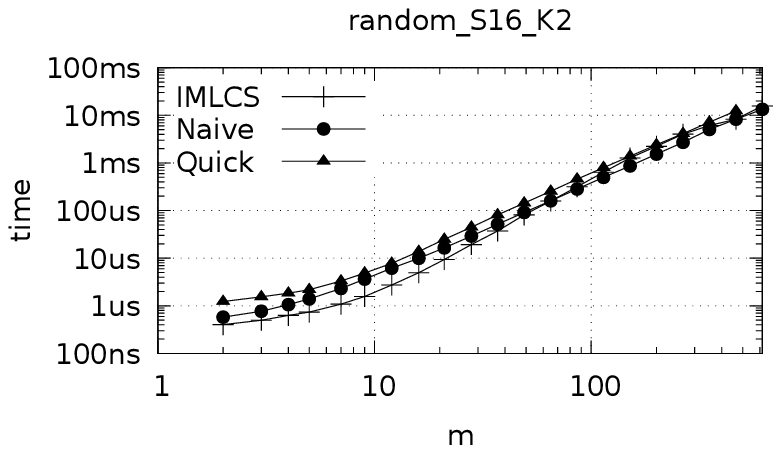}
\includegraphics{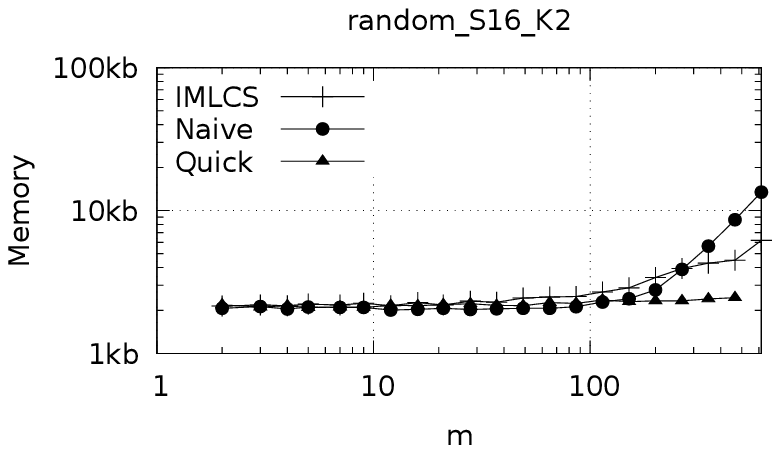}
\includegraphics{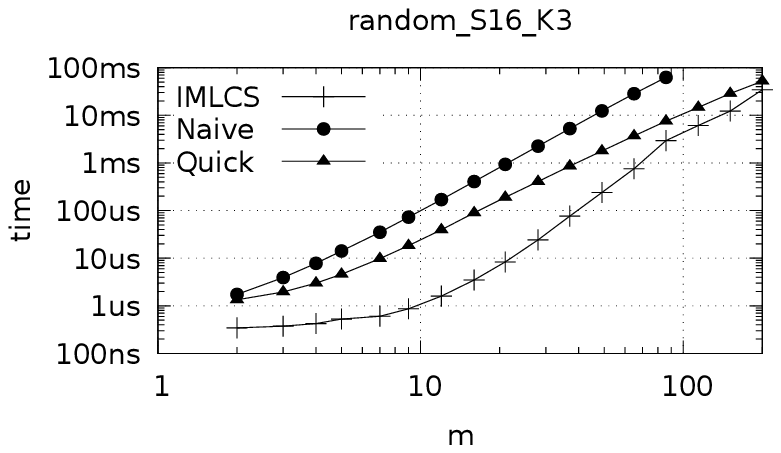}
\includegraphics{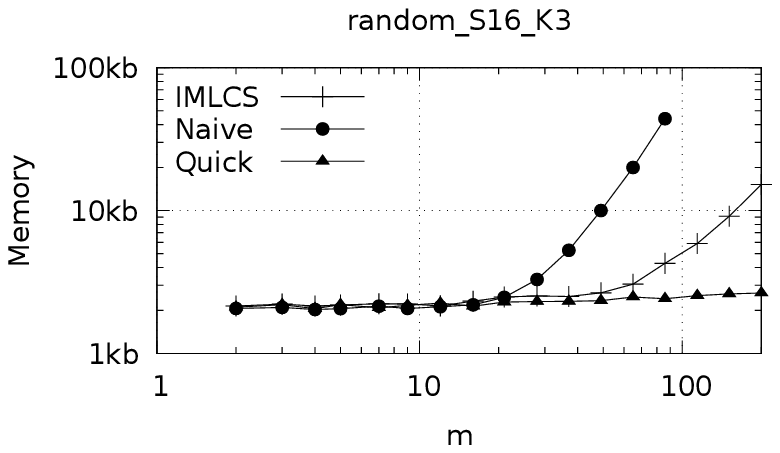}
\includegraphics{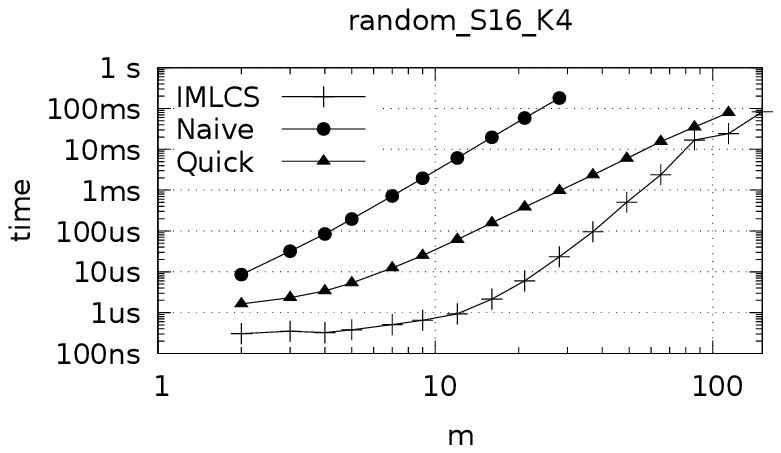}
\includegraphics{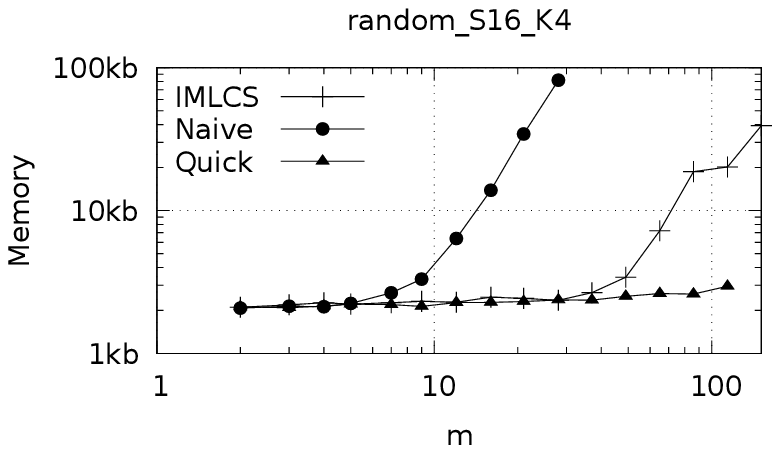}
\includegraphics{random_S16_K5.eps}
\includegraphics{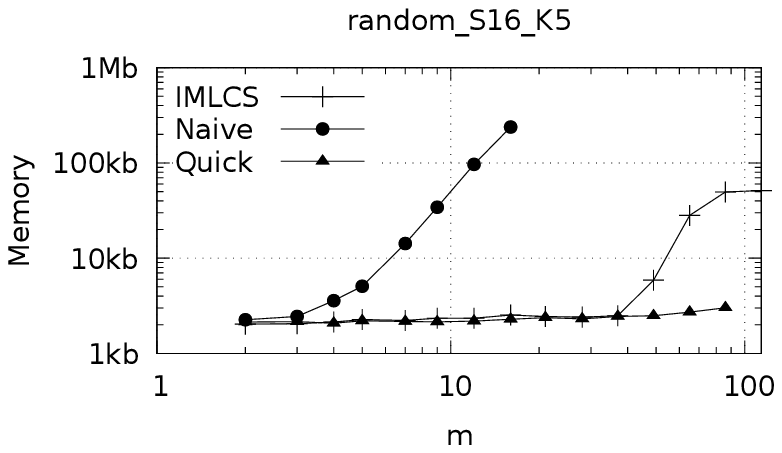}
\includegraphics{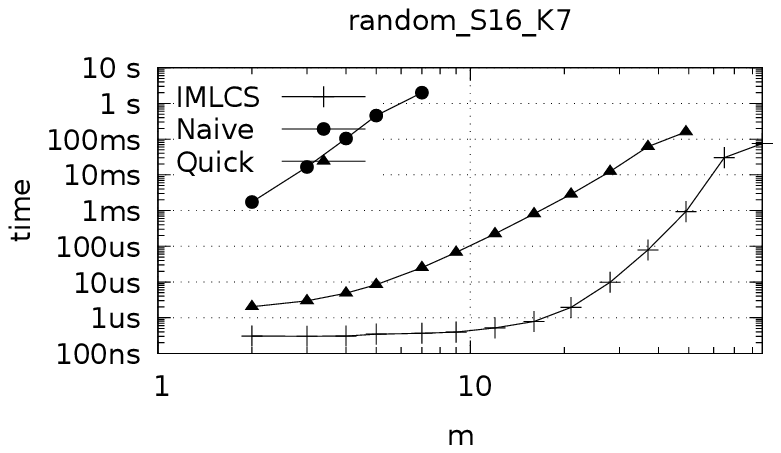}
\includegraphics{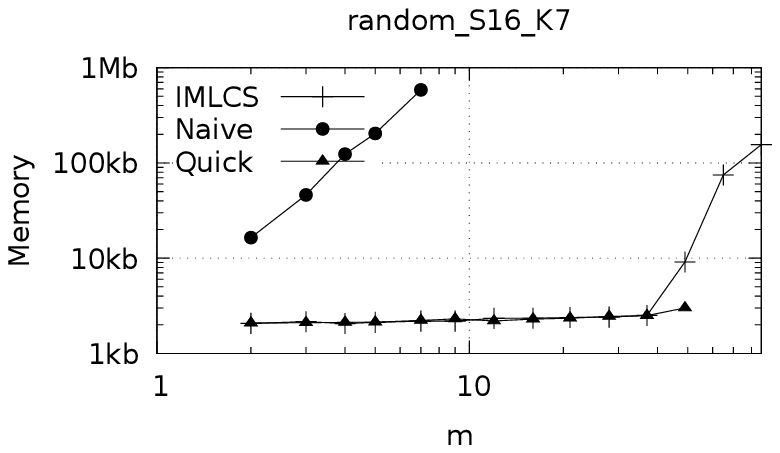}
\includegraphics{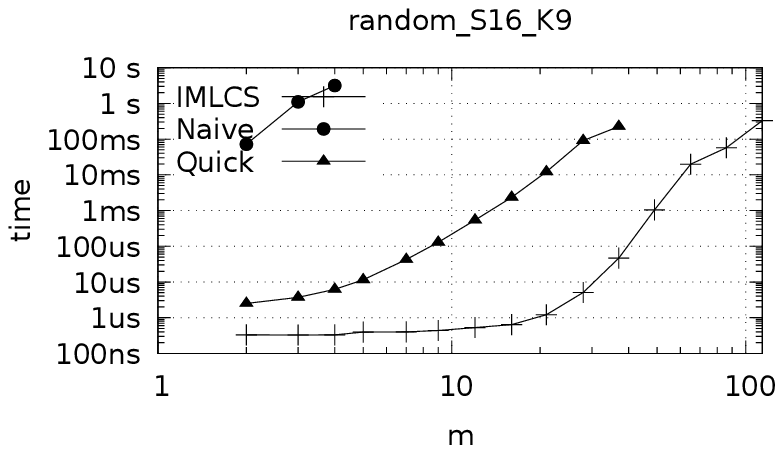}
\includegraphics{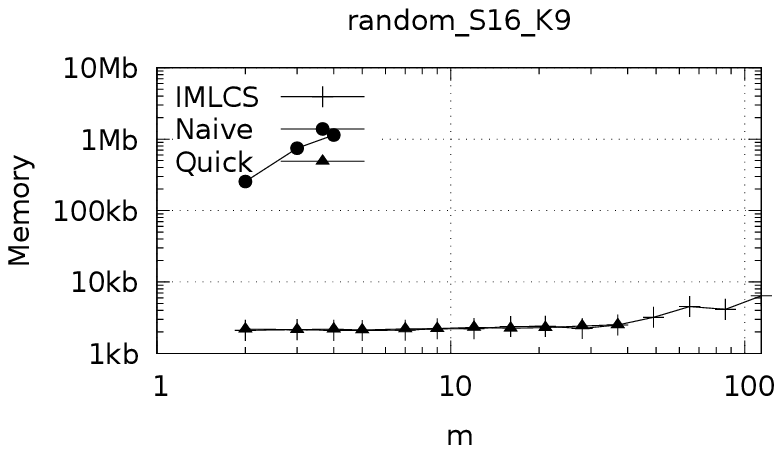}
\includegraphics{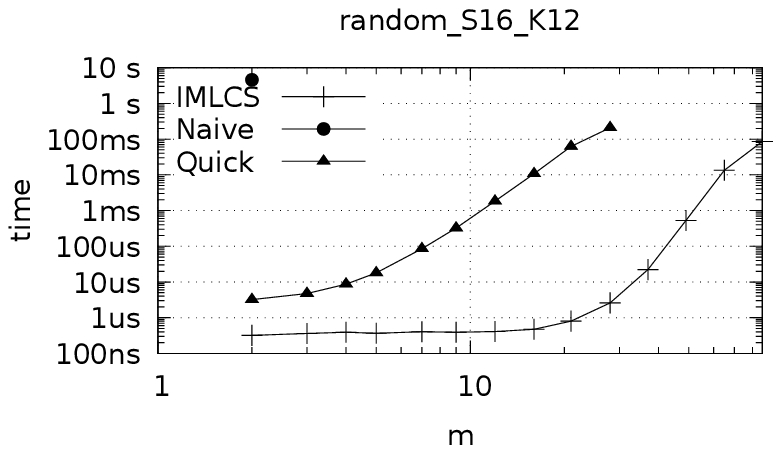}
\includegraphics{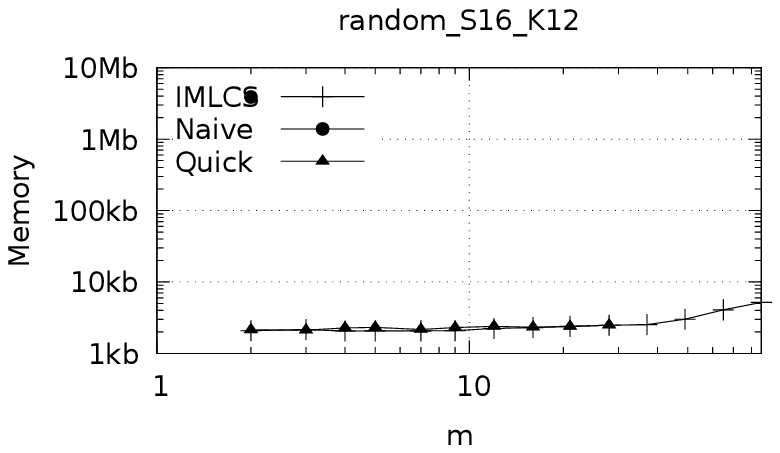}
\includegraphics{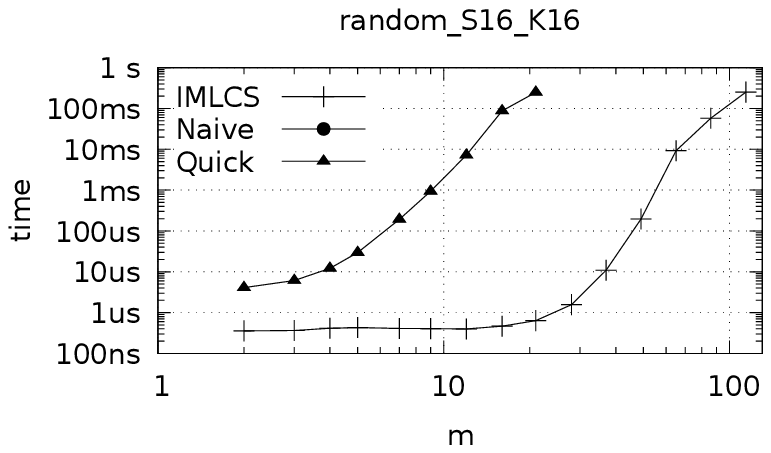}
\includegraphics{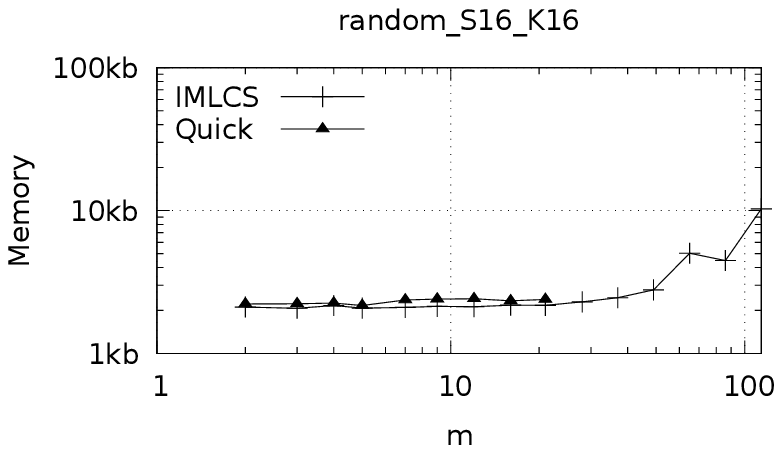}

\includegraphics{random_S1_K2M.eps}
\includegraphics{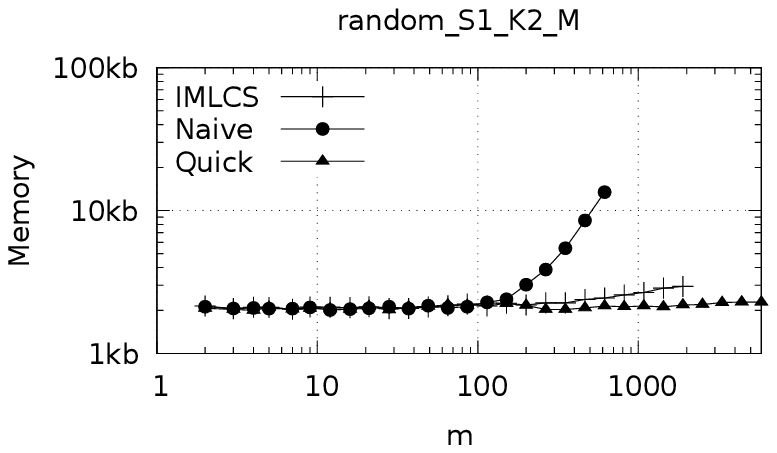}
\includegraphics{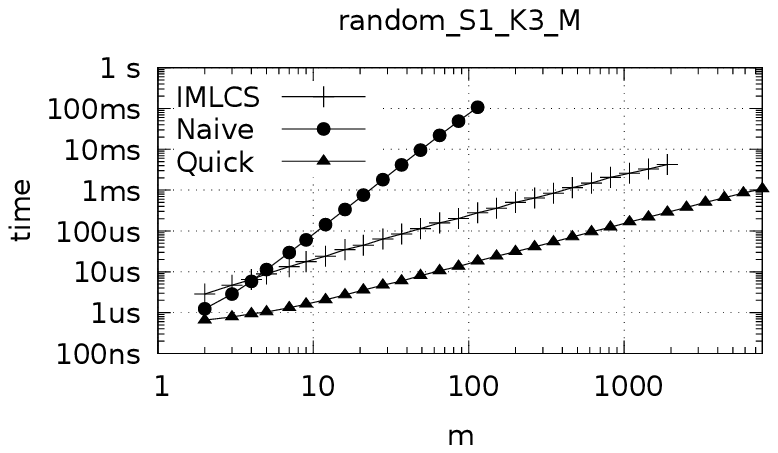}
\includegraphics{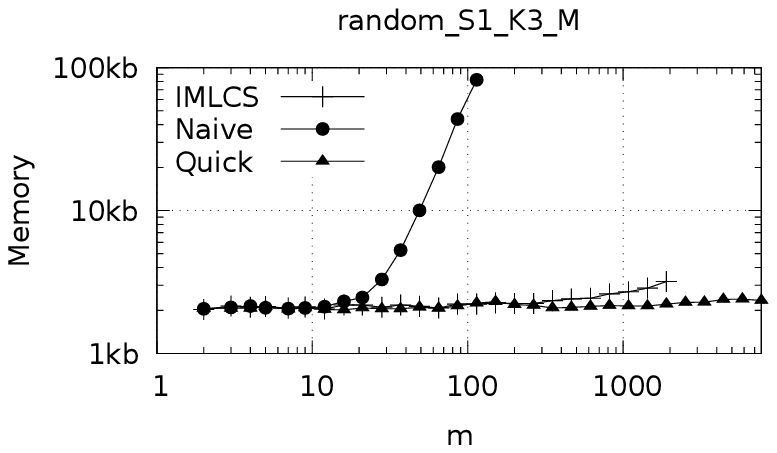}
\includegraphics{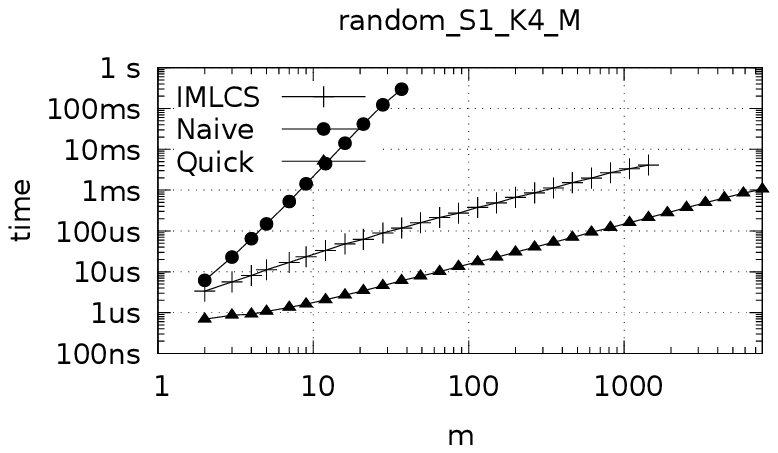}
\includegraphics{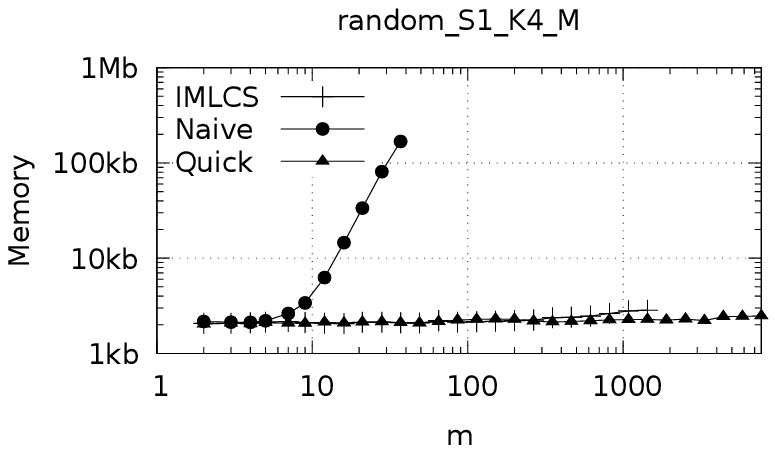}
\includegraphics{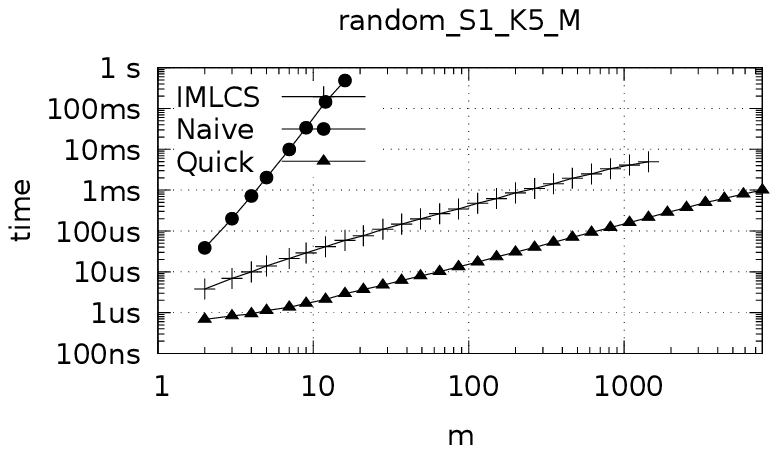}
\includegraphics{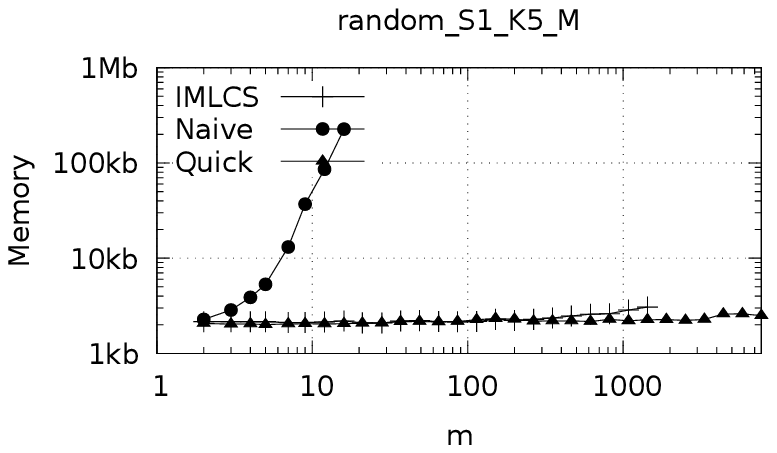}
\includegraphics{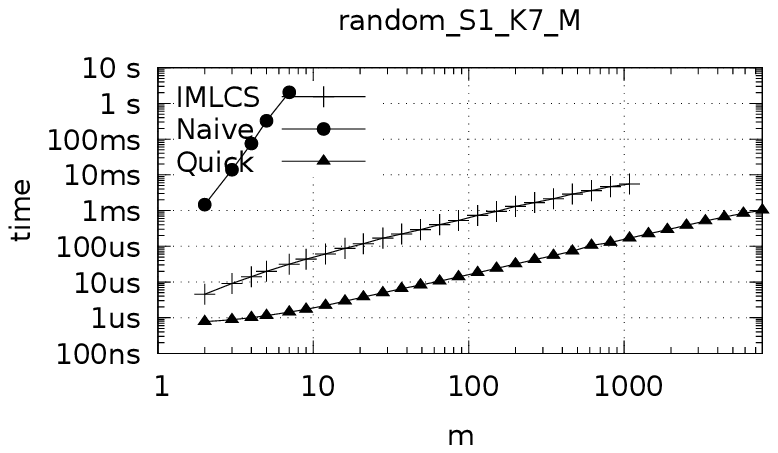}
\includegraphics{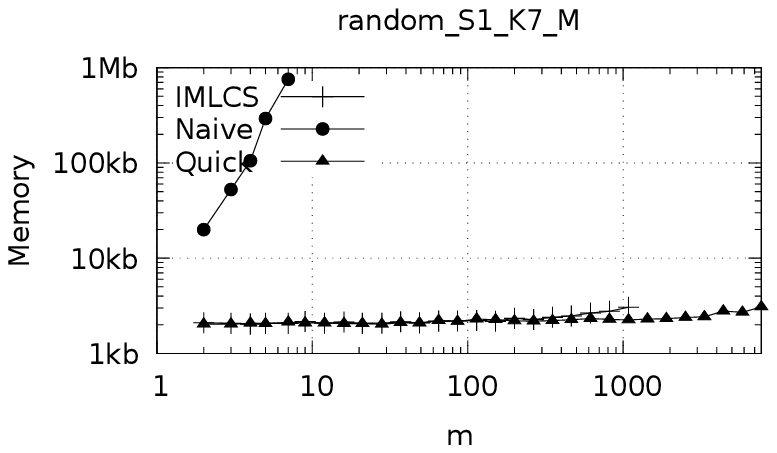}
\includegraphics{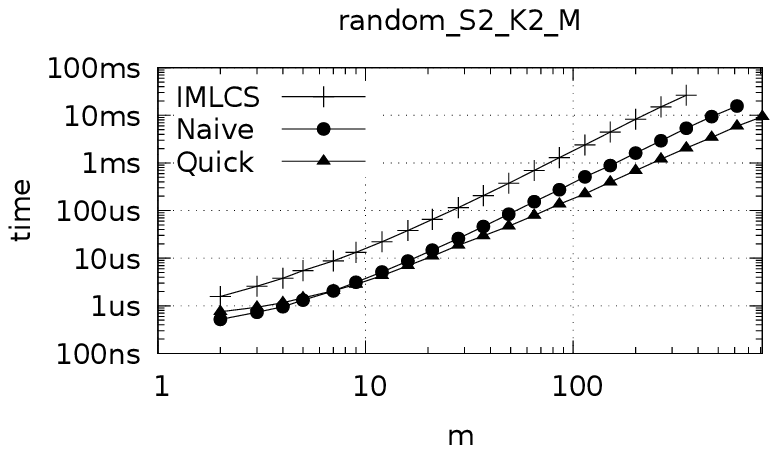}
\includegraphics{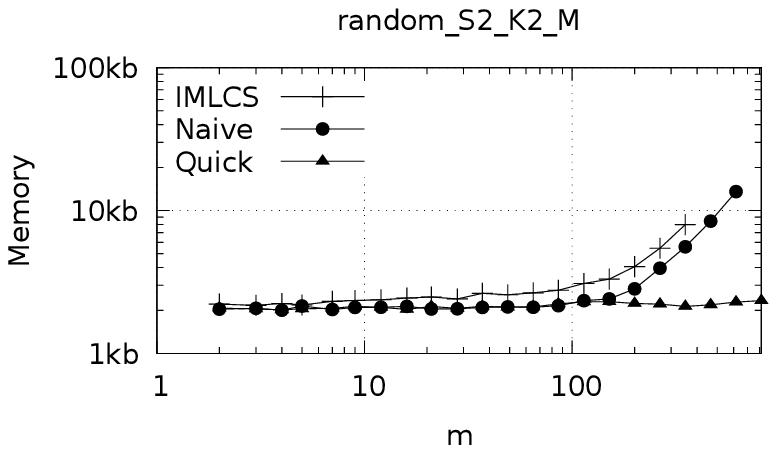}
\includegraphics{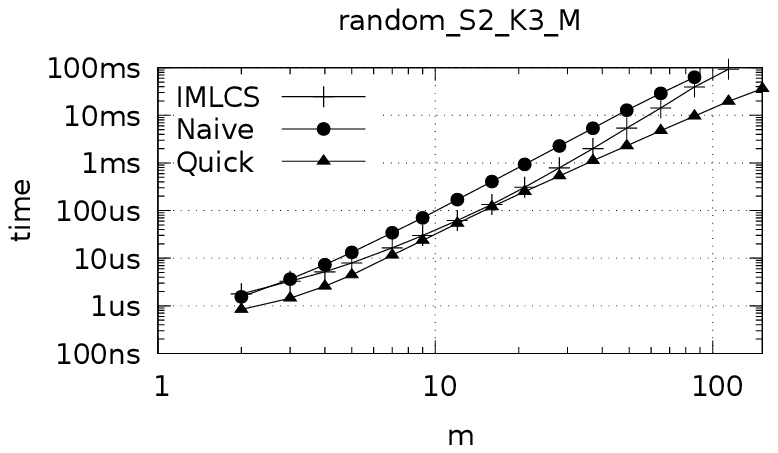}
\includegraphics{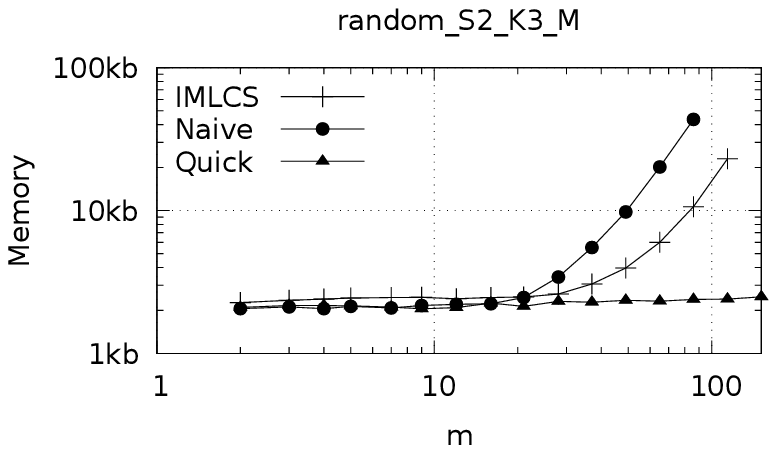}
\includegraphics{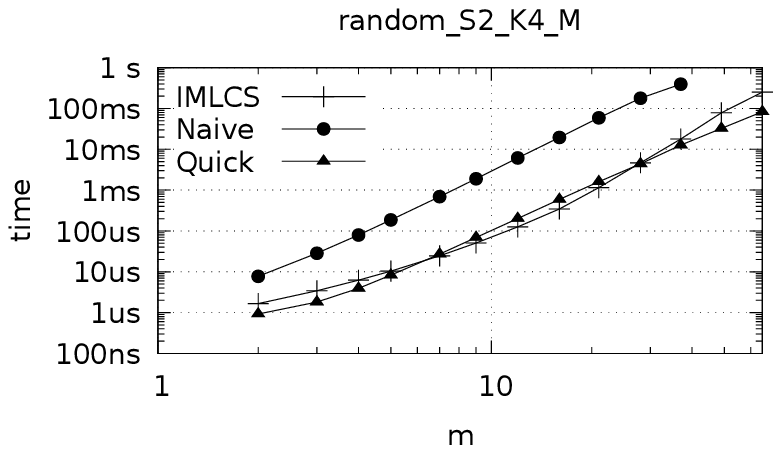}
\includegraphics{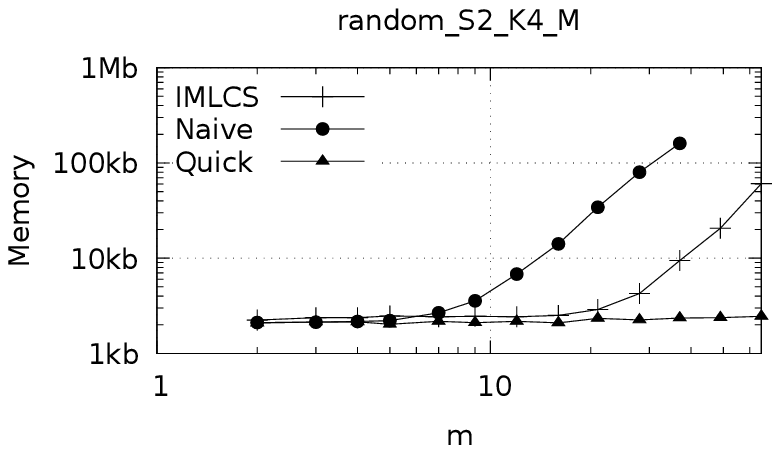}
\includegraphics{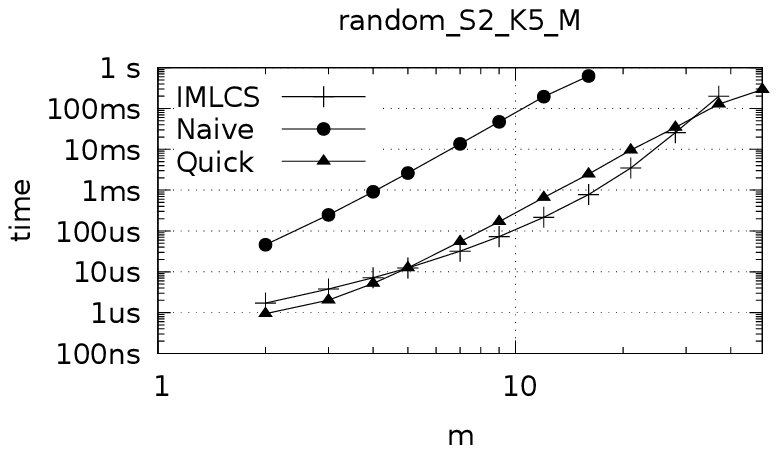}
\includegraphics{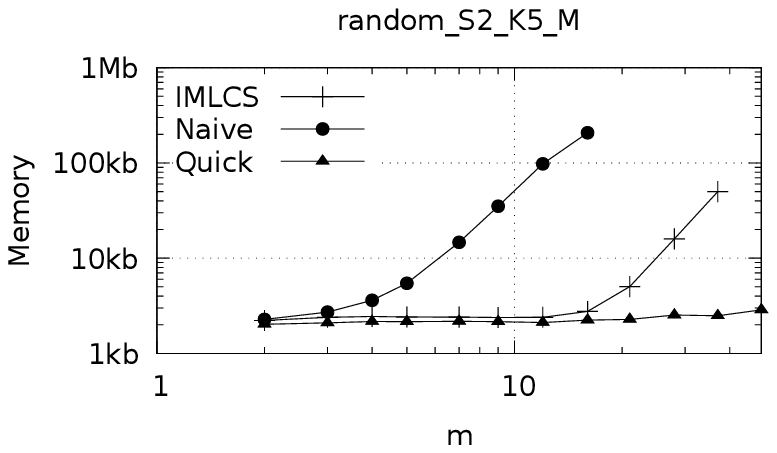}
\includegraphics{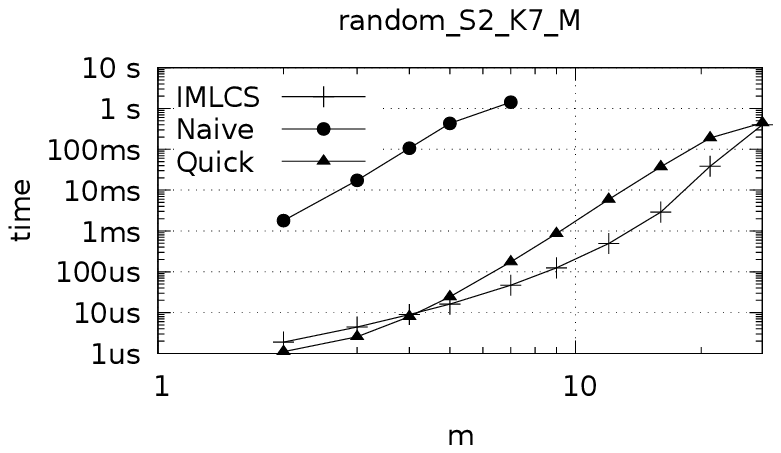}
\includegraphics{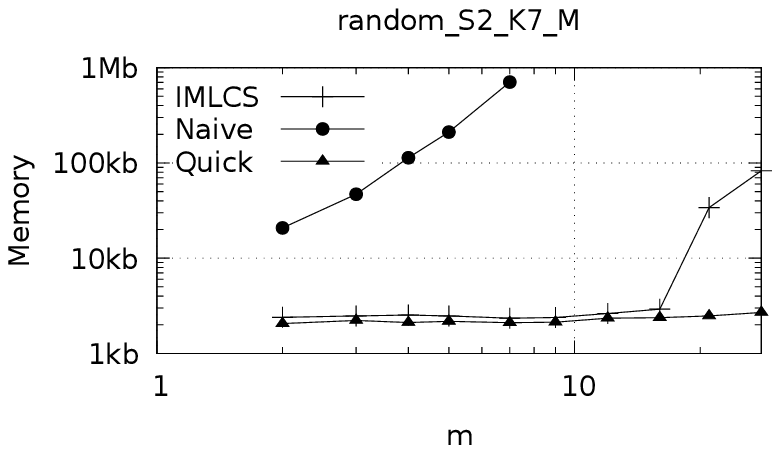}
\includegraphics{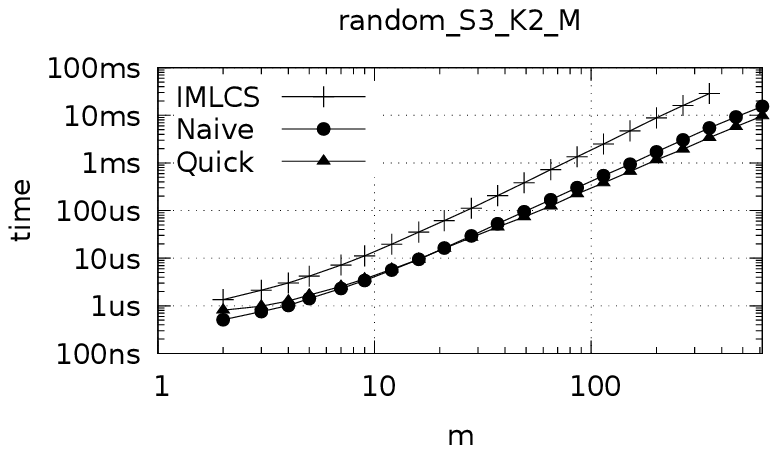}
\includegraphics{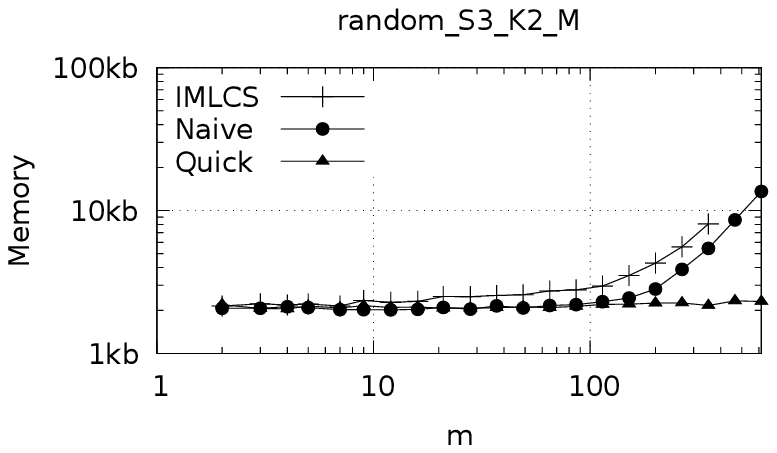}
\includegraphics{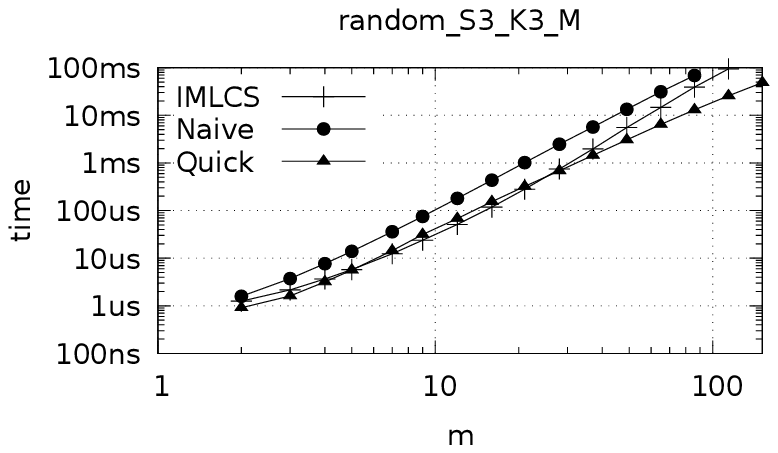}
\includegraphics{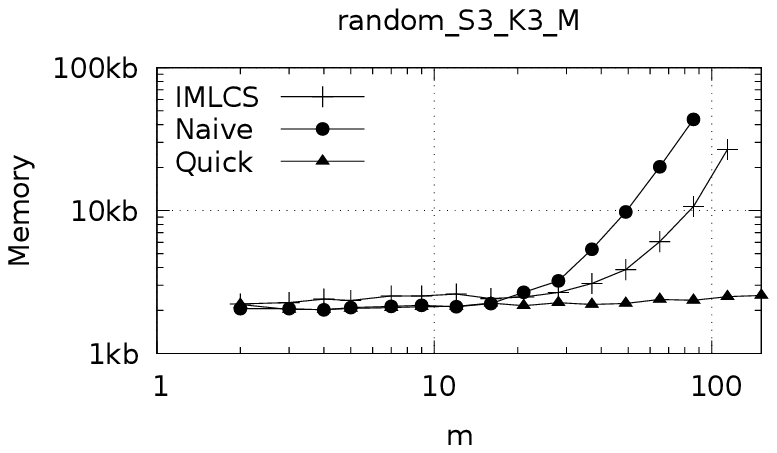}
\includegraphics{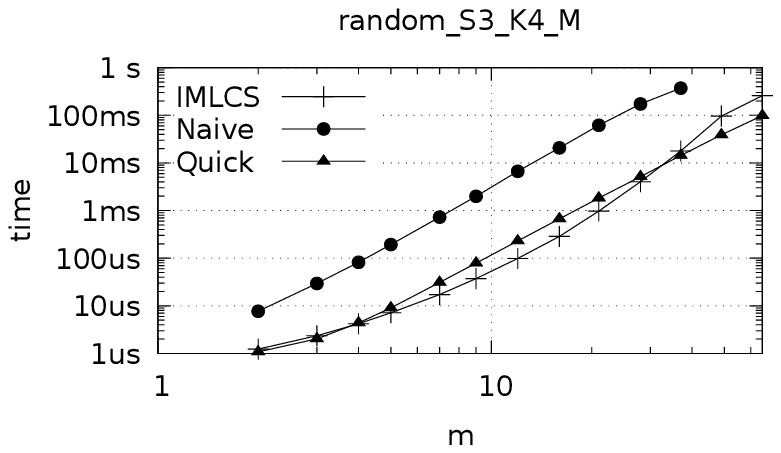}
\includegraphics{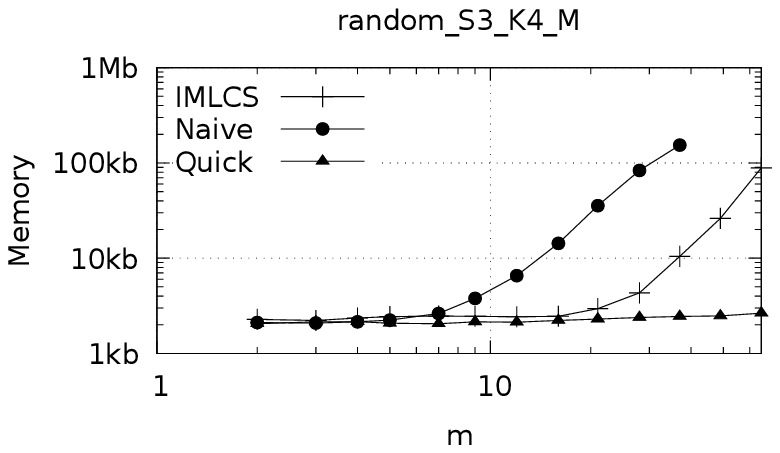}
\includegraphics{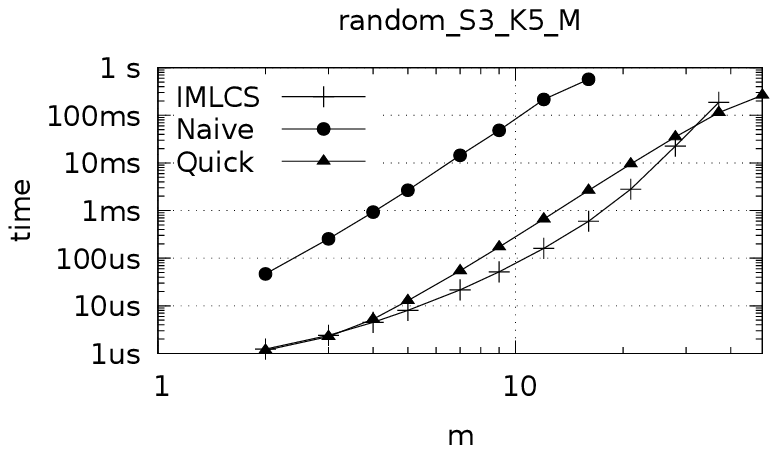}
\includegraphics{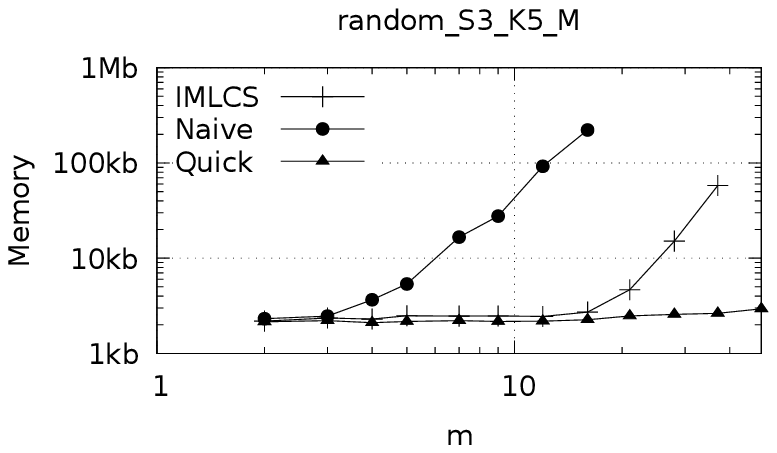}
\includegraphics{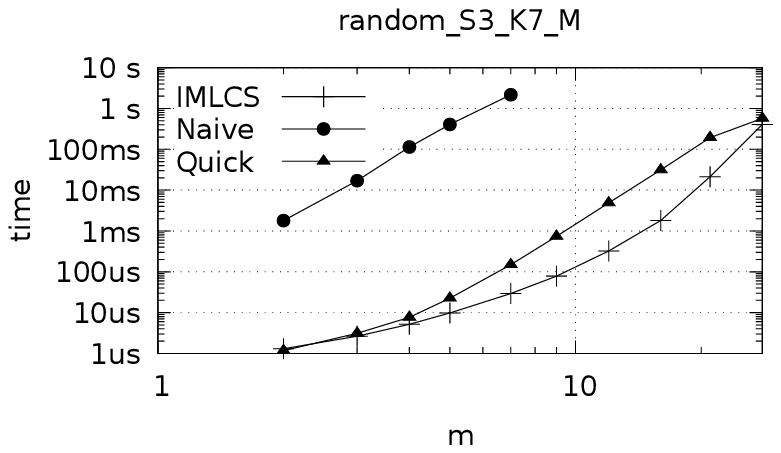}
\includegraphics{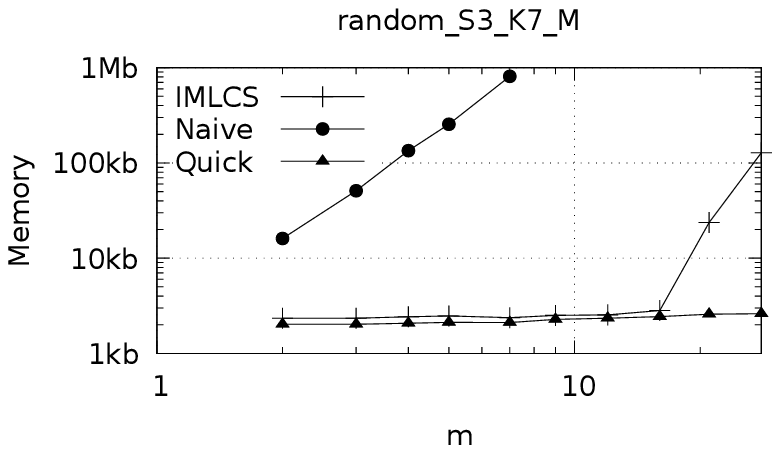}
\includegraphics{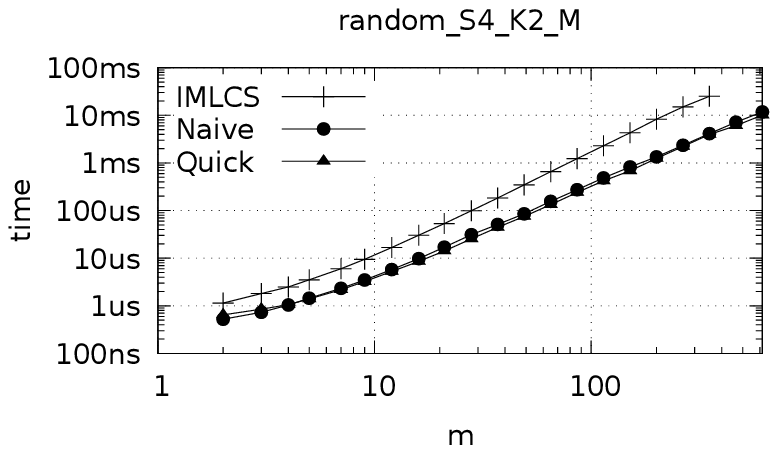}
\includegraphics{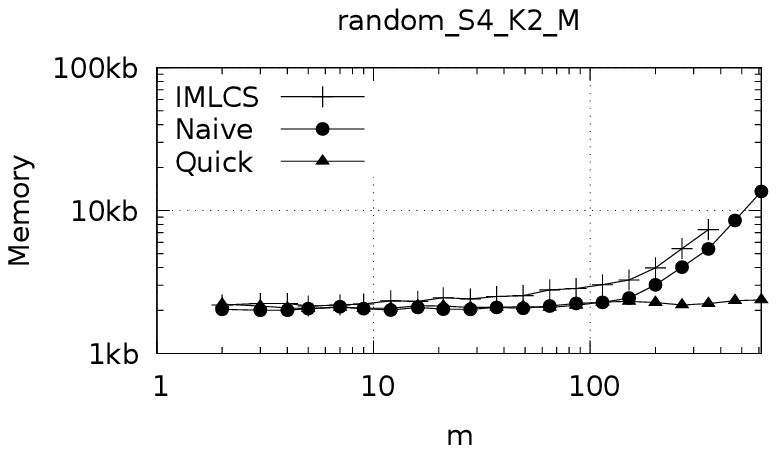}
\includegraphics{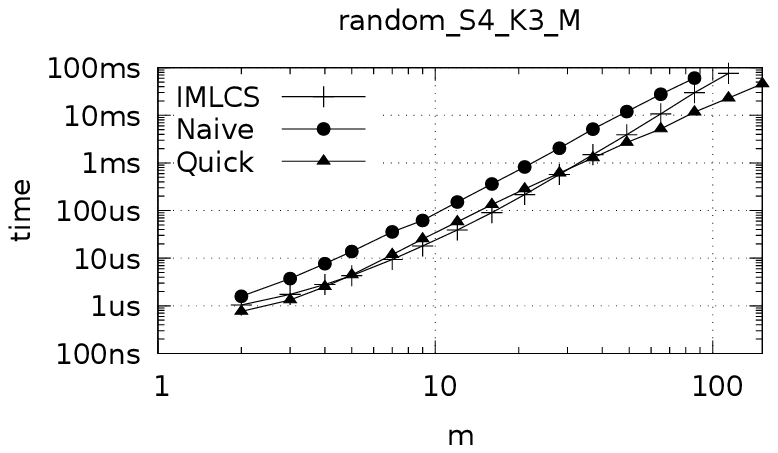}
\includegraphics{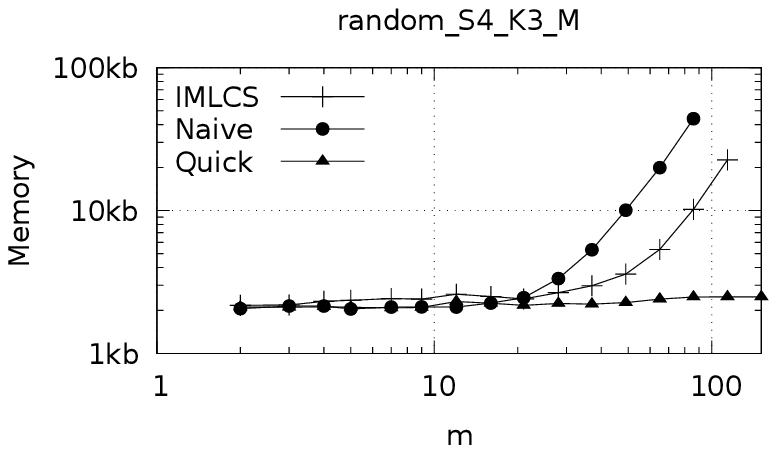}
\includegraphics{random_S4_K4M.eps}
\includegraphics{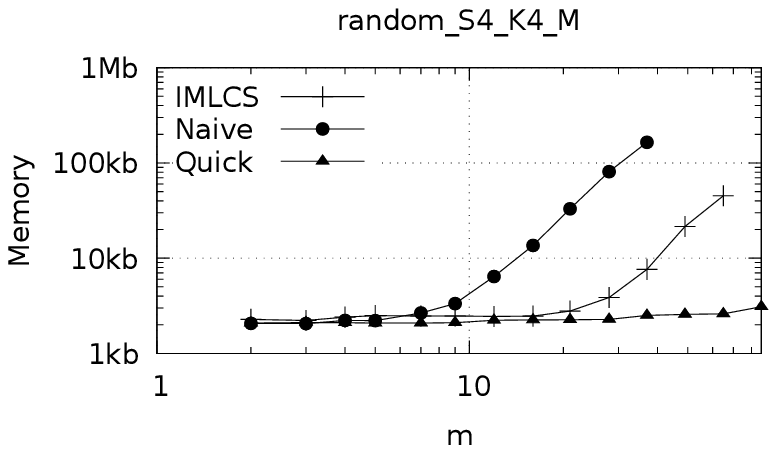}
\includegraphics{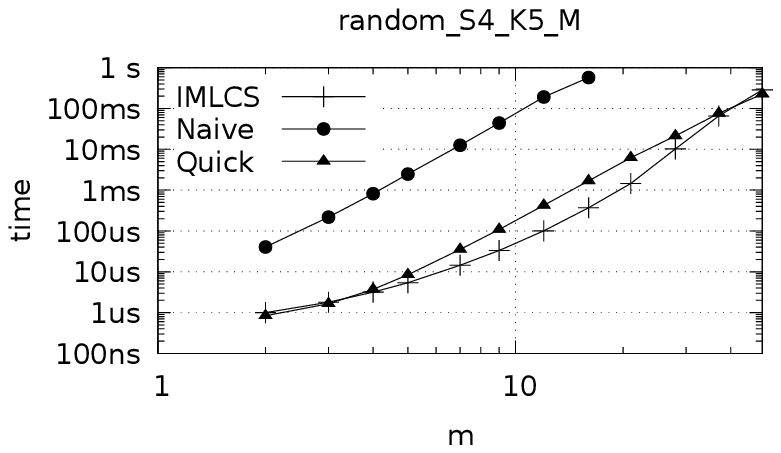}
\includegraphics{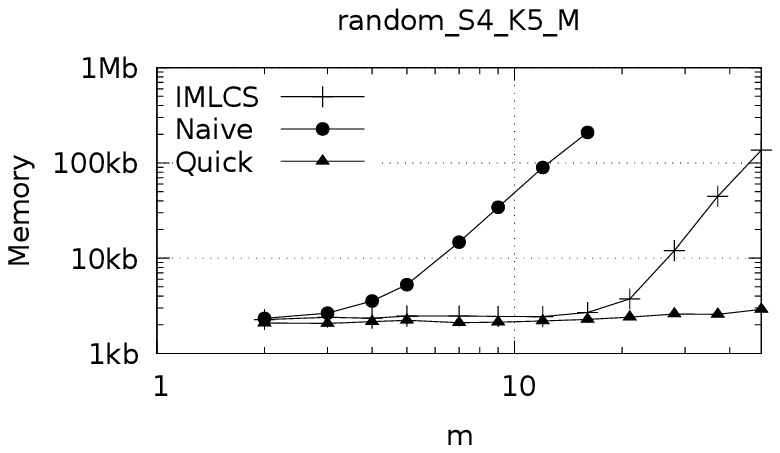}
\includegraphics{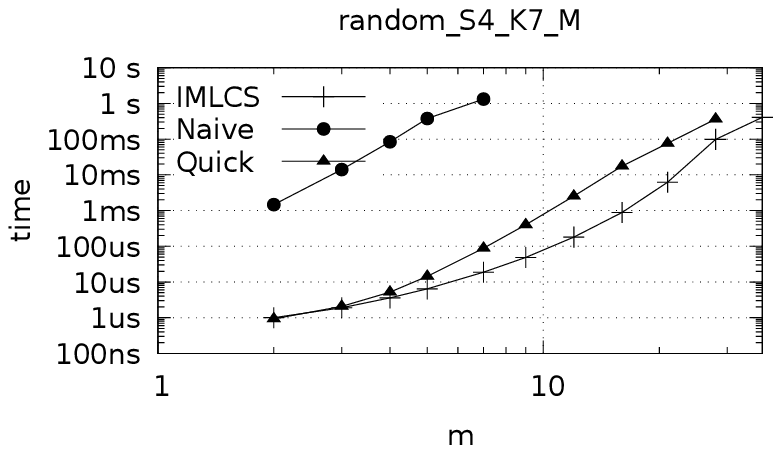}
\includegraphics{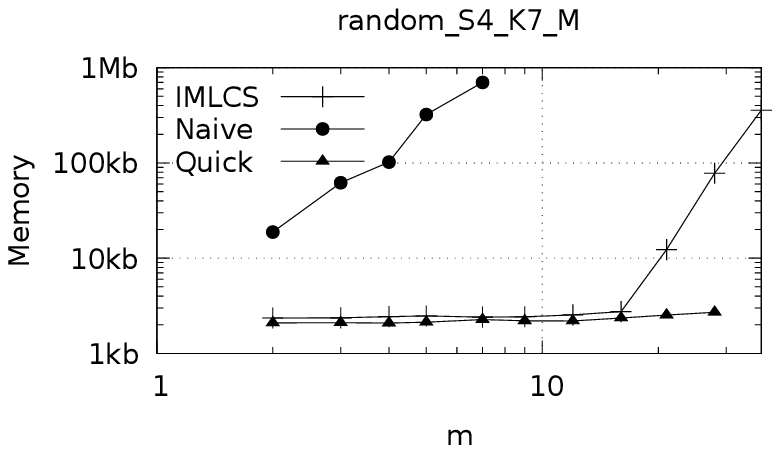}
\includegraphics{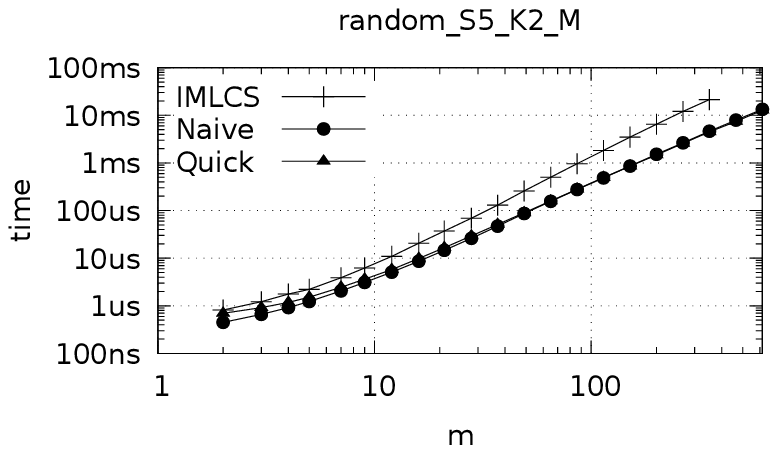}
\includegraphics{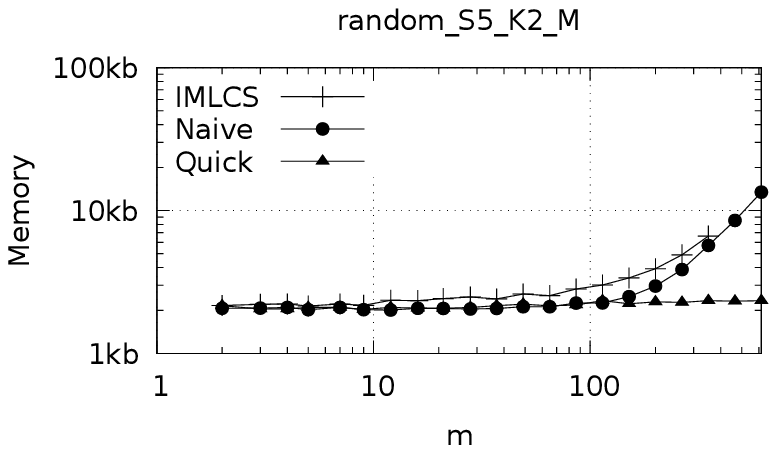}
\includegraphics{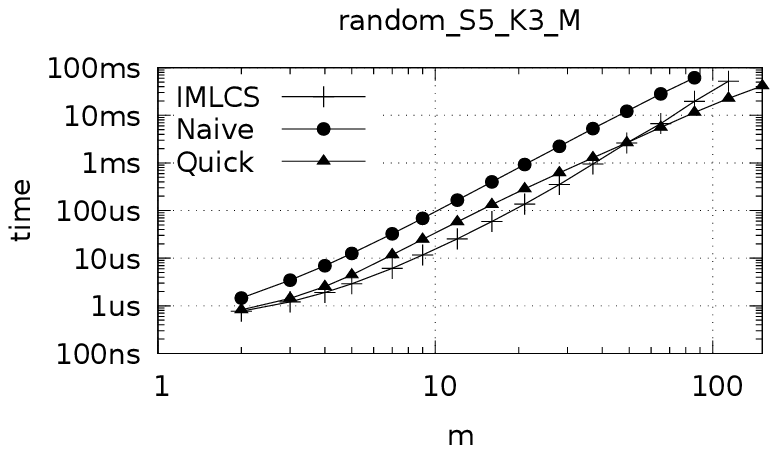}
\includegraphics{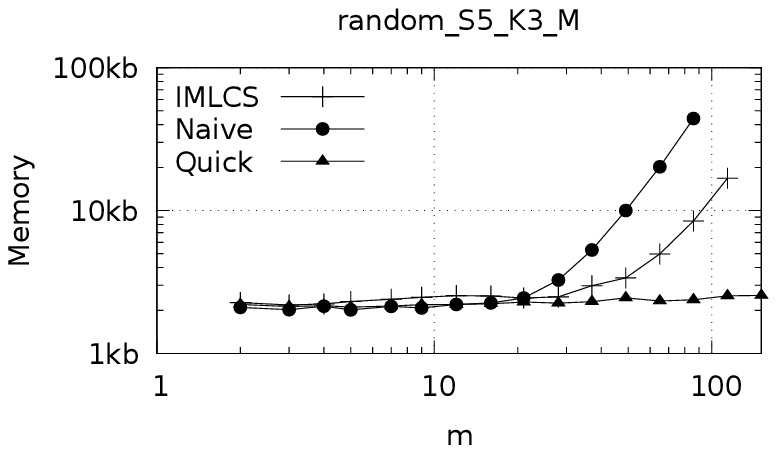}
\includegraphics{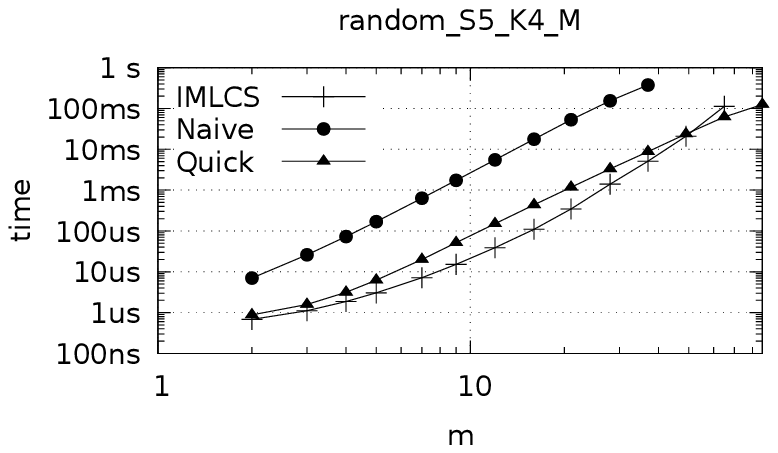}
\includegraphics{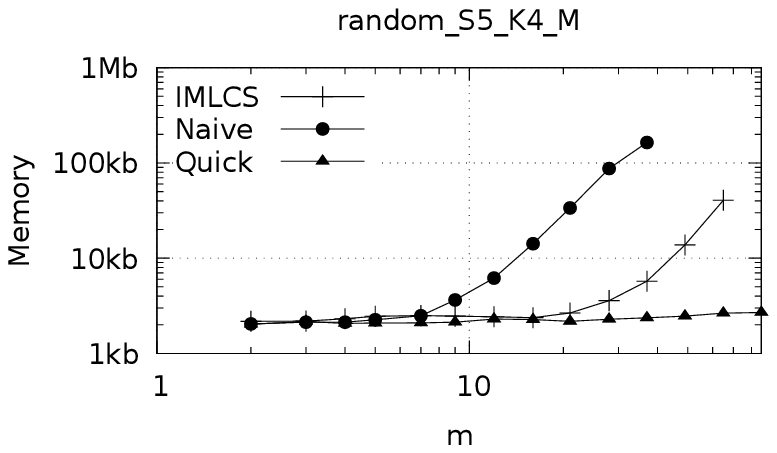}
\includegraphics{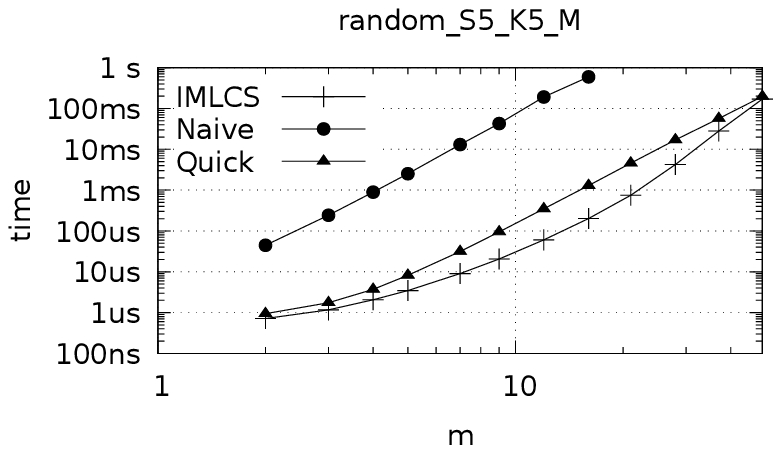}
\includegraphics{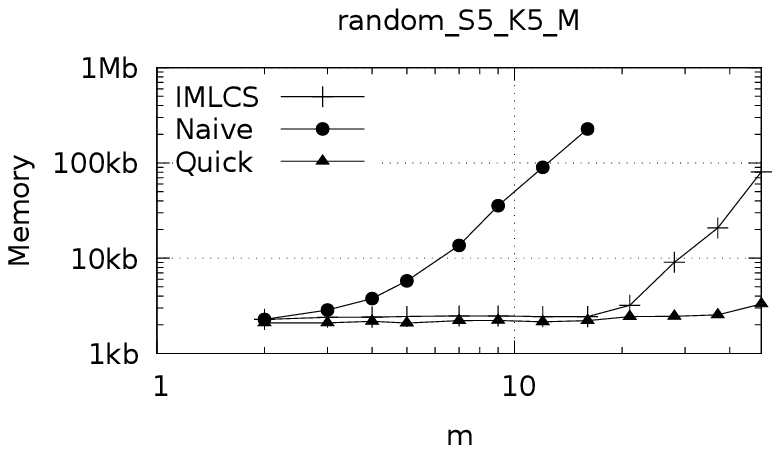}
\includegraphics{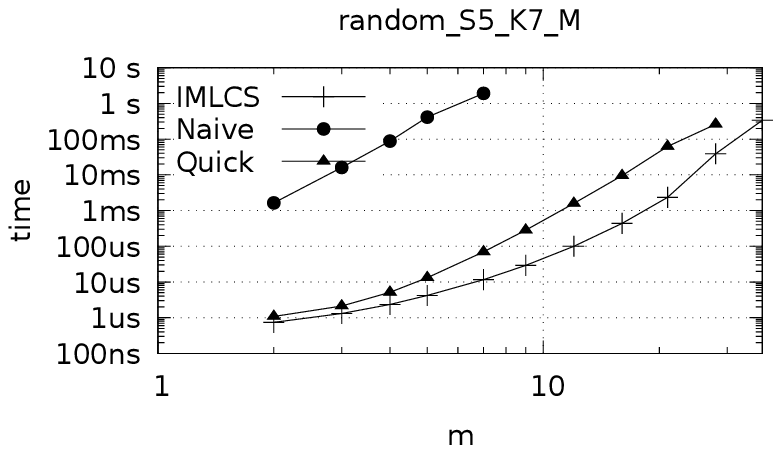}
\includegraphics{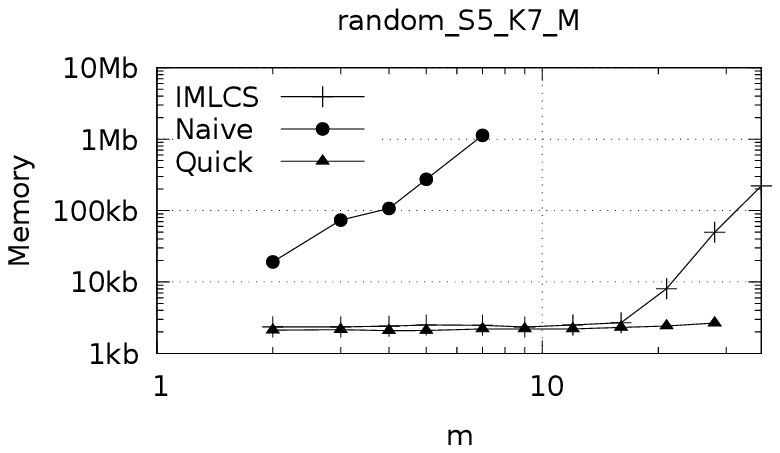}
\includegraphics{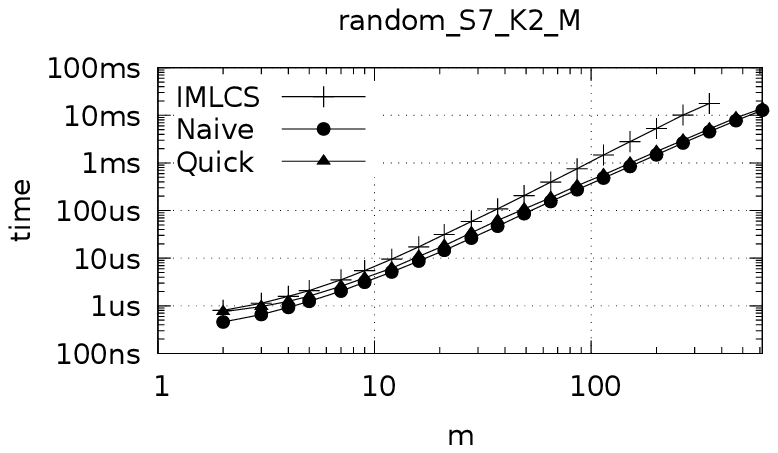}
\includegraphics{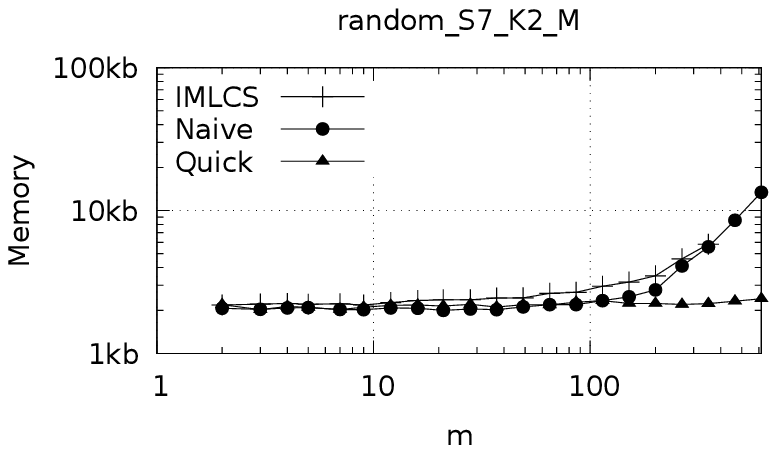}
\includegraphics{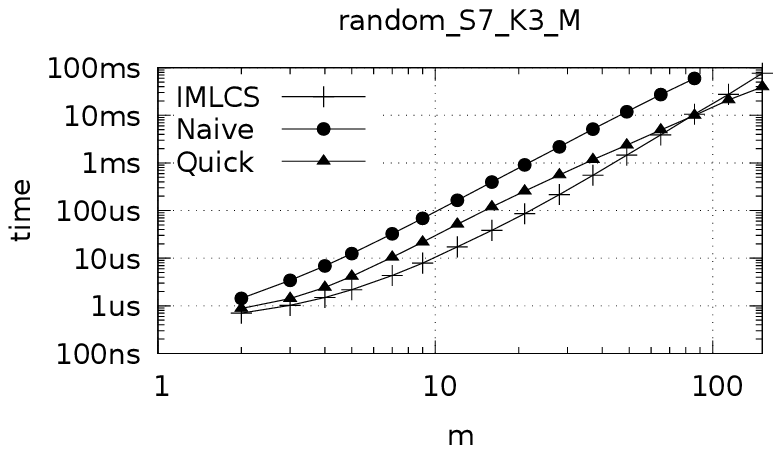}
\includegraphics{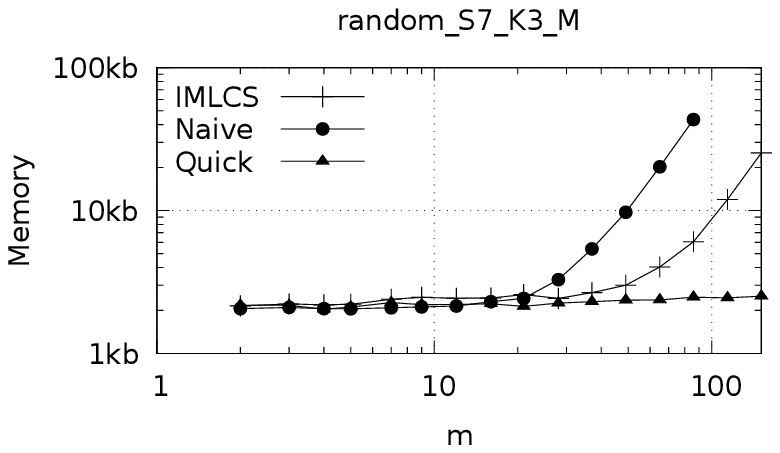}
\includegraphics{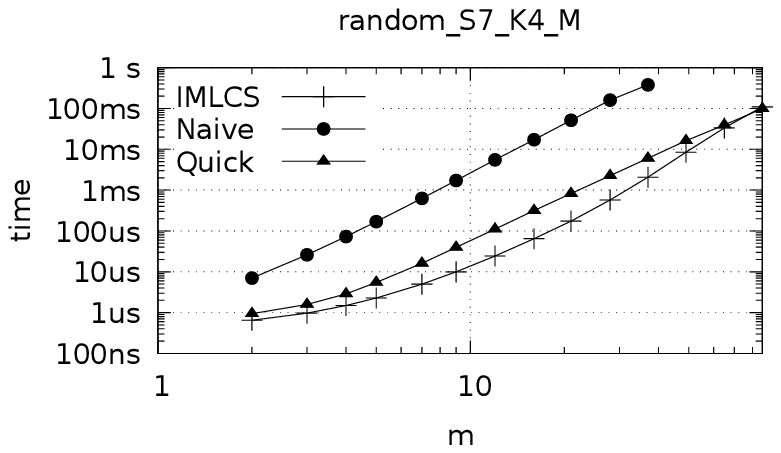}
\includegraphics{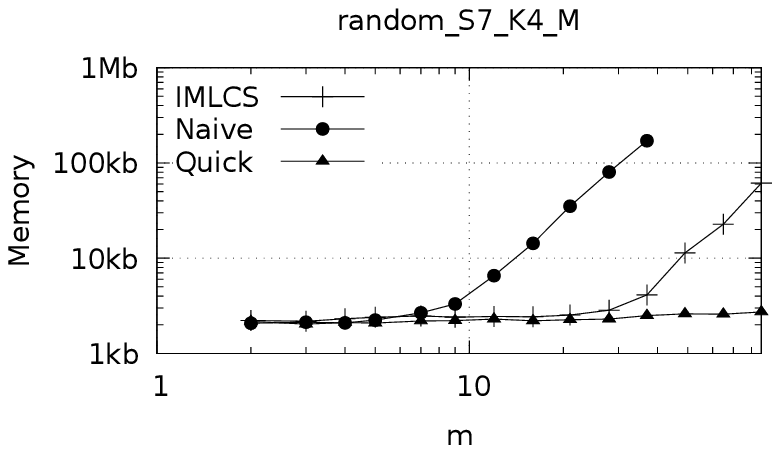}
\includegraphics{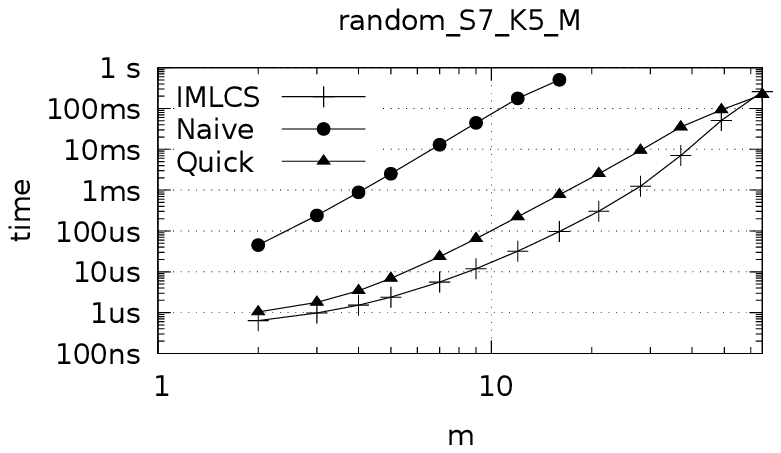}
\includegraphics{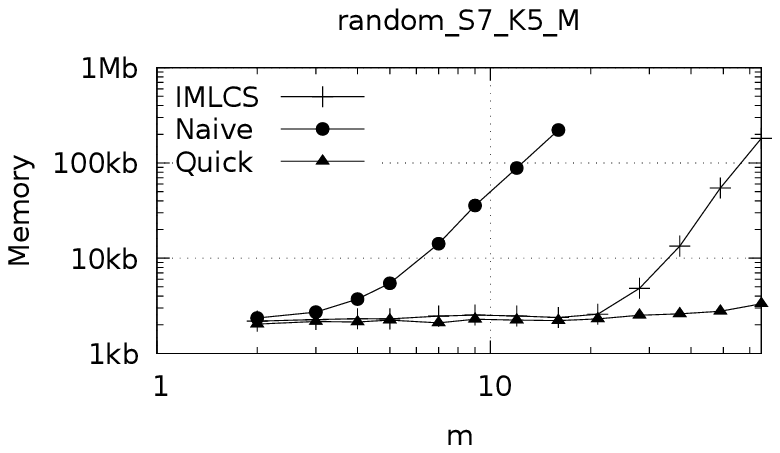}
\includegraphics{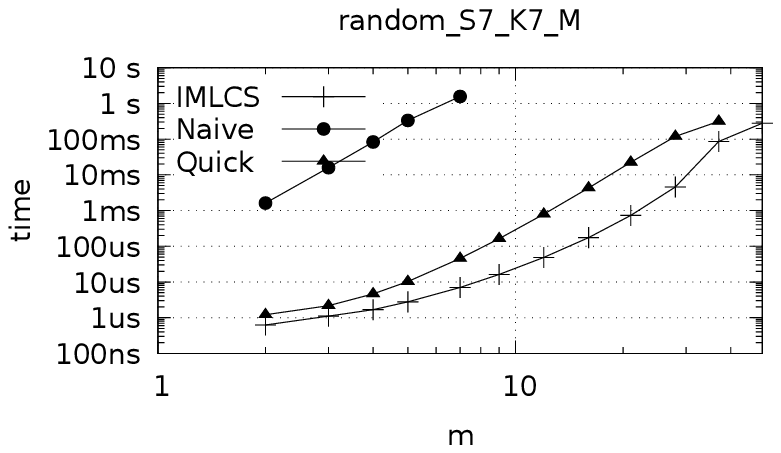}
\includegraphics{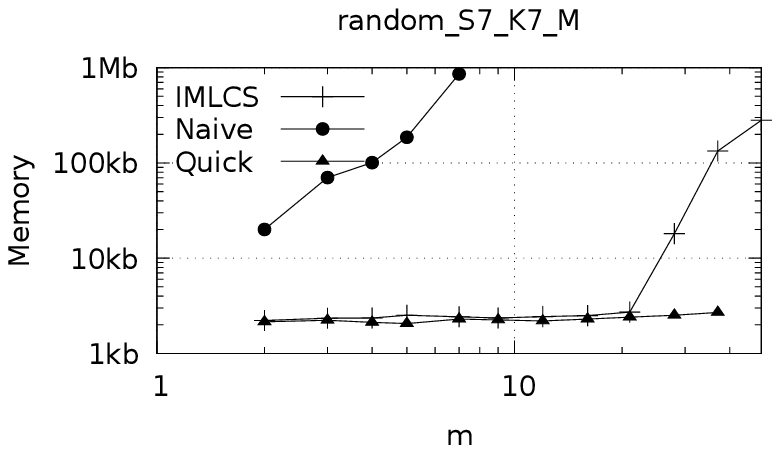}
\includegraphics{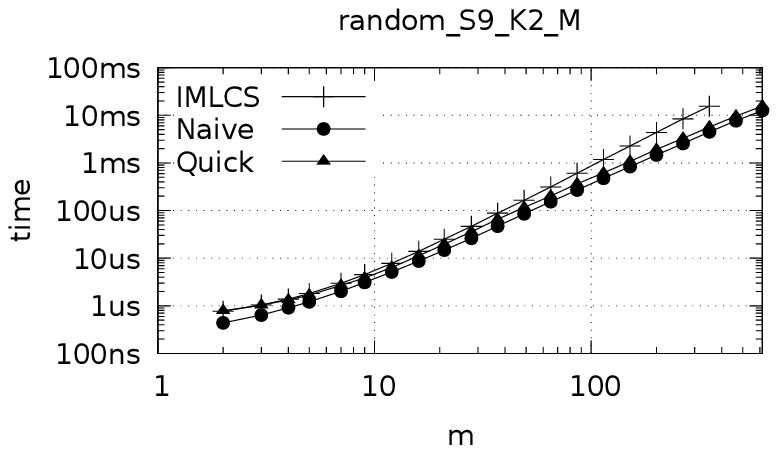}
\includegraphics{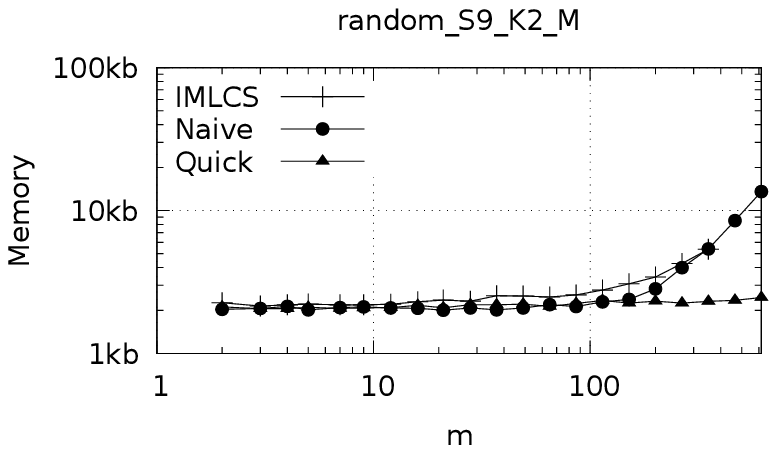}
\includegraphics{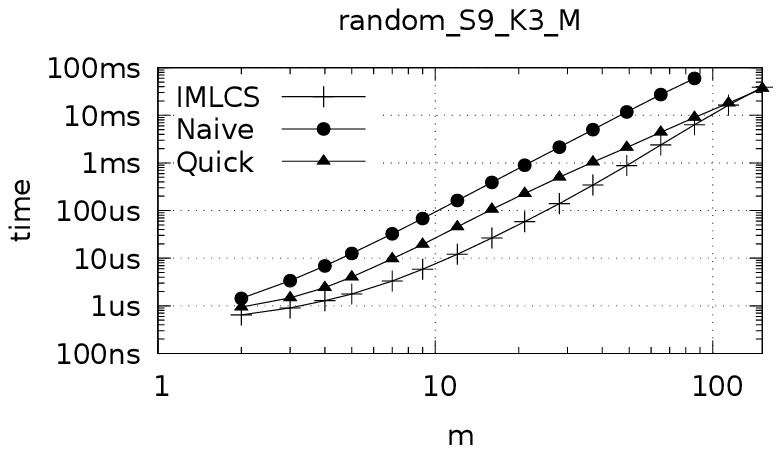}
\includegraphics{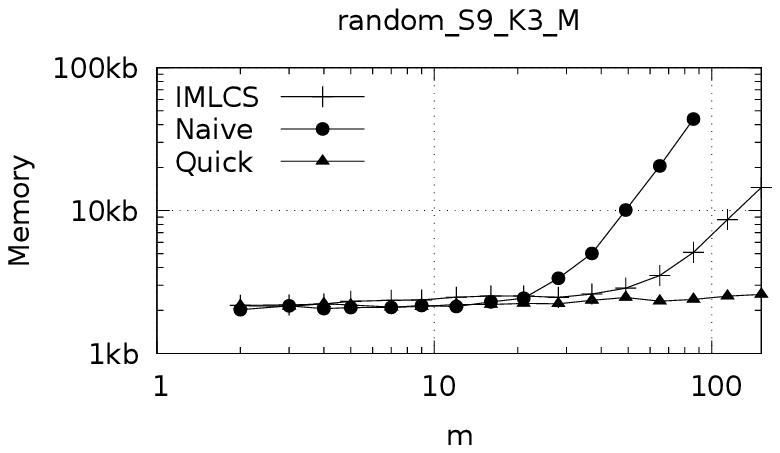}
\includegraphics{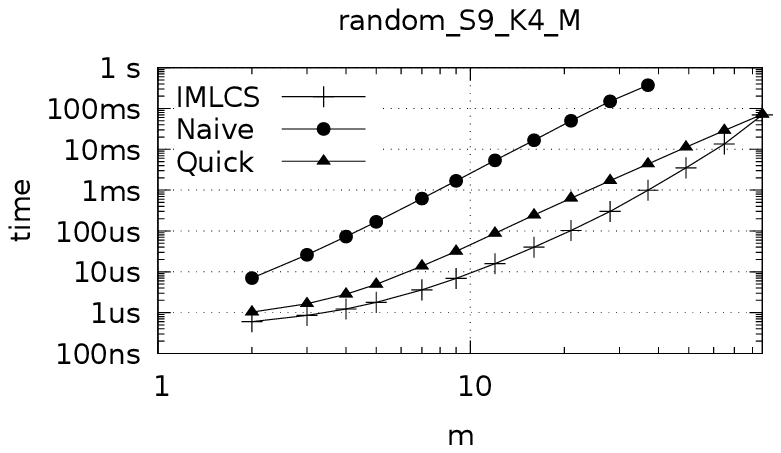}
\includegraphics{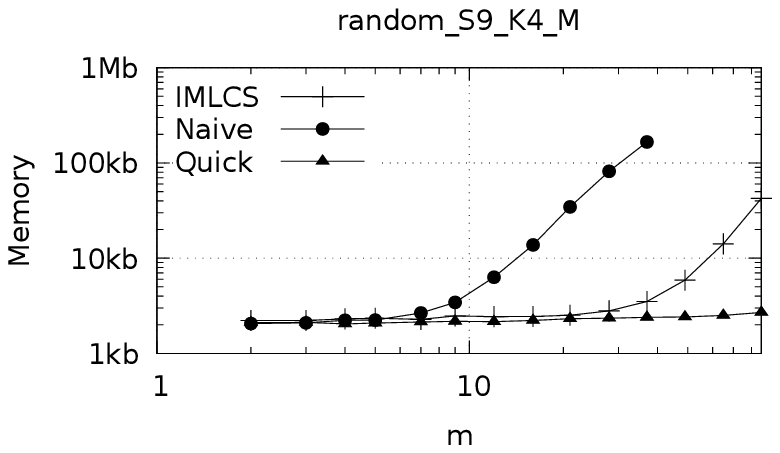}
\includegraphics{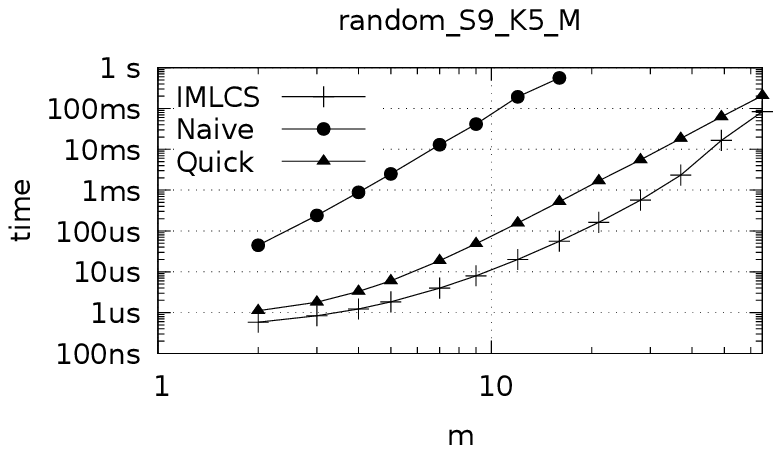}
\includegraphics{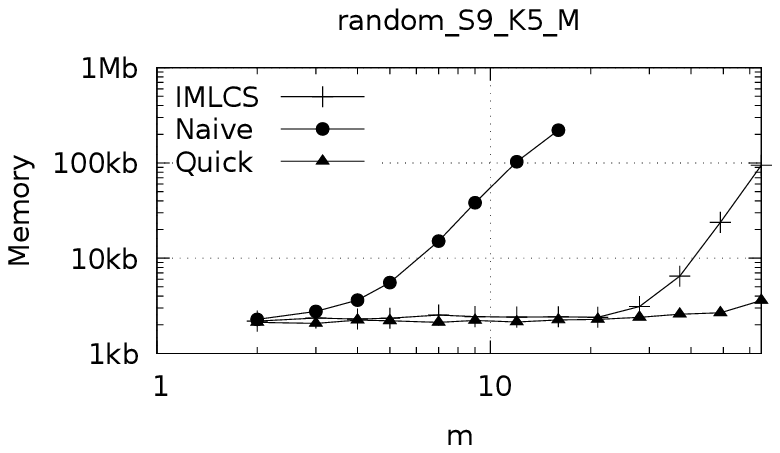}
\includegraphics{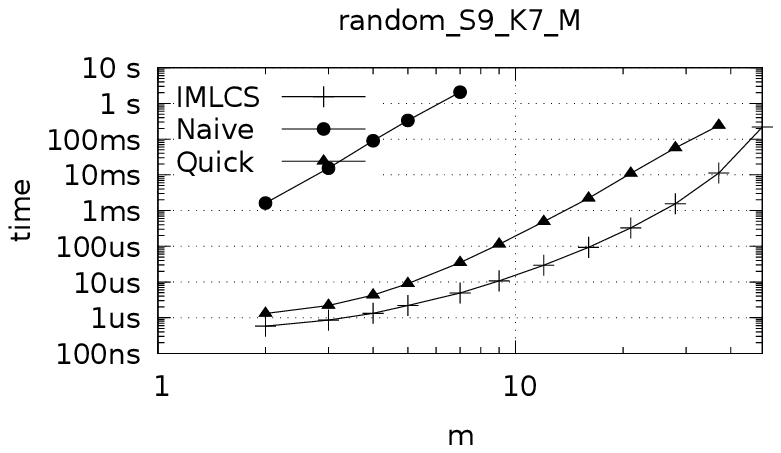}
\includegraphics{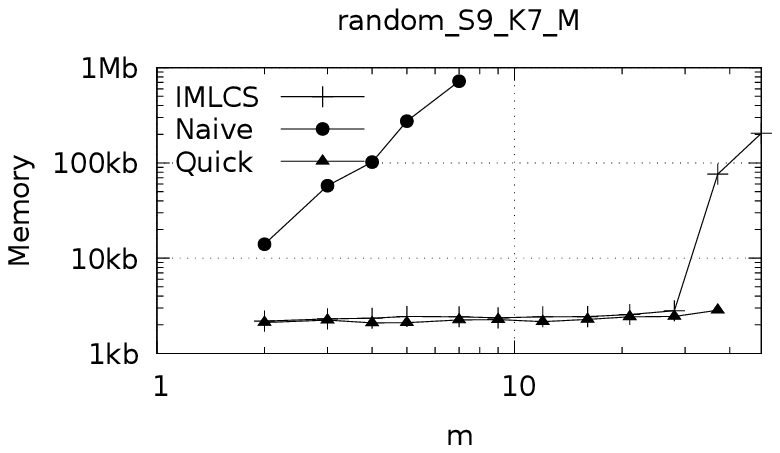}
\includegraphics{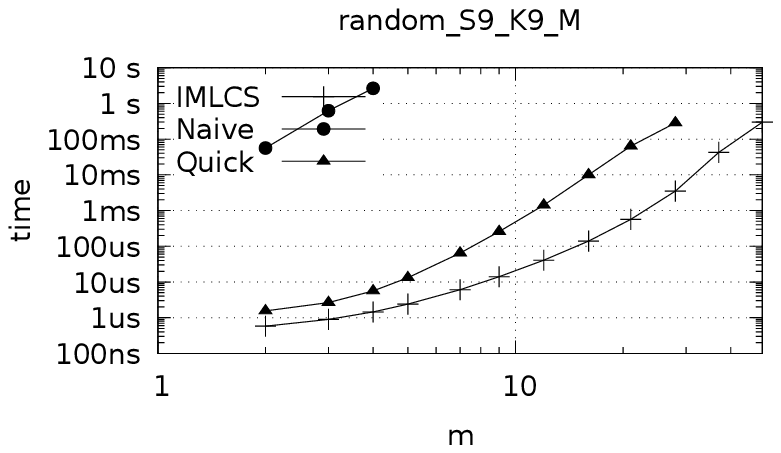}
\includegraphics{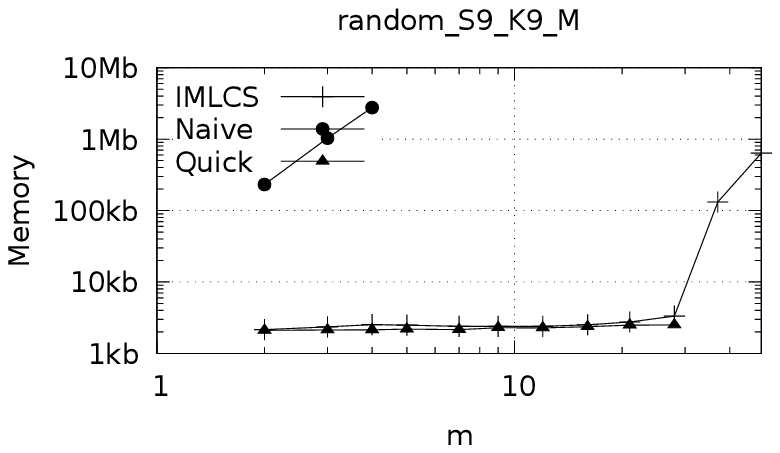}
\includegraphics{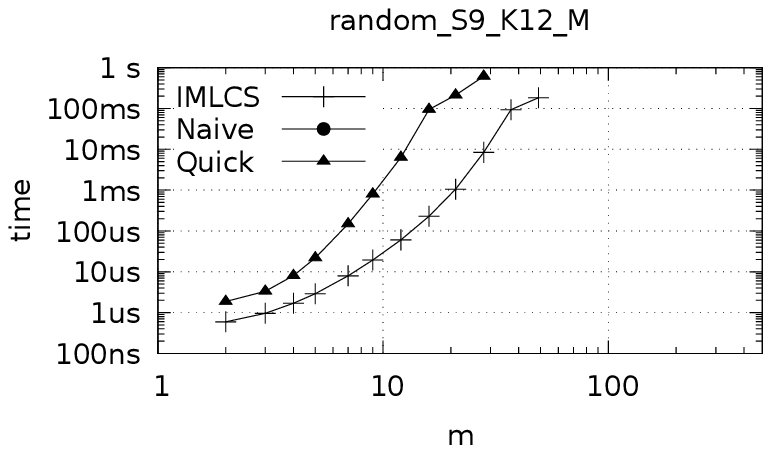}
\includegraphics{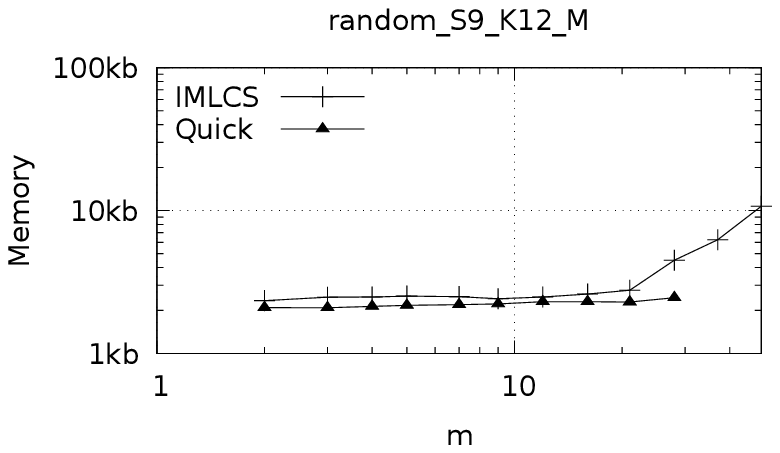}
\includegraphics{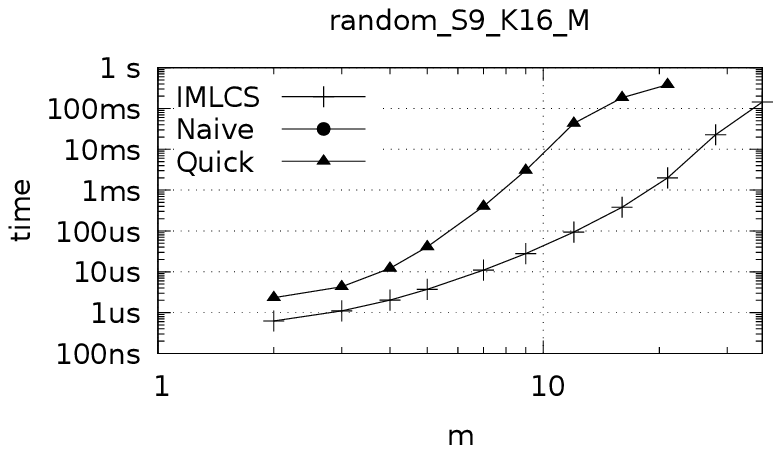}
\includegraphics{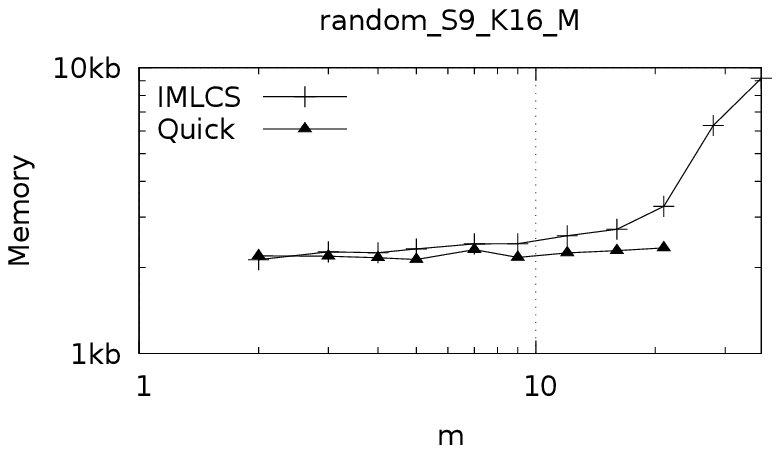}
\includegraphics{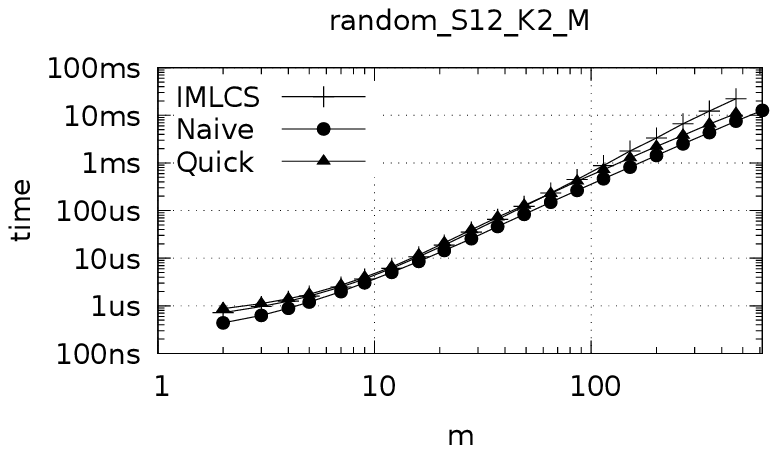}
\includegraphics{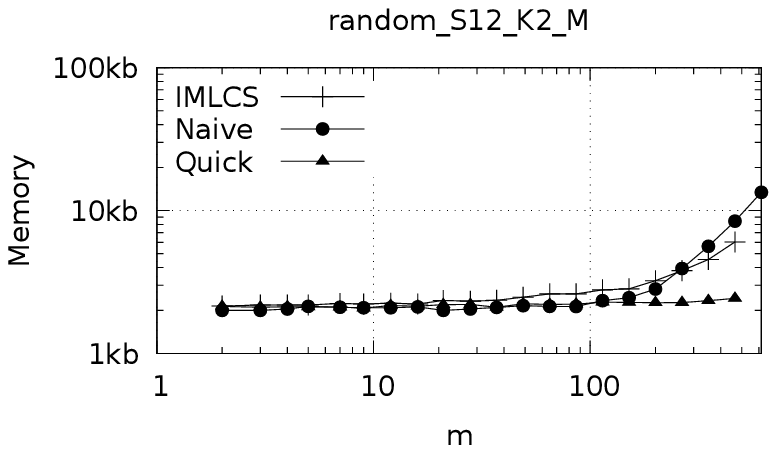}
\includegraphics{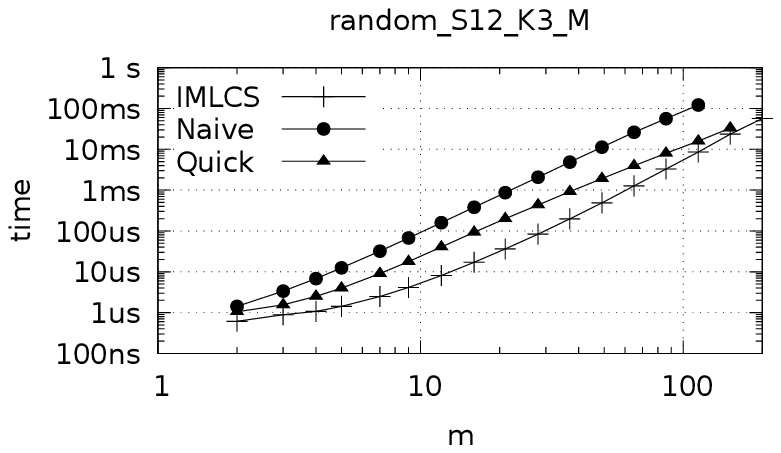}
\includegraphics{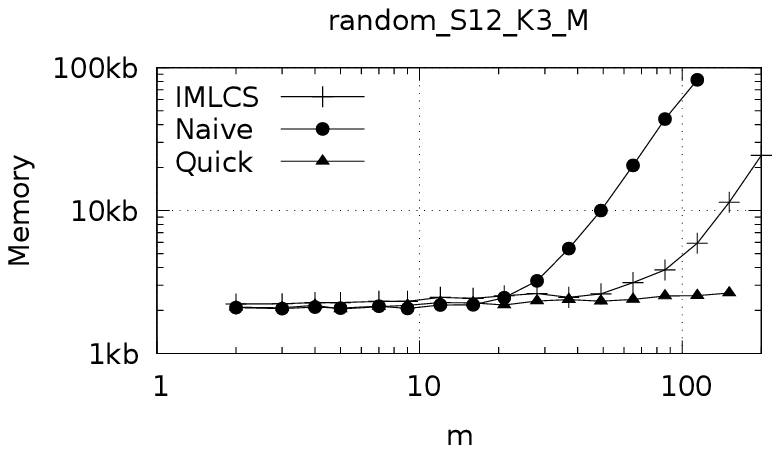}
\includegraphics{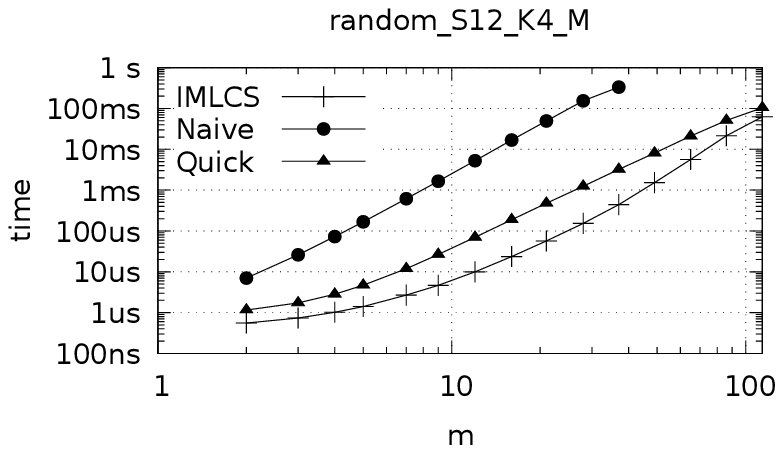}
\includegraphics{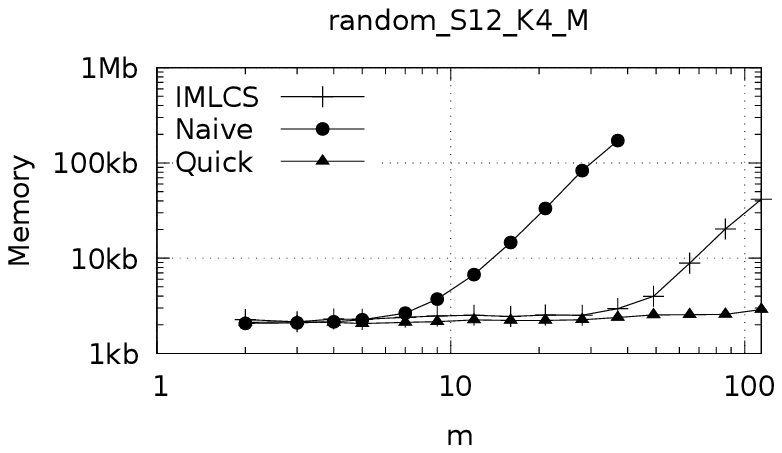}
\includegraphics{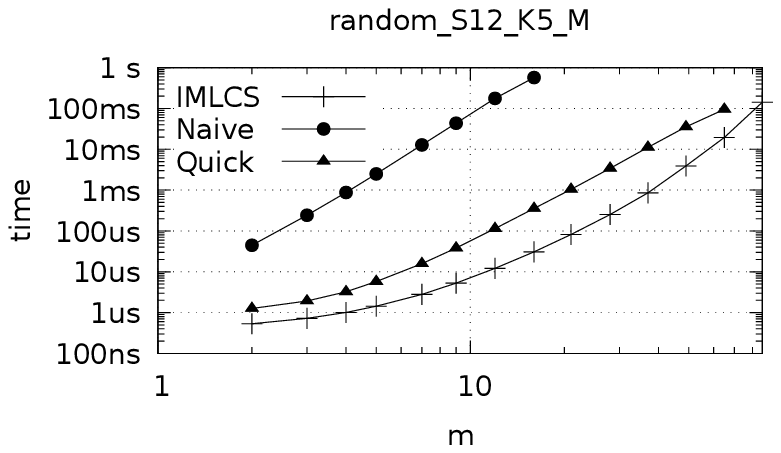}
\includegraphics{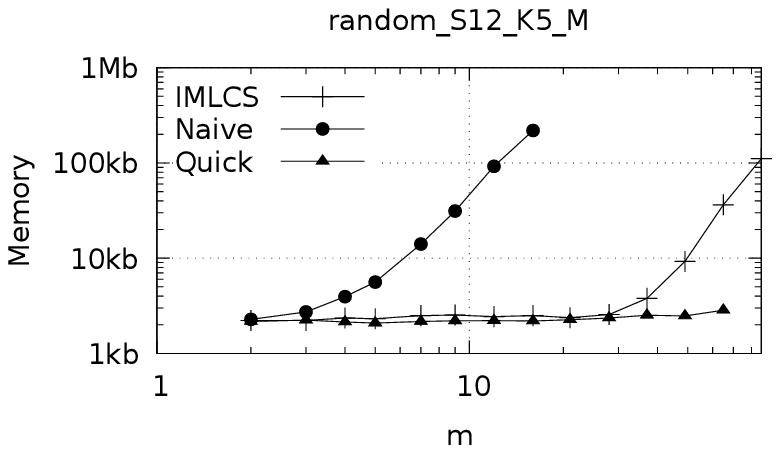}
\includegraphics{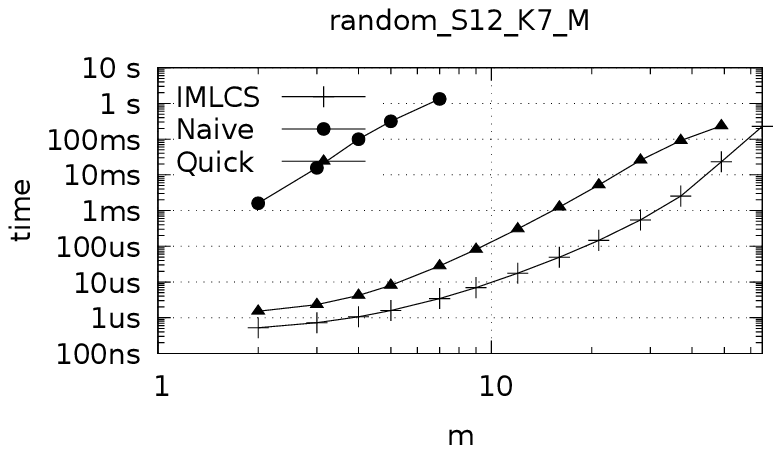}
\includegraphics{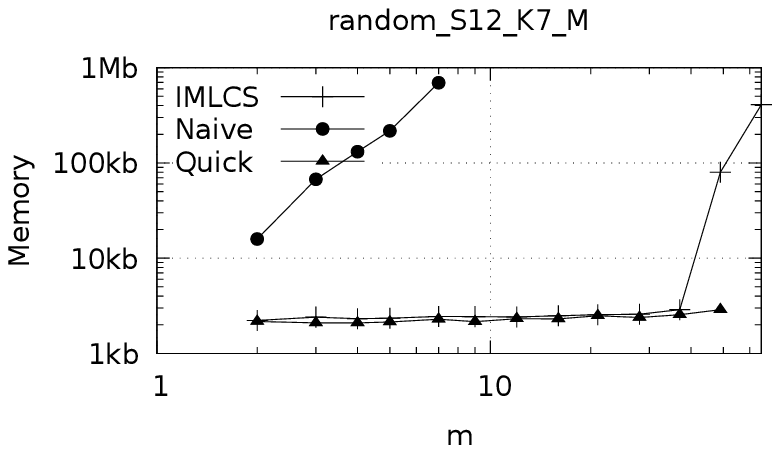}
\includegraphics{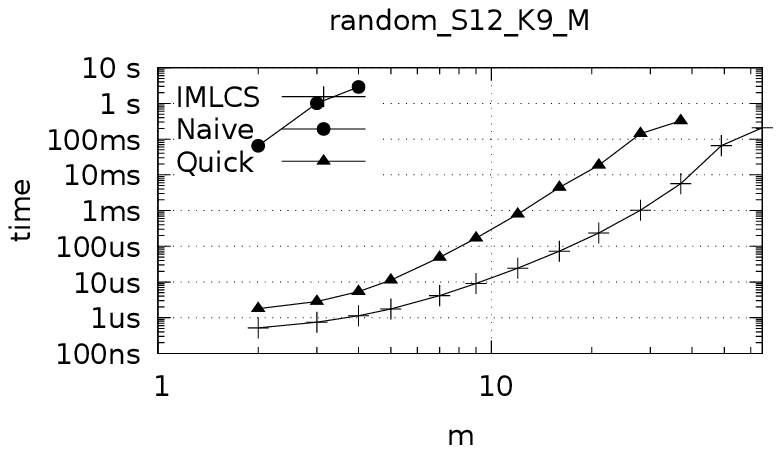}
\includegraphics{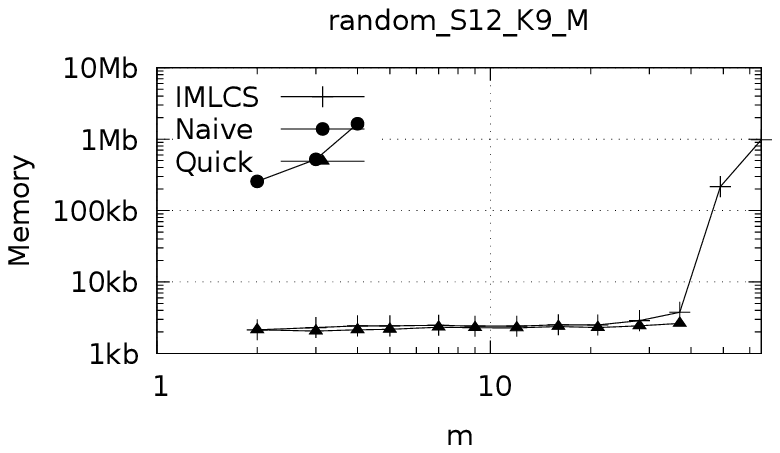}
\includegraphics{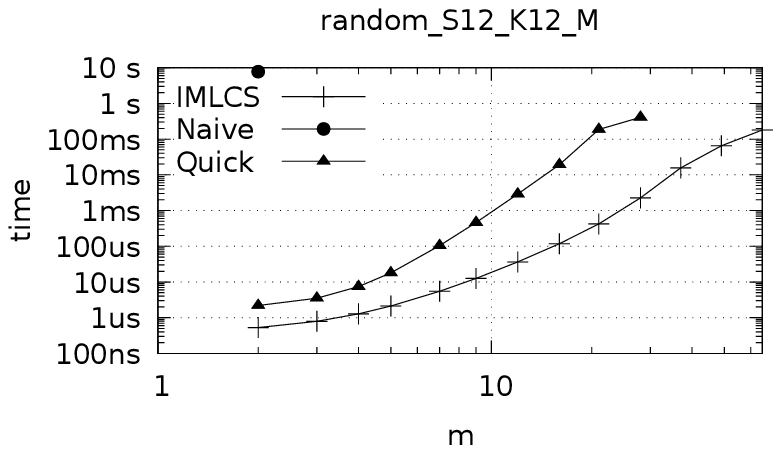}
\includegraphics{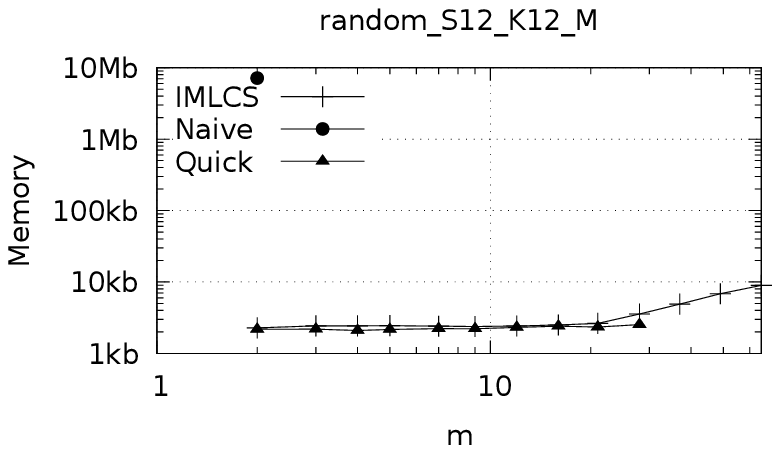}
\includegraphics{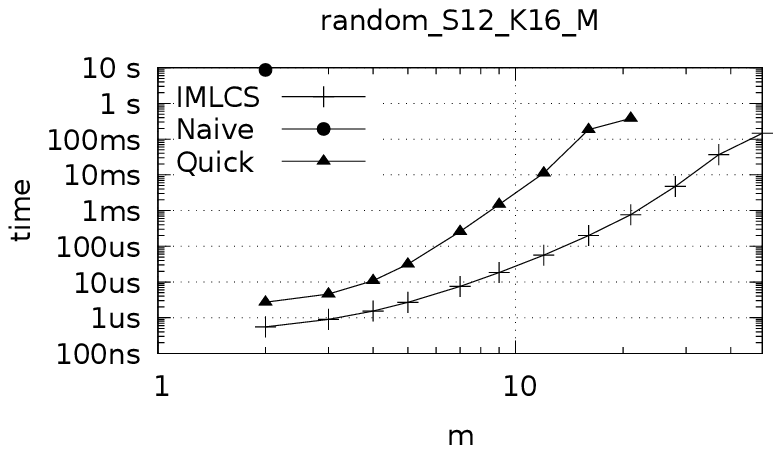}
\includegraphics{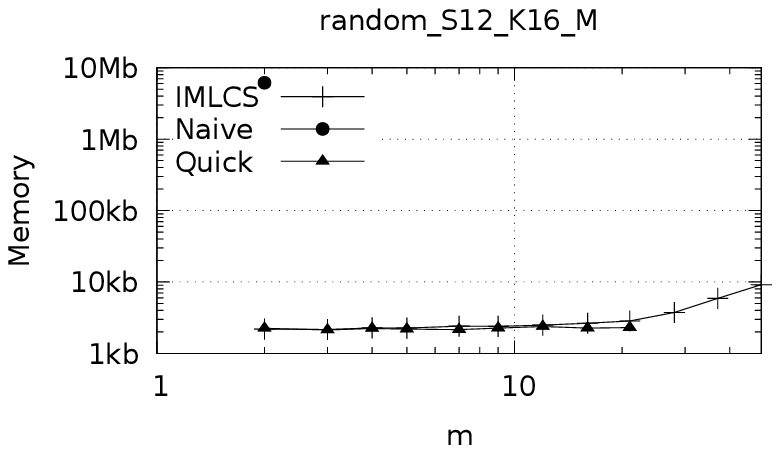}
\includegraphics{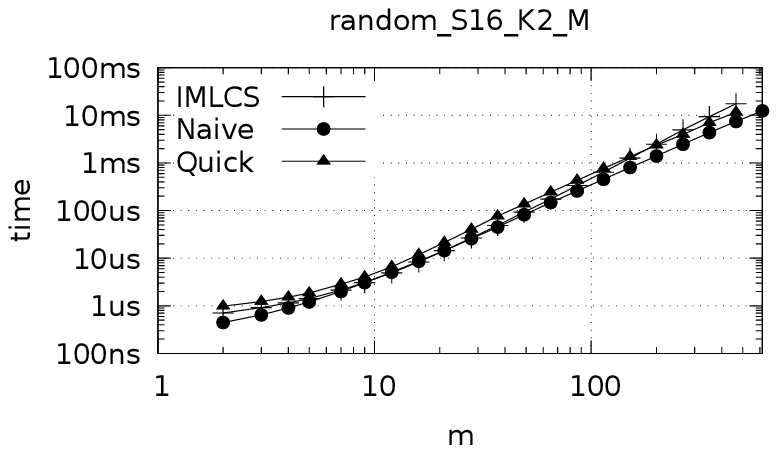}
\includegraphics{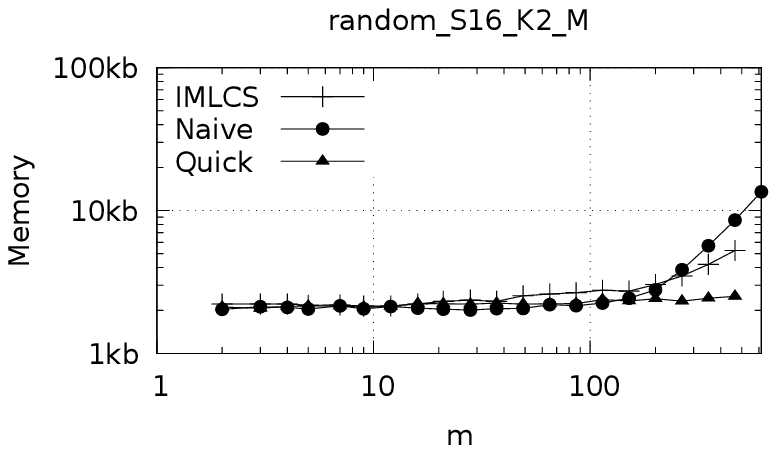}
\includegraphics{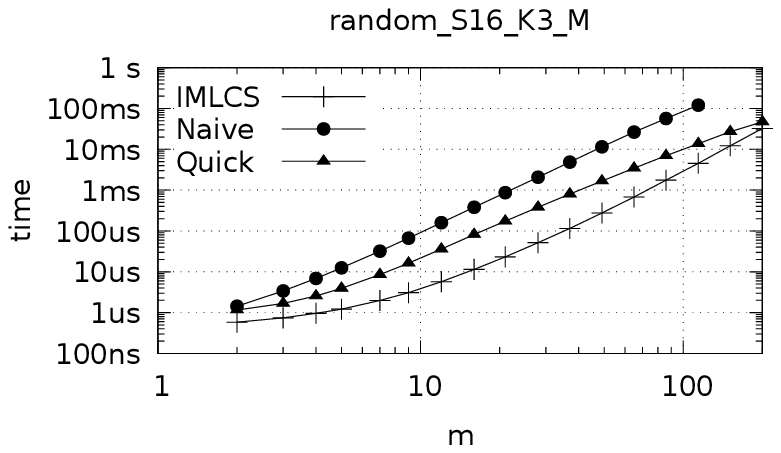}
\includegraphics{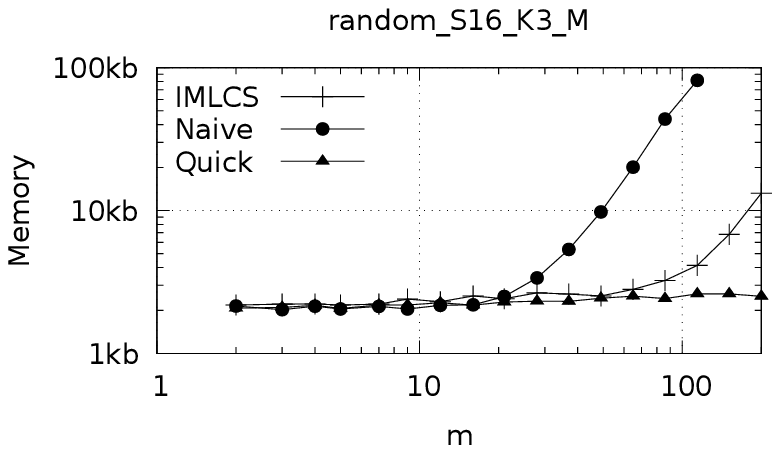}
\includegraphics{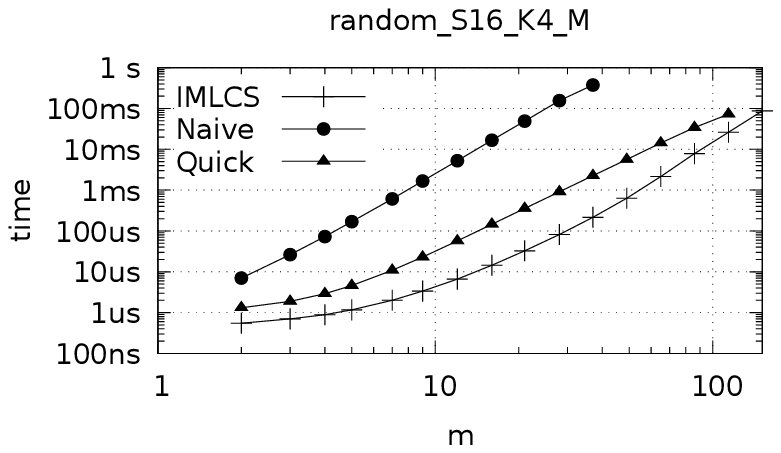}
\includegraphics{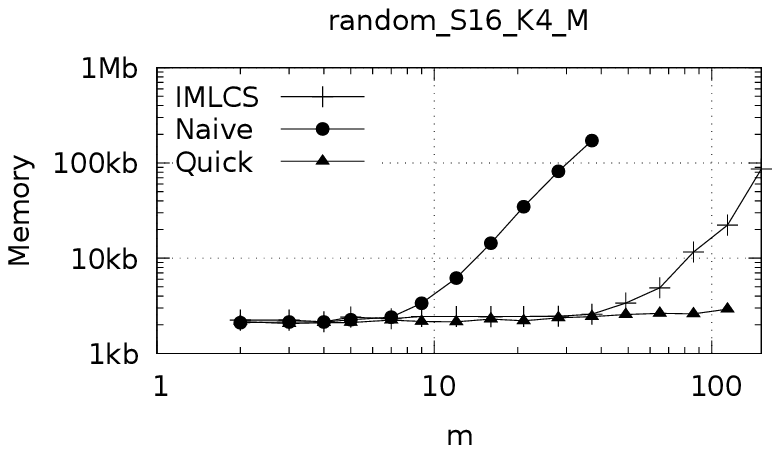}
\includegraphics{random_S16_K5M.eps}
\includegraphics{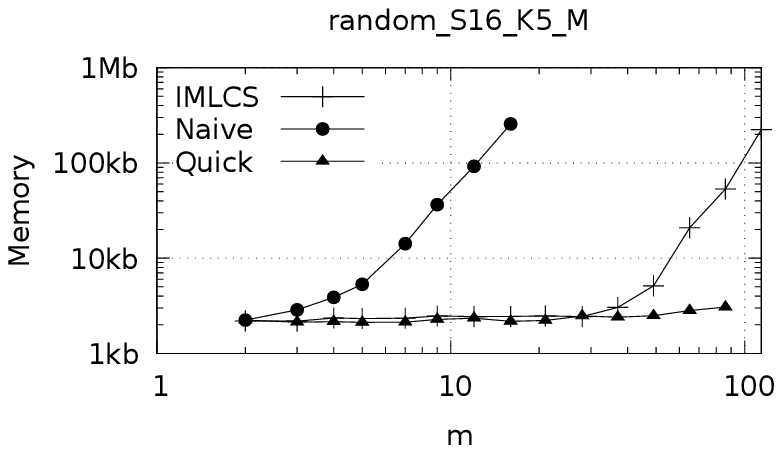}
\includegraphics{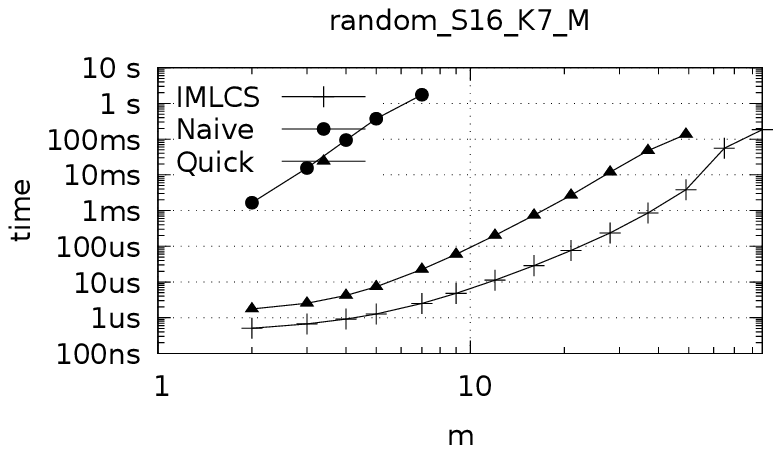}
\includegraphics{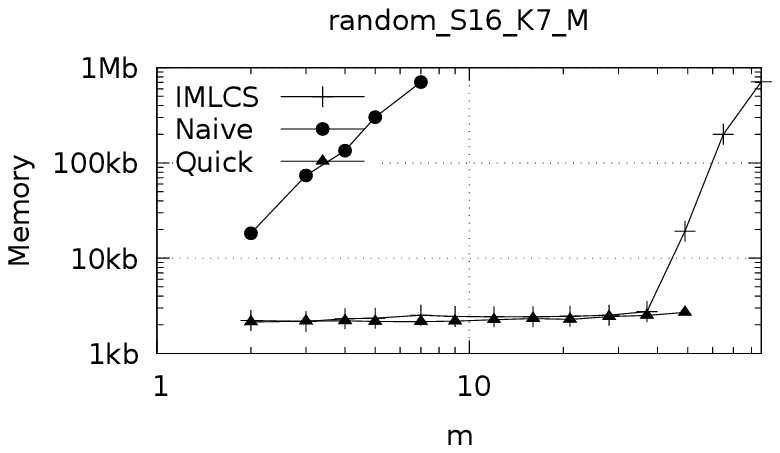}
\includegraphics{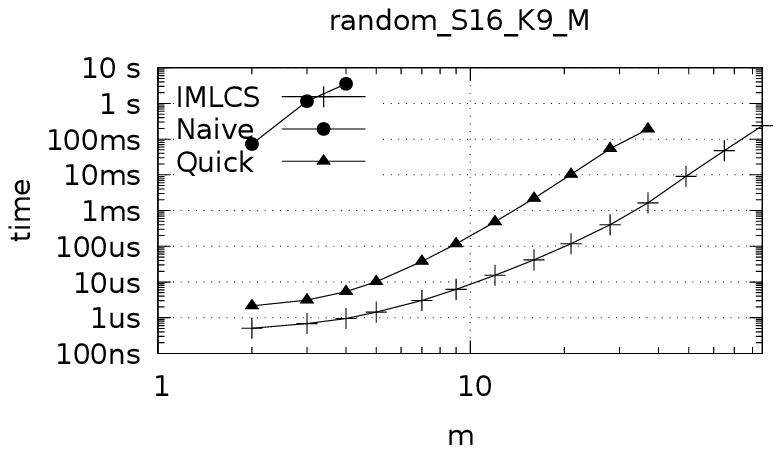}
\includegraphics{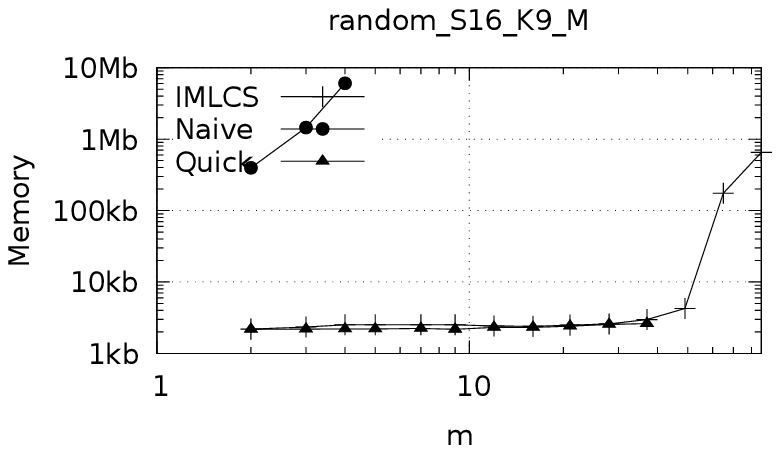}
\includegraphics{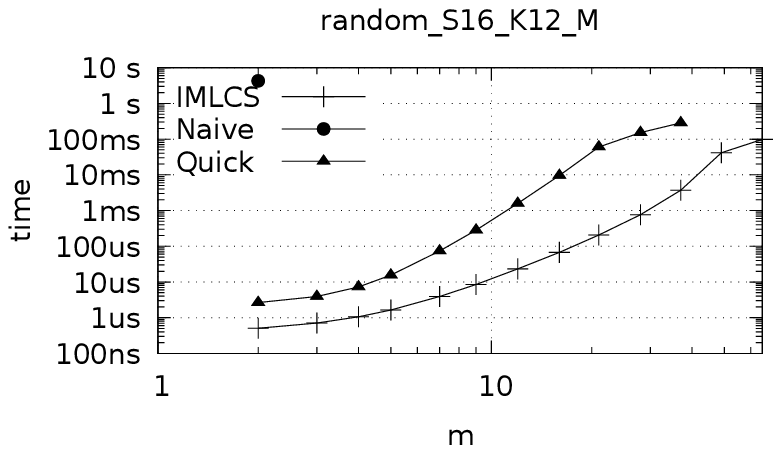}
\includegraphics{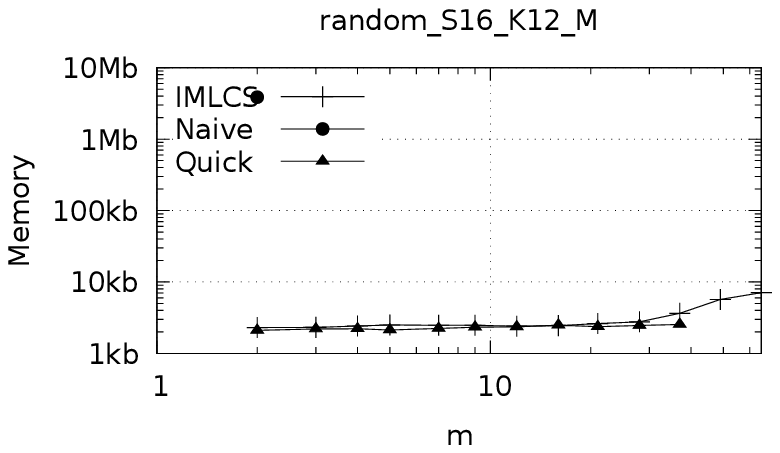}
\includegraphics{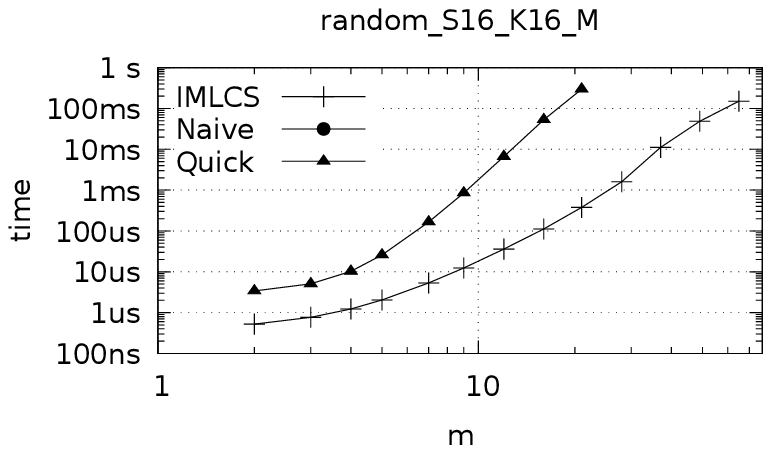}
\includegraphics{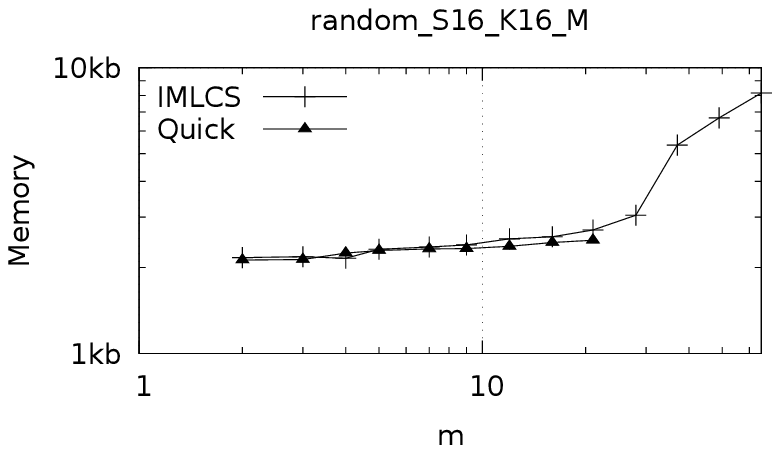}

\includegraphics{proteins_S21_K2.eps}
\includegraphics{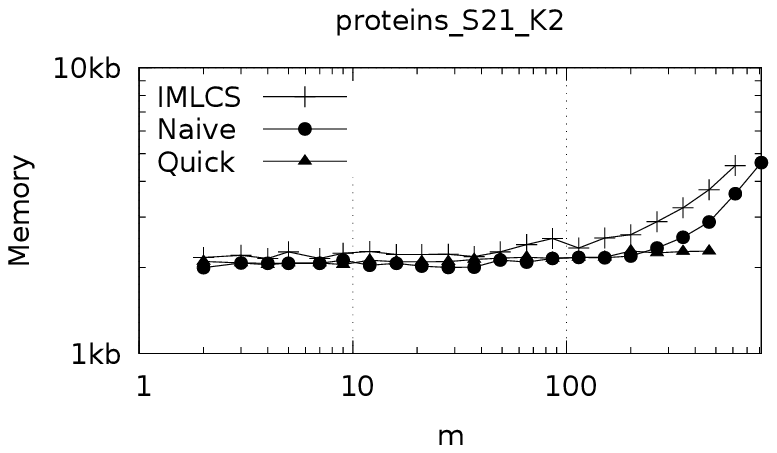}
\includegraphics{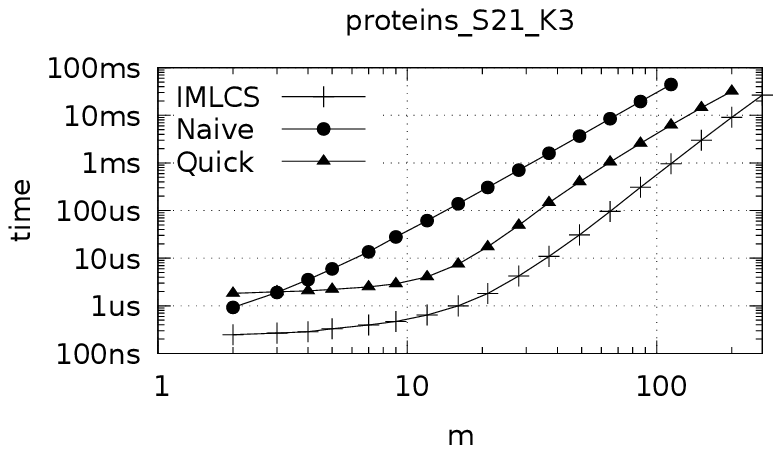}
\includegraphics{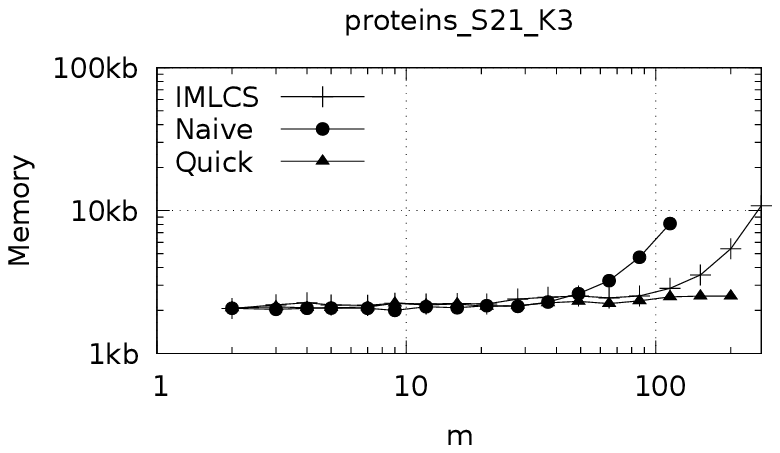}
\includegraphics{proteins_S21_K4.eps}
\includegraphics{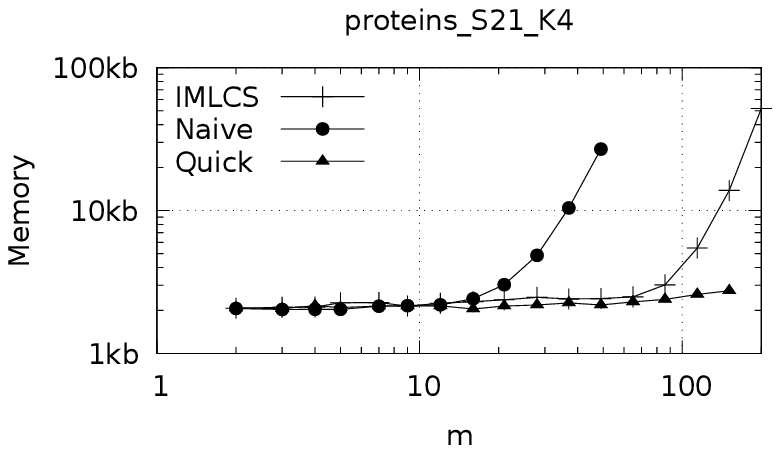}
\includegraphics{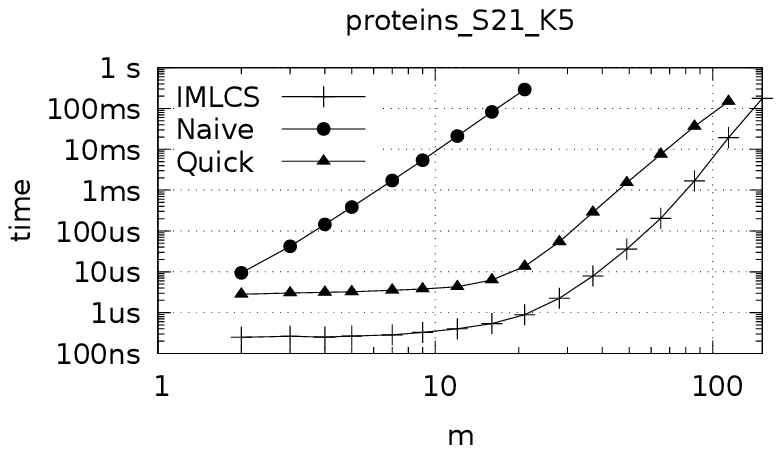}
\includegraphics{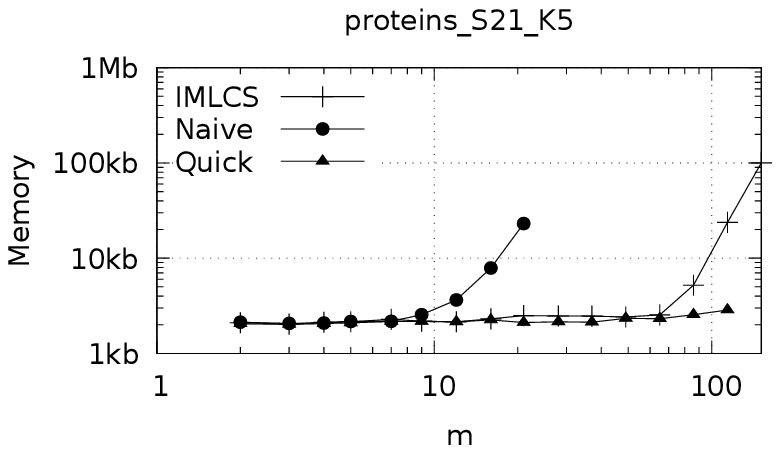}
\includegraphics{proteins_S21_K7.eps}
\includegraphics{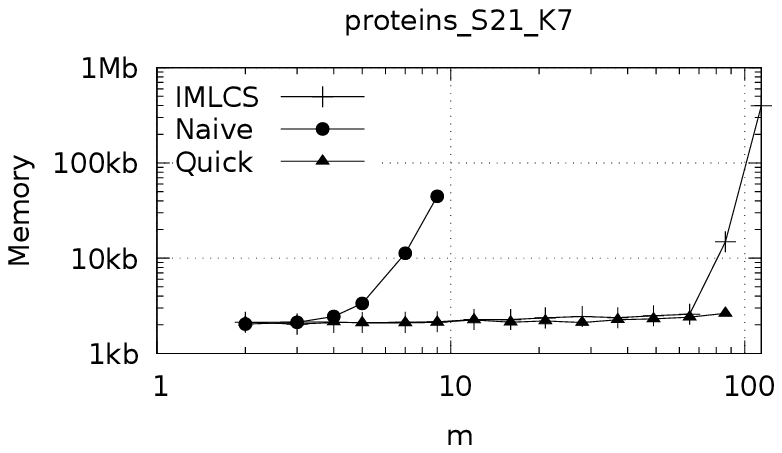}

\ifCLASSOPTIONcompsoc
  \section*{Acknowledgments}
\else
  \section*{Acknowledgment}
\fi
We are grateful to Hideo Bannai for interesting discussions on this topic, at
StringMasters, Lisbon 2018.

The work reported in this article was supported by national funds through
Funda\c{c}\~ao para a Ci\^encia e Tecnologia (FCT) through projects NGPHYLO
PTDC/CCI-BIO/29676/2017 and UID/CEC/50021/2019. Funded in part by European
Union’s Horizon 2020 research and innovation programme under the Marie
Sk{\l}odowska-Curie Actions grant agreement No 690941.

\ifCLASSOPTIONcaptionsoff
  \newpage
\fi

\bibliographystyle{plainnat.bst}
\bibliography{IEEEabrv,journal}

\begin{IEEEbiography}[{\includegraphics[width=1in,height=1.25in,clip,keepaspectratio]{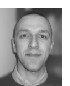}}]{Lu\'{i}s M. S. Russo}
  Luís M. S. Russo received a Ph.D. degree from Instituto Superior
  Técnico, Lisbon, Portugal, in 2007. He is currently an Assistant
  Professor with Instituto Superior Técnico. His current research interests
  include algorithms and data structures for string processing and
  optimization.
\end{IEEEbiography}

\begin{IEEEbiography}[{\includegraphics[width=1in,height=1.25in,clip,keepaspectratio]{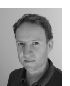}}]{Alexandre
  P. Francisco}
Alexandre P. Francisco has a Ph.D. degree in Computer Science and
Engineering and he is currently an Assistant Professor at the CSE Dept,
IST, University of Lisbon. His current research interests include the design
and analysis of data structures and algorithms, with applications on
network mining and large data processing.
\end{IEEEbiography}

\begin{IEEEbiography}[{\includegraphics[width=1in,height=1.25in,clip,keepaspectratio]{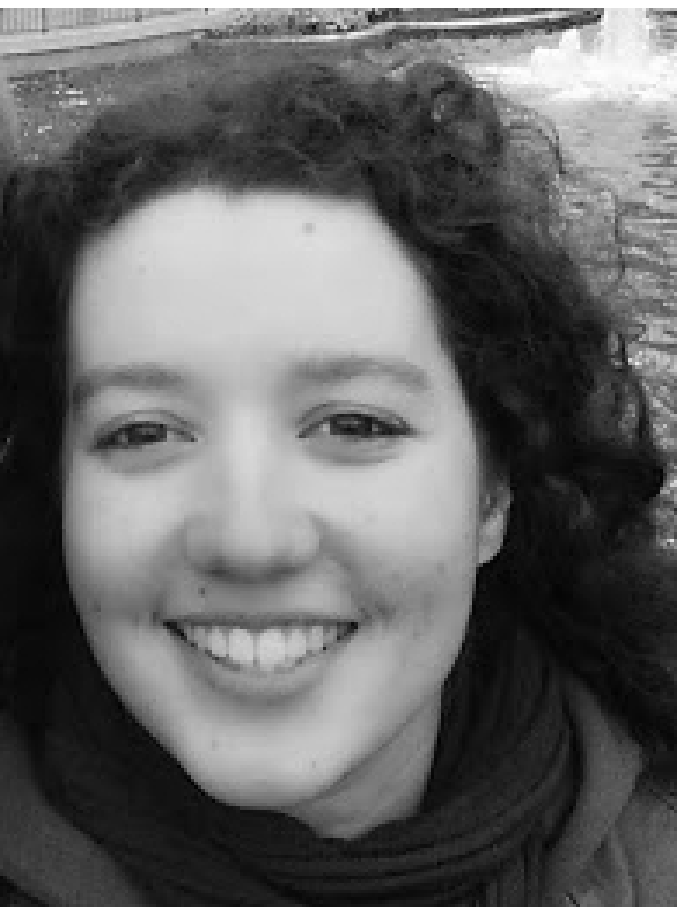}}]{Tatiana Rocher}
  Tatiana Rocher received a Ph.D. in computer science in 2018 in Lille,
  France. She currently has a post-doctoral position in INESC-ID in
  Lisbon. Her research interests include string algorithms and data
  structures.
\end{IEEEbiography}

\end{document}